\documentclass[twocolumn,prb,showpacs,preprintnumbers,superscriptaddress,amsmath,amssymb,citeautoscript]{revtex4-1}
\usepackage{graphicx}
\usepackage{dcolumn}
\usepackage{color}
\usepackage{bm}
\usepackage{epstopdf}

\newcommand{\bpm}{\begin{pmatrix}}
\newcommand{\epm}{\end{pmatrix}}

\DeclareMathOperator{\sgn}{sgn}

\newcommand{\be}{\begin{equation}}
\newcommand{\ee}{\end{equation}}
\newcommand{\ba}{\begin{eqnarray}}
\newcommand{\ea}{\end{eqnarray}}
\newcommand{\baa}{\begin{eqnarray*}}
\newcommand{\eaa}{\end{eqnarray*}}

\newcommand{\bs}{\boldsymbol}

\begin{document}

\title{Determining the spin-orbit coupling via spin-polarized spectroscopy of magnetic impurities}

\author{V. Kaladzhyan}
\email{vardan.kaladzhyan@cea.fr}
\affiliation{Institut de Physique Th\'eorique, CEA/Saclay,
Orme des Merisiers, 91190 Gif-sur-Yvette Cedex, France}
\affiliation{Laboratoire de Physique des Solides, CNRS, Univ. Paris-Sud, Universit\'e Paris-Saclay, 91405 Orsay Cedex, France}
\author{P. Simon}
\affiliation{Laboratoire de Physique des Solides, CNRS, Univ. Paris-Sud, Universit\'e Paris-Saclay, 91405 Orsay Cedex, France}
\author{C. Bena}
\affiliation{Institut de Physique Th\'eorique, CEA/Saclay,
Orme des Merisiers, 91190 Gif-sur-Yvette Cedex, France}
\affiliation{Laboratoire de Physique des Solides, CNRS, Univ. Paris-Sud, Universit\'e Paris-Saclay, 91405 Orsay Cedex, France}

\date{\today}

\begin{abstract}
We study the spin-resolved spectral properties of the impurity states associated to the presence of magnetic impurities in two-dimensional, as well as one-dimensional systems with Rashba spin-orbit coupling. We focus on Shiba bound states in superconducting materials, as well as on impurity states in metallic systems.
Using a combination of a numerical T-matrix approximation and a direct analytical calculation of the bound state wave function, we compute the local density of states  (LDOS) together with its Fourier transform (FT).
We find that the FT of the spin-polarized LDOS, a quantity accessible via spin-polarized STM, allows to accurately extract  the strength of the spin-orbit coupling. Also we confirm that the presence of magnetic impurities is strictly necessary for such measurement, and that non-spin-polarized experiments cannot have access to the value of the spin-orbit coupling.

\end{abstract}


\maketitle
\section{Introduction}
The electronic bands of materials that lack an inversion center are split by the spin-orbit (SO) coupling. A strong SO coupling implies that the spin of the electron is tied to to the direction of its momentum. 
Materials with strong SO coupling have been receiving a considerable attention in the past decade partly because 
SO is playing an important role for the  discovery of new topological classes of materials \cite{Hasan2010,Qi2011}.
Two-dimensional  topological insulators, first predicted   in graphene \cite{Kane2005}, have been discovered in HgTe/CdTe heterostructures \cite{Konig2007} following a theoretical prediction by Bernevig {\it et al.} \cite{Bernevig2006}. They are characterized by one-dimensional helical 
edge states where the spin is locked to the direction of propagation due to the strong SO coupling.
Similar features occur for the surface states of  3D topological insulators which also have a strong bulk SO coupling \cite{Hasan2010}. The spin-to-momentum locking was directly  observed by angle-resolved photoemission spectroscopy (ARPES) experiments \cite{Hsieh2009a,Hsieh2009b}.

Topological superconductors share many  properties with topological insulators. They possess exotic edge states called Majorana fermions, particles which are their own antiparticles \cite{Hasan2010}. Topological superconductivity can be either induced by the proximity with a standard s-wave superconductor or be intrinsic. 
In the former case, Majorana states have thus been proposed to form in one-dimensional \cite{Oreg2010,Lutchyn2010} and two-dimensional semiconductors \cite{Sau2010,Alicea2010} with strong SO coupling when proximitized with a s-wave superconductor, and in the presence of a Zeeman  field. 
Following this strategy, many experiments have reported signatures of Majorana fermions through transport spectroscopy in one dimensional topological  wires \cite{Mourik2012,Deng2012,Das2012,Marcus2013,Marcus2016}. However, there are presently only a few material candidates such as strontium ruthenate \cite{Maeno2003}, certain heavy fermion superconductors 
\cite{Taillefer2002}, or some doped topological insulators such as 
$\rm Cu_xBi_2Se_3$ \cite{Sasaki2011}, that may host intrinsic topological superconductivity. 

Although SO coupling has been playing an essential role in the discovery of  new topological materials, it is also of crucial importance in the physics of spin Hall effect \cite{Sinova2015}, in spintronics \cite{Zutic2004} and quantum (spin) computation since it allows to electrically detect and manipulate spin currents  in confined nanostructures (see Ref. \onlinecite{Manchon2015} for a recent review).

Based on the prominent role played by SO in the past decades, it is thus of great interest to be able to evaluate the SO coupling value in a given material accurately, though in general this is a very difficult task. Inferences can be made from ARPES measurements \cite{LaShell1996,Ast2007,Varykhalov2012}; in particular spin-polarized ARPES measurements have been used to evaluate the SO coupling in  various materials \cite{Hoesch2002,Hochstrasser2002,Hoesch2004,Hirahara2007,Hsieh2009,Yaji2010,Jozwiak2011}. Other possibilities involve magneto-transport measurements in confined nanostructures: this technique has been used to measure the SO coupling in clean carbon nanotubes \cite{Kuemmeth2008} or in InAs nanowires \cite{Fasth2007}.

Here we propose  a method to measure the magnitude of the SO coupling directly using spin-polarized scanning tunnelling microscopy (STM) \cite{Wiesendanger2009}, and the Fourier transform (FT) of the local density of states (LDOS) near magnetic impurities (FT-STM). The FT-STM technique has been used in the past in metals, where it helped in mapping the band structure and the shape and the properties of the Fermi surface \cite{Hasegawa1993,Crommie1993,Sprunger1997,Petersen2000,Hofmann1997,Vonau2004,Vonau2005,Simon2011}, as well as in extracting information about the spin properties of the quasiparticles \cite{Pascual2004}. More spectacularly, it was used successfully in high-temperature SCs to map with high resolution the particular d-wave structure of the Fermi surface, as well as to investigate the properties of the pseudogap \cite{Hoffman2002,McElroy2003,Vershinin2004}. 

In this paper, on one hand, we calculate the Fourier transform of the spin-polarized  local density of states (SP LDOS) of the so-called Shiba bound state \cite{Yu1965,Shiba1968,Rusinov1969,Balatsky2006} associated with a magnetic impurity in a superconductor. Shiba bound states have been measured experimentally by STM \cite{Yazdani1997,Shuai-Hua2008,Franke2015} and it has actually been shown that the extent of the Shiba wave function can reach tens of nanometers
in 2D superconductors, which allows one to measure the spatial dependence of the LDOS of such states with high resolution \cite{Menard2015}.
We consider both one-dimensional and two-dimensional superconductors with SO coupling.  While two-dimensional systems  such as e.g. 
Sr$_2$RuO$_4$,\cite{Maeno2003}  or NbSe$_2$ \cite{Ugeda2015,Menard2015}  become superconducting when brought at low temperature, one-dimensional wires such as InAs and InSb are not superconducting at low temperature.  In order to see the formation of Shiba states one would need to proximitize them by a SC substrate. The formation of Shiba states in such systems \cite{Brydon2015,Lutchyn2015}, as well as in p-wave superconductors \cite{Kaladzhyan2015,Kaladzhyan2016}, has been recently touched upon, but the effect of the SO coupling on the FT of the SP LDOS in the presence of magnetic impurities has not previously been analyzed. 

On the other hand we focus on the effects of the spin-orbit coupling on the impurity states of a classical magnetic impurity in one-dimensional and two-dimensional metallic systems such as Pb\cite{Brun2014} and Bi, as well as InAs and InSb semiconducting wires that can be also modeled as metals in the energy range that we consider. We should note that for these systems no bound state forms at a specific energy, but the impurity is affecting equally the entire energy spectrum. 

By studying the two classes of systems described above we show that the SO coupling can directly be read-off from the FT features of the SP LDOS in the vicinity of the magnetic impurity. We note that such a signature appears only for magnetic impurities, and only when the system is investigated using spin-polarized STM measurements, the non-spin-polarized measurements do not provide information on the SO, as it has also been previously noted \cite{Petersen1999}. The main difference between the SC and metallic systems, beyond the existence of a bound state in the former case, is that the spin-polarized Friedel oscillations around the impurity have additional features in the SC phase, the most important one being the existence of oscillations with a wavelength exactly equal to the SO coupling length scale; such oscillations are not present in the metallic phase.
Another difference is the broadening of the FT features in the superconducting phase compared to the non-SC phase in which the sole broadening is due to the quasiparticle lifetime.

We focus on Rashba SO coupling as assumed to be the most relevant for the systems considered, but we have checked that our conclusion holds for other types of SO. To obtain the SP LDOS we use a T-matrix approximation\cite{Simon2011,Mahan2000,Bruus2004}, and we present both numerical and analytical results which allow us to obtain a full understanding of the observed features, of the splittings due to the SO, as well as of the spin-polarization of the impurity states and of the symmetry of the FT features.

In Sec. II, we present the general model for two-dimensional and one-dimensional cases and the basics of the T-matrix technique. In Sec. III we show our results for the SP LDOS, calculated both numerically and analytically, for 2D systems,  both in the SC and metallic phase. Sec. IV is devoted to SP LDOS of impurity in one-dimensional systems.
Our Conclusions are presented in section V. Details of the analytical calculations are given in the Appendices.

\section{Model} We consider an s-wave superconductor  with a SC paring $\Delta_s$, and Rashba SO coupling $\lambda$, for which the Hamiltonian, written in the Nambu basis $\Psi_{\bm p}=(\psi_{\uparrow {\bm p}},\psi_{\downarrow {\bm p}},\psi^{\dag}_{\downarrow{-\bm p}},-\psi^{\dag}_{\uparrow {-\bm p}})^T$, is given by:
\begin{equation}
\mathcal{H}_{0} = 
	\begin{pmatrix}
		\xi_{\bm p} \bm{\sigma}_0 & \Delta_s \bm{\sigma}_0 \\
		\Delta_s \bm{\sigma}_0 & -\xi_{\bm p} \bm{\sigma}_0.
	\end{pmatrix} +
	\mathcal{H}_{SO}.
\label{H0}
\end{equation}
The energy spectrum is $\xi_{\bm p} \equiv \frac{\bm{p}^2}{2m}-\varepsilon_F$, 
where $\varepsilon_F$ is the Fermi energy.
The operator $\psi^\dag_{\sigma {\bm p}}$ creates a particle of spin $\sigma=\uparrow,\downarrow$ of momentum ${\bm p} \equiv (p_x,p_y)$ in 2D and ${\bm p} \equiv p_x$ in 1D. Below we set $\hbar$ to unity. The system is considered to lay in the $(x,y)$ plane in 2D case, whereas in 1D case we set $p_y$ to zero in the expressions above, and we consider a system lying along the $x$-axis. The metallic limit is recovered by setting $\Delta_s=0$.
The Rashba Hamiltonian can be written as
\be
\mathcal{H}_{SO} =  \lambda \left( p_y {\bm \sigma}_x - p_x {\bm\sigma}_y \right) \otimes \bm{\tau}_z,
\label{HSO}
\ee
in 2D and simply as $\mathcal{H}_{SO} =  \lambda p_x {\bm\sigma}_y  \otimes \bm{\tau}_z$ in 1D. We have introduced
${\bm\sigma}$ and ${\bm\tau}$, the Pauli matrices acting respectively in the spin and the particle-hole subspaces.
The unperturbed retarded Green's function can be obtained from the above Hamiltonian via $G_0(E,{\bm p})=\left[(E+i\delta)\mathbb{I}_4-\mathcal{H}_0({\bm p})\right]^{-1}$, where $\delta$ is the inverse quasiparticle lifetime. 

In what follows we study what happens when a single localized impurity is introduced in  this system. 
We consider magnetic impurities of spin ${\bm J} = (J_x,J_y,J_z)$ described by the following Hamiltonian:
\begin{equation}
\mathcal{H}_{imp} =  \bm{J} \cdot \bm{\sigma} \otimes \bm{\tau}_0 \cdot \delta(\bm{r}) \;\equiv V \cdot \delta(\bm{r}),
\label{hm}
\end{equation}
where ${\bm J}$ is the magnetic strength.  We only consider here classical impurities oriented 
either along the $z$-axis, ${\bm J} = (0,0,J_z)$, or along the $x$-axis, ${\bm J} = (J_x,0,0)$. This is justified provided the Kondo temperature is much smaller than the superconducting gap \cite{Balatsky2006}.

To find the impurity states in the model described above we use the T-matrix approximation described in [\onlinecite{Mahan2000,Bruus2004,Balatsky2006}] and [\onlinecite{Simon2011}]. We also neglect the  renormalization of the superconducting gap because it is mainly local\cite{Balatsky2006,Meng2015} and therefore only introduces minor effects for our purposes. Since the impurity is localized, the T-matrix is given by:
\be
T(E) = \left[1-V \int \frac{d^2\mathbf{p}}{(2\pi)^2} G_0(E,\mathbf{p})  \right]^{-1} V.
\label{tm0}
\ee

The real-space dependence of the non-polarized, $\delta\rho({\bm r},E)$, and SP LDOS, $S_{\hat n}({\bm r},E)$, with $\hat{n}=x,y,z$, can be found as
\ba
S_x({\bm r},E) &= -\frac{1}{\pi} \Im \left[  \Delta G_{12} + \Delta G_{21}\right], \nonumber\\
S_y({\bm r},E) &= -\frac{1}{\pi} \Re \left[ \Delta G_{12} -  \Delta G_{21}\right], \nonumber\\
S_z({\bm r},E) &= -\frac{1}{\pi} \Im \left[ \Delta G_{11} - \Delta G_{22}\right], \nonumber\\
\delta\rho({\bm r},E) &= -\frac{1}{\pi} \Im \left[ \Delta G_{11} + \Delta G_{22}\right], \nonumber
\ea
with
\begin{align*}
\Delta G (E, \mathbf{r}) \equiv G_0(E,-\mathbf{r}) T(E) G_0(E, \mathbf{r}), 
\end{align*}
where $\Delta G_{ij}$ denotes the $ij$-th component of the matrix $\Delta G$, and $G_0 (E, \mathbf{r})$ is the unperturbed retarded Green's function in real space, given by the Fourier transform
\begin{align}
G_0 (E, \mathbf{r})= \int \frac{d \bm p}{(2\pi)^2} G_0 (E, \mathbf{p}) e^{i \bm{pr}}.
\label{FTdef}
\end{align}

The FT of the SP LDOS components in momentum space, $S_{\hat n}({\bm p},E) = \int d \bm r \, S_{\hat n}({\bm r},E) e^{-i \bm{pr}}$, with $\hat{n}=x,y,z$, as well as the FT of the non-polarized LDOS, $\delta\rho({\bm p},E) = \int d \bm r \, \delta\rho({\bm r},E) e^{-i \bm{pr}} $ are given by 
\ba
S_x({\bm p},E) &=& \frac{i}{2 \pi} \negthickspace \int \negthickspace \frac{d\mathbf{q}}{(2\pi)^2} [\tilde{g}_{12}(E,\mathbf{q},\mathbf{p})+\tilde{g}_{21}(E,\mathbf{q},\mathbf{p})],\\
S_y({\bm p},E) &=& \frac{1}{2\pi} \negthickspace\int \negthickspace\frac{d\mathbf{q}}{(2\pi)^2} [g_{21}(E,\mathbf{q},\mathbf{p})-g_{12}(E,\mathbf{q},\mathbf{p})],
\\
S_z({\bm p},E) &=& \frac{i}{2 \pi} \negthickspace\int\negthickspace \frac{d\mathbf{q}}{(2\pi)^2} [\tilde{g}_{11}(E,\mathbf{q},\mathbf{p})-\tilde{g}_{22}(E,\mathbf{q},\mathbf{p})],\\
\delta\rho({\bm p},E)&=& \frac{i}{2 \pi} \negthickspace\int \negthickspace\frac{d\mathbf{q}}{(2\pi)^2} [\tilde{g}_{11}(E,\mathbf{q},\mathbf{p})+\tilde{g}_{22}(E,\mathbf{q},\mathbf{p})],
\label{tm}
\ea
where $d\bm q \equiv dq_x dq_y$, 
\ba
g(E,\mathbf{q},\mathbf{p}) &=& G_0 (E,\mathbf{q}) T(E) G_0(E, \mathbf{p+q}) \nonumber\\
&+& G^*_0(E, \mathbf{p+q}) T^*(E) G^*_0 (E,\mathbf{q}), \nonumber\\
\tilde{g}(E,\mathbf{q},\mathbf{p}) &=& G_0 (E,\mathbf{q}) T(E) G_0(E, \mathbf{p+q}) \nonumber\\
&-& G^*_0(E, \mathbf{p+q}) T^*(E) G^*_0 (E,\mathbf{q}), \nonumber
\ea 
and $g_{ij}$, $\tilde{g}_{ij}$ denote the corresponding components of the matrices $g$ and $\tilde{g}$.
Note that while the non-polarized and the SP LDOS are of course real functions when evaluated in position space, their Fourier transforms need not be. Sometimes we  get either or both real and imaginary components for the FT, depending on their corresponding symmetries. In the figures we shall indicate each time if we plot the real or the imaginary component of the FT.

To obtain the FT of the non-polarized and the SP LDOS, we first evaluate the momentum integrals in Eqs.~(\ref{tm0}-\ref{tm}) numerically.
For this we  use a square lattice version of the Hamiltonians (\ref{H0}) and (\ref{HSO}), where we take 
the tight-binding spectrum $\Xi_{\bm p} \equiv \mu -2 t (\cos p_x + \cos p_y)$ with chemical potential $\mu$ and hopping parameter $t$. 
We set the lattice constant  to unity. It is also worth noting that all the numerical integrations are performed over the first Brillouin zone and that we use dimensionless units by setting $t=1$.

Alternatively, as detailed in the appendices, we find the exact form for the non-polarized and SP LDOS in the continuum limit by performing the integrals in the FT of the Green's functions analytically. 
Moreover, when considering the SC systems, the energies $E$ of the Shiba states together with the corresponding eigenstates for the Shiba wave functions $\Phi$ at the origin 
can be obtained from the corresponding eigenvalue equation\cite{Pientka2013} 
\begin{equation}
	\left[\mathbb{I}_4-V G_0(E,{\bm r}={\bm 0}) \right]\Phi(\bm 0) = 0.
\label{eigenv}
\end{equation}
The spatial dependence of the Shiba state wave function is determined using
\begin{equation}
	\Phi (\bm r) = G_0(E, \bm r) V \Phi(\bm 0).
\label{eigenf}
\end{equation}
The real-space Green's function is obtained simply by a Fourier transform of  the unperturbed Green's function in momentum space, $G_0(E,{\bm p})$. The non-polarized and the SP LDOS are given  by
\begin{equation}
	\rho(E, \bm r) = \Phi^\dag(\bm r) 
		\begin{pmatrix} 	
			0 & 0 \\ 
			0 & \sigma_0
		\end{pmatrix} \Phi (\bm r),
\label{LDOS}
\end{equation}
and
\begin{equation}
	\bm S(E, \bm r) = \Phi^\dag(\bm r) 
		\begin{pmatrix} 	
			0 & 0 \\ 
			0 & \bm \sigma
		\end{pmatrix} \Phi (\bm r),
\label{SPLDOS}
\end{equation}
where we take into account only the hole components of the wave function, and not the electron ones. This is because the physical observables are related to only one of the two components, for example in a STM measurement one injects an electron at a given energy and thus have access to the allowed number of electronic states, not to both the electronic and hole states simultaneously. The Bogoliubov-de Gennes Hamiltonian contains the so-called particle-hole redundancy, and the electron and the hole components can be simply recovered from each other by overall changes of sign, and/or changing the sign of the energy. Below we compute only the hole components, but there would have been no qualitative differenced had we computed the electron component.

\section{Results for two dimensional systems}  
\subsection{Real and momentum space dependence of the 2D Shiba bound states}
\begin{figure}[h]
\includegraphics*[width=0.7\columnwidth]{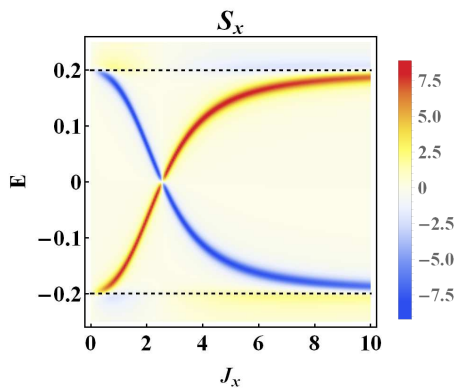}
\caption{(Color) The averaged SP LDOS induced by an impurity as a function of the impurity strength for an in-plane magnetic impurity. The dashed line shows the superconducting gap. A similar result is obtained when the impurity spin is perpendicular to  the plane. Note that the two Shiba states with opposite energies have opposite spin. We set $t=1, \mu=3, \delta=0.01, \lambda=0.5, \Delta_s=0.2$. }
\label{figSPDOS}
\end{figure}

For a 2D superconductor with SO coupling in the presence of a magnetic impurity one expects the formation of a single pair of Shiba states \cite{Brydon2015,Lutchyn2015}. The energies of the particle-hole symmetric Shiba states\footnote{In this paper, we use the plural when refering to Shiba states in order to facilitate the discussion. However, it should be kept in mind that for a given localized magnetic impurity, there is a single Shiba state with particle and hole components whose  real space wave function can actually behave differently.} are given by (independent of the direction of the impurity):
\begin{eqnarray*}
E_{1,\bar 1} &=& \pm \frac{1-\alpha^2}{1+\alpha^2} \Delta_s, 
\end{eqnarray*}
where $\alpha = \pi \nu J$ and $\nu = \frac{m}{2\pi}$. (See Appendix A for details of how the energies of the Shiba states are calculated.)
Up to the critical value $\alpha_c = 1$ these energies are ordered the following way: $E_1 > E_{\bar 1}$. 
As soon as $\alpha > \alpha_c$, energy levels $E_1$ and $E_{\bar 1}$ exchange places, making the order the following: $E_{\bar 1} > E_1$. 
This corresponds to a change of the ground state parity \cite{Sakurai1970,Schrieffer1997,Balatsky2006}.
For $\alpha \gg 1$ the subgap states approach the gap edge and eventually merge with the continuum. 
For the type of impurities considered here,   there is no dependence of these energies on the SO coupling in the low-energy approximation, though a weak dependence is introduced when one takes into account the non-linear form of the spectrum. The dependence of energy of the Shiba states on the impurity strength $J $ is depicted in Fig.~\ref{figSPDOS} where we plot the total spin of the impurity state $S(\mathbf{p}=0)$ as a function of energy and impurity strength. Note that the two opposite-energy Shiba states have opposite spins.

\begin{figure}[h!]
	\centering
	\begin{tabular}{cc}
		\textbf{z-impurity} & \textbf{x-impurity} \\
		\includegraphics*[width=0.48\columnwidth]{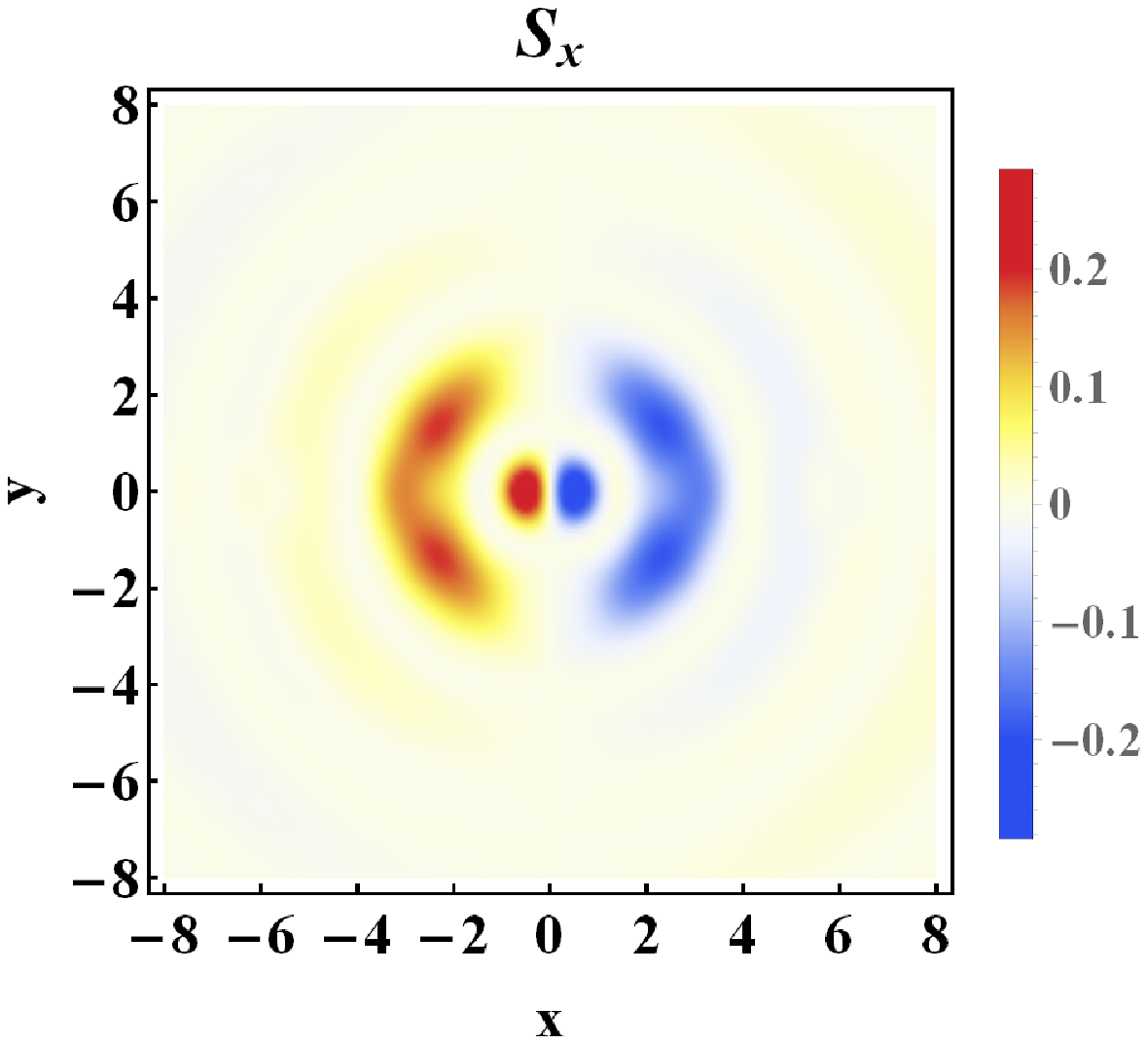} & 
		\includegraphics*[width=0.48\columnwidth]{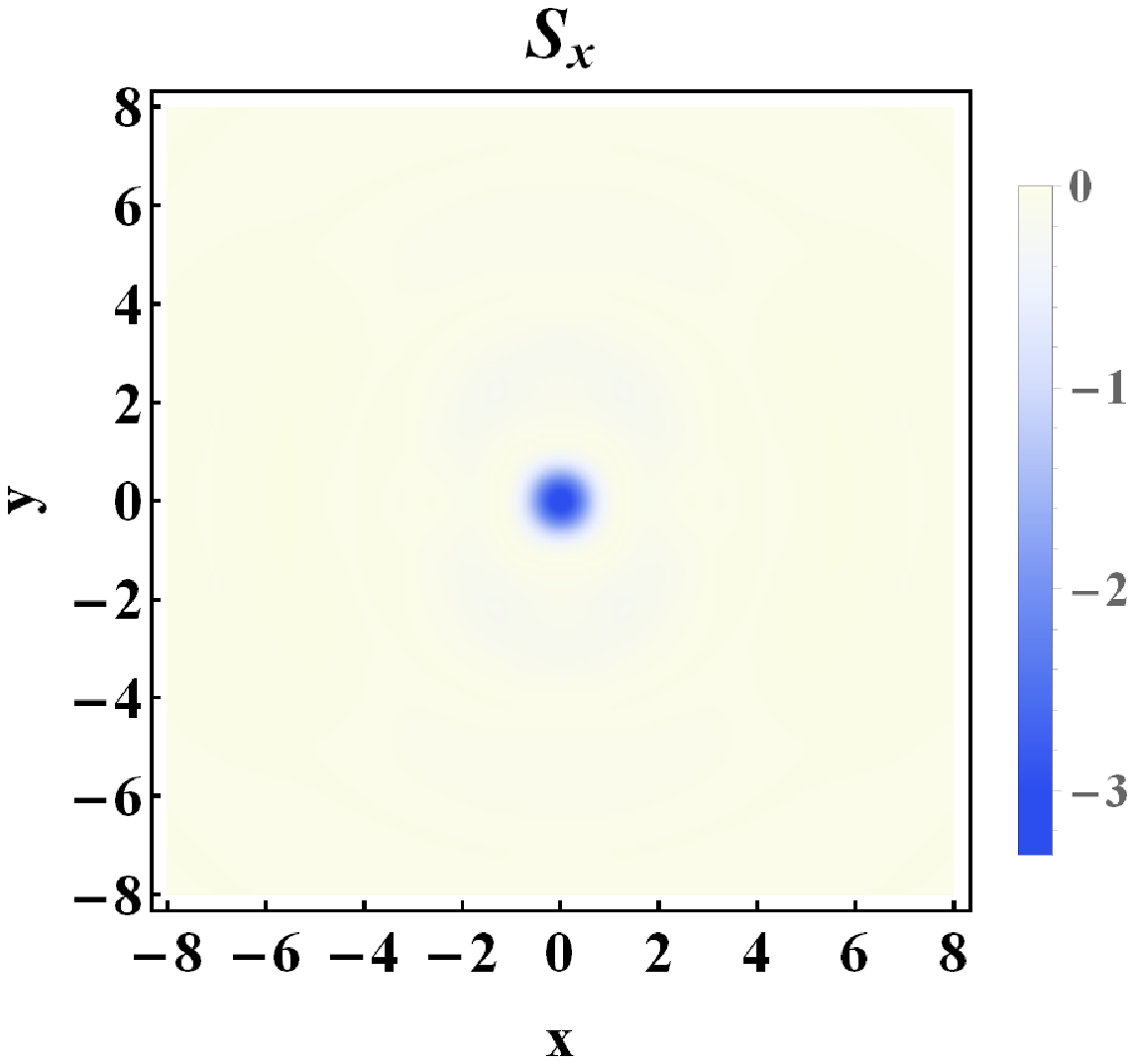}  \\
		\includegraphics*[width=0.48\columnwidth]{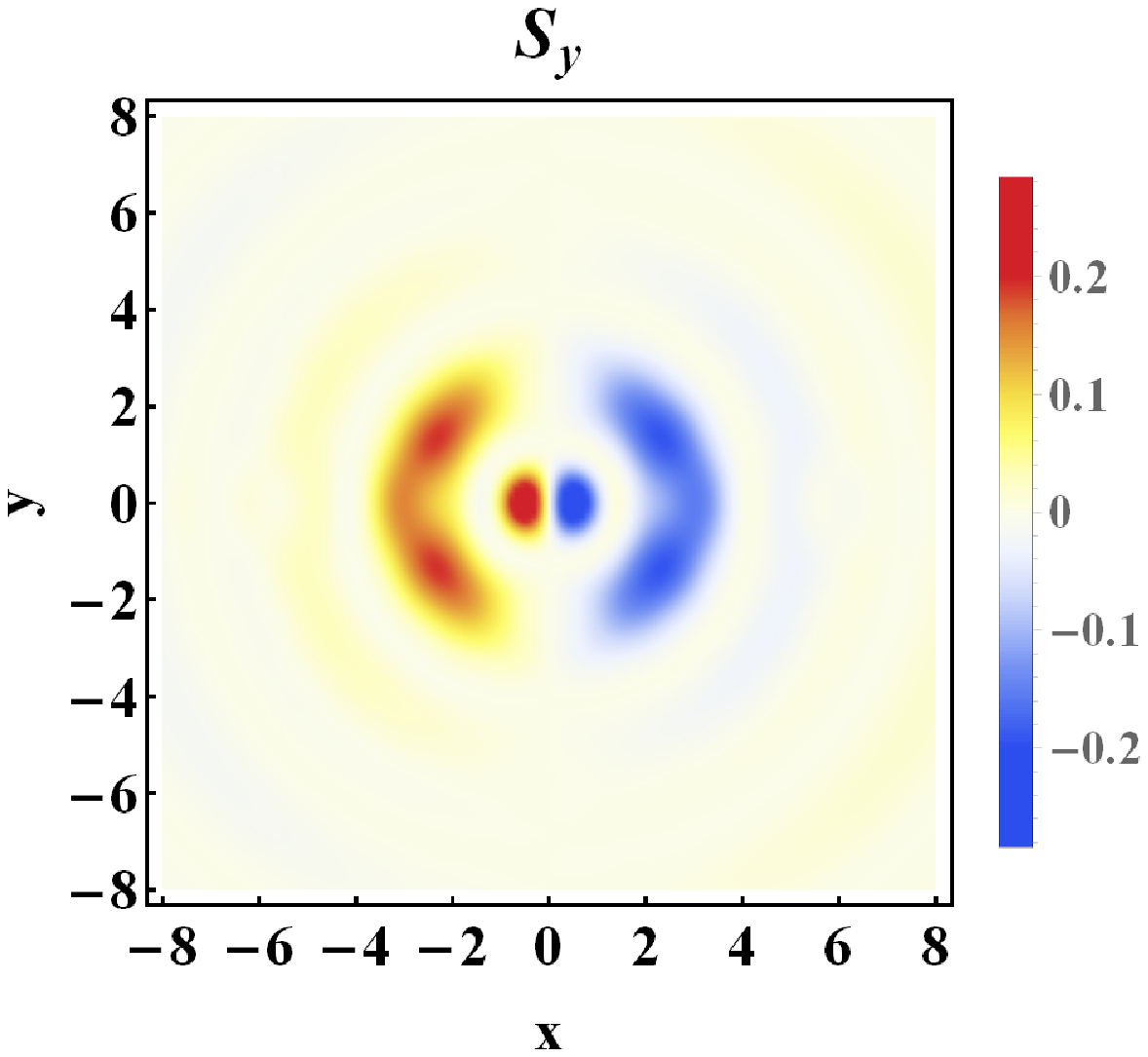} & 
		\includegraphics*[width=0.49\columnwidth]{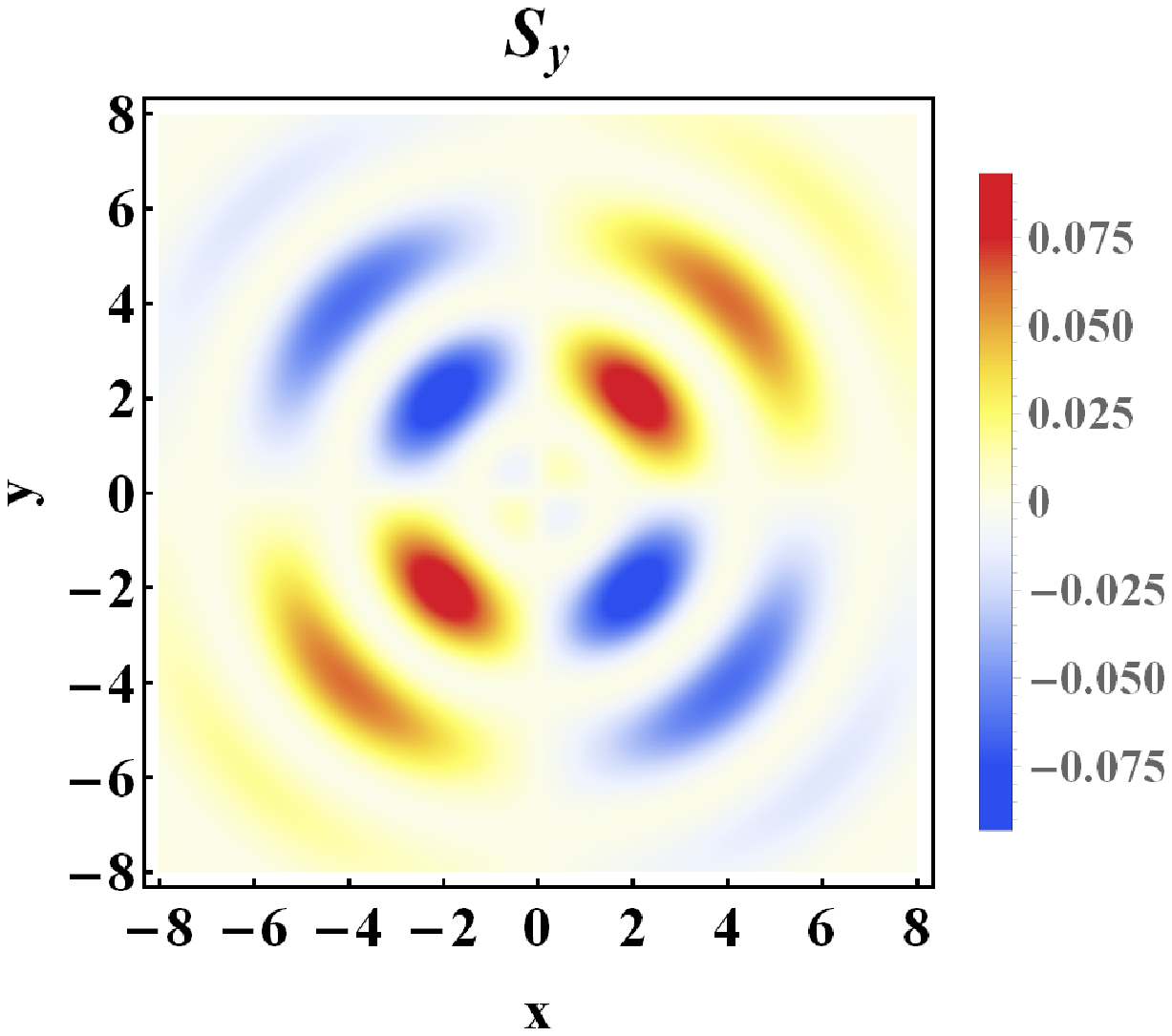}  \\
		\includegraphics*[width=0.48\columnwidth]{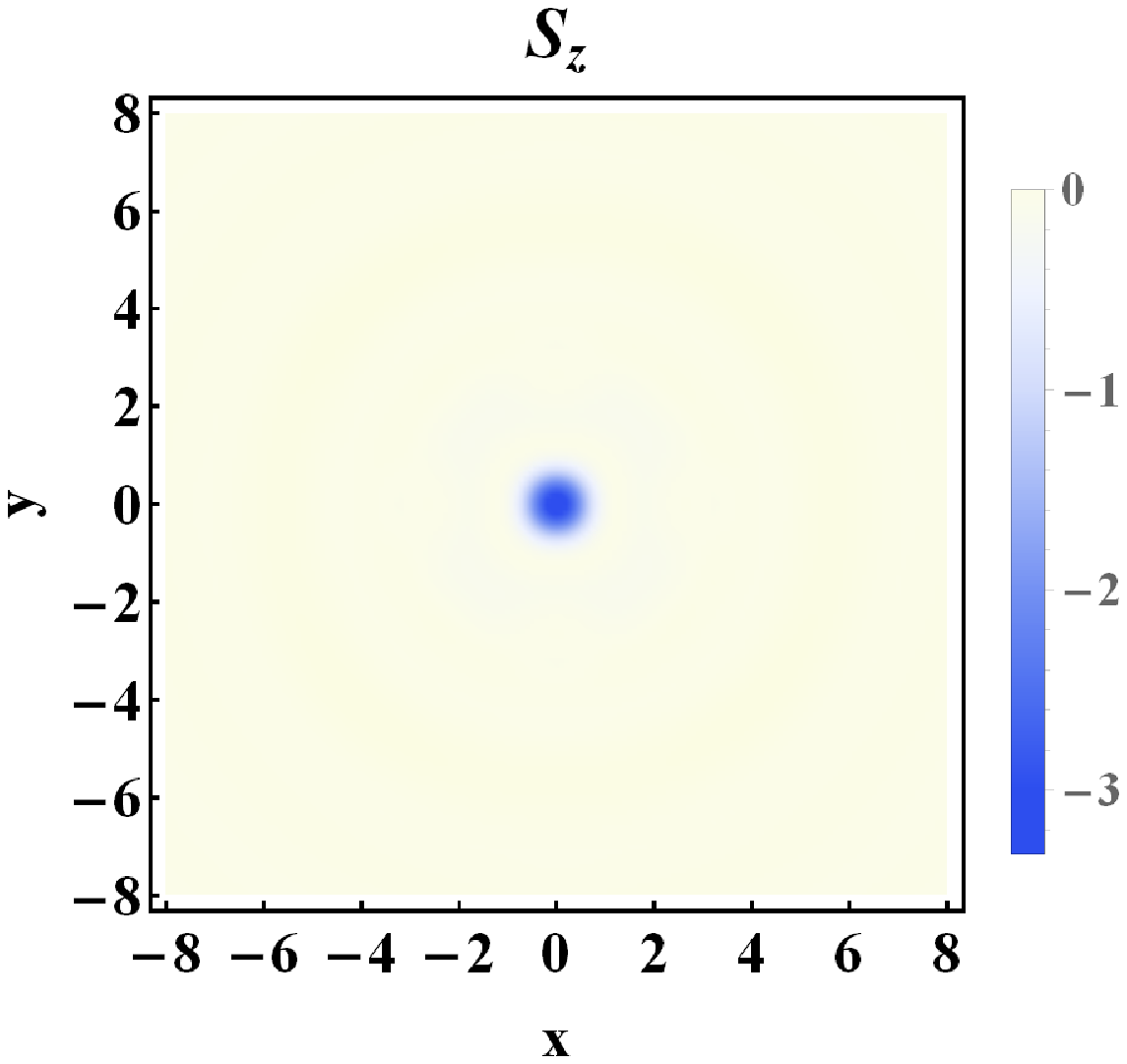} &
		\includegraphics*[width=0.48\columnwidth]{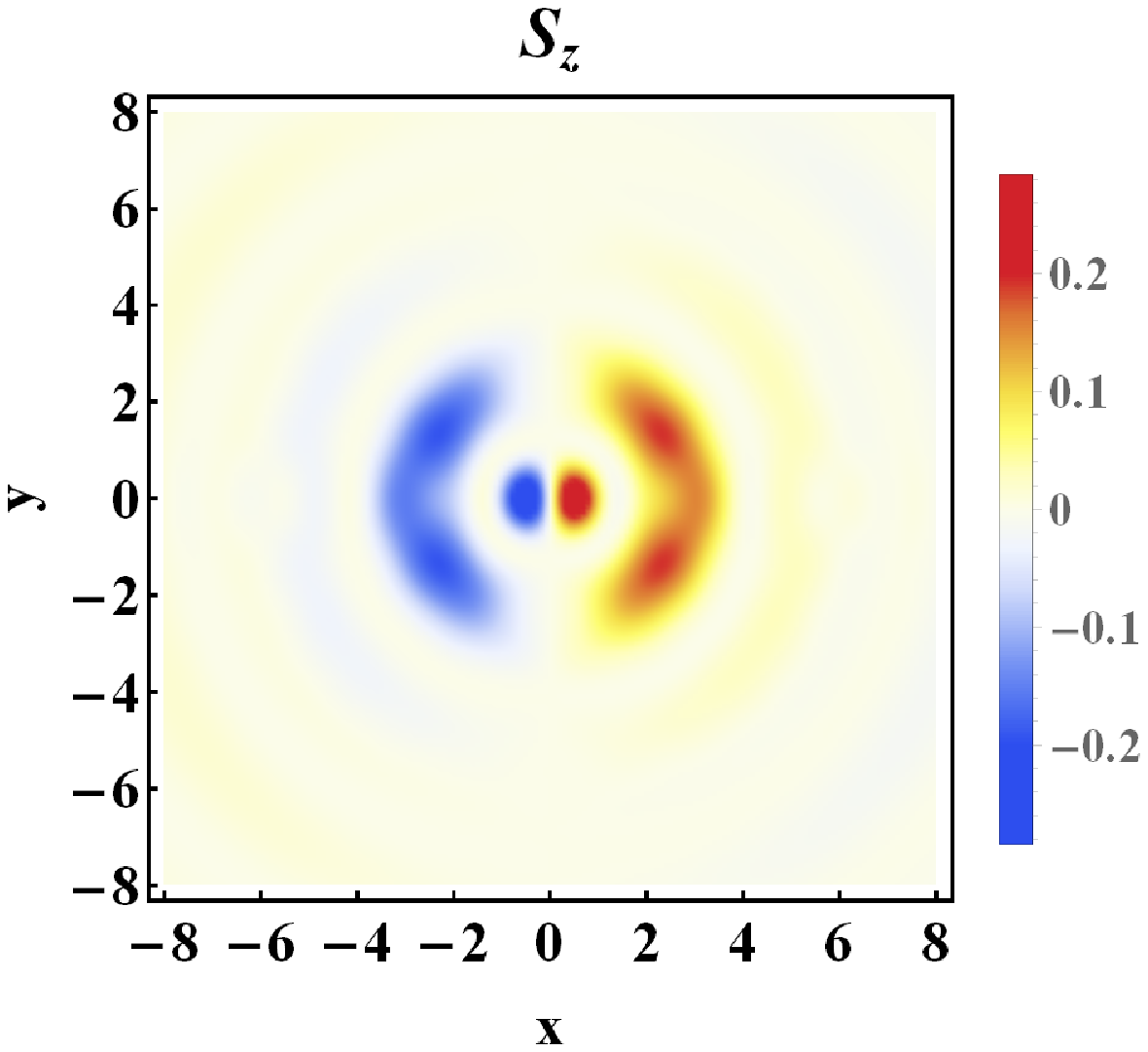}  \\
		\includegraphics*[width=0.47\columnwidth]{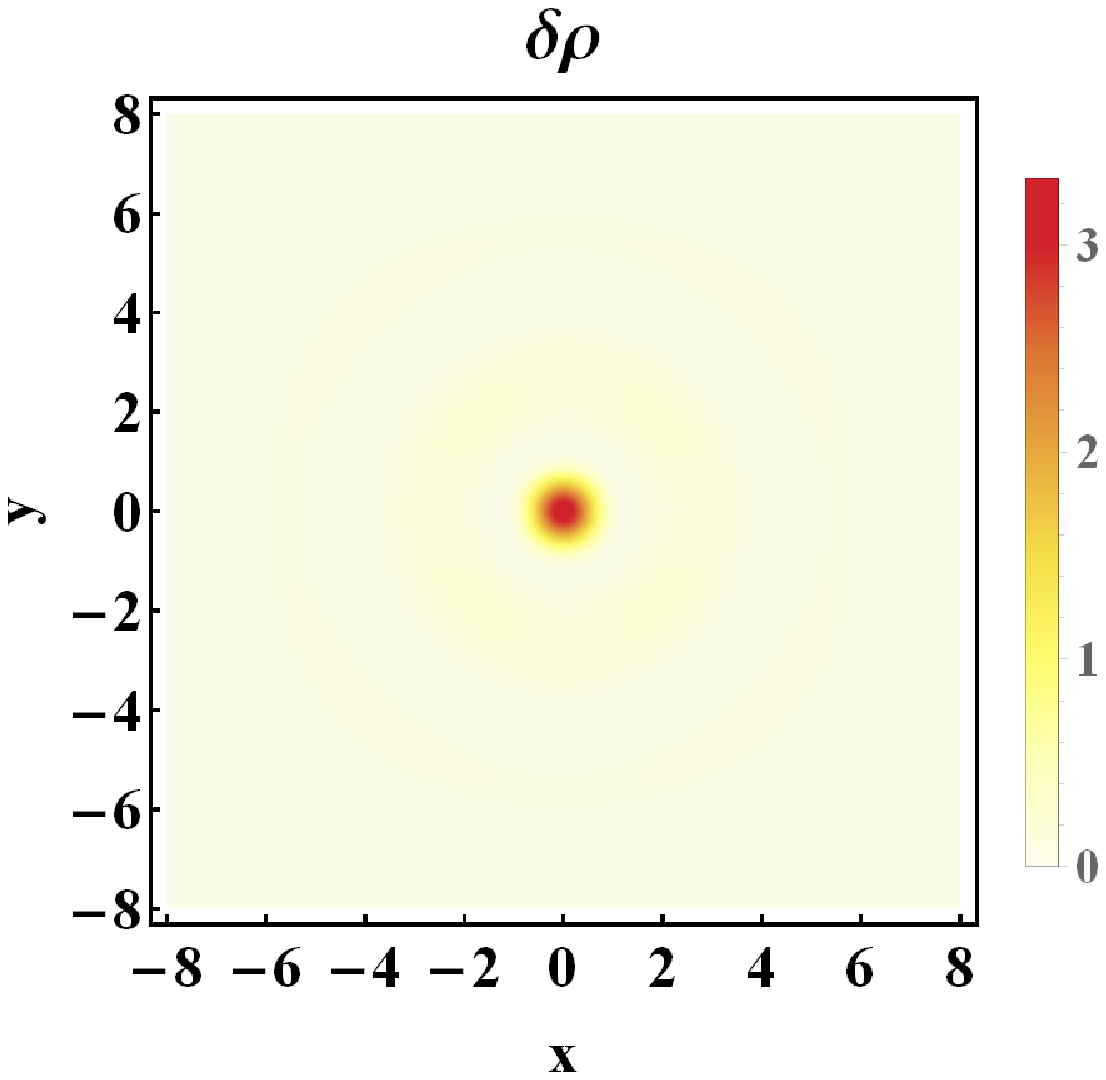} &
		\includegraphics*[width=0.47\columnwidth]{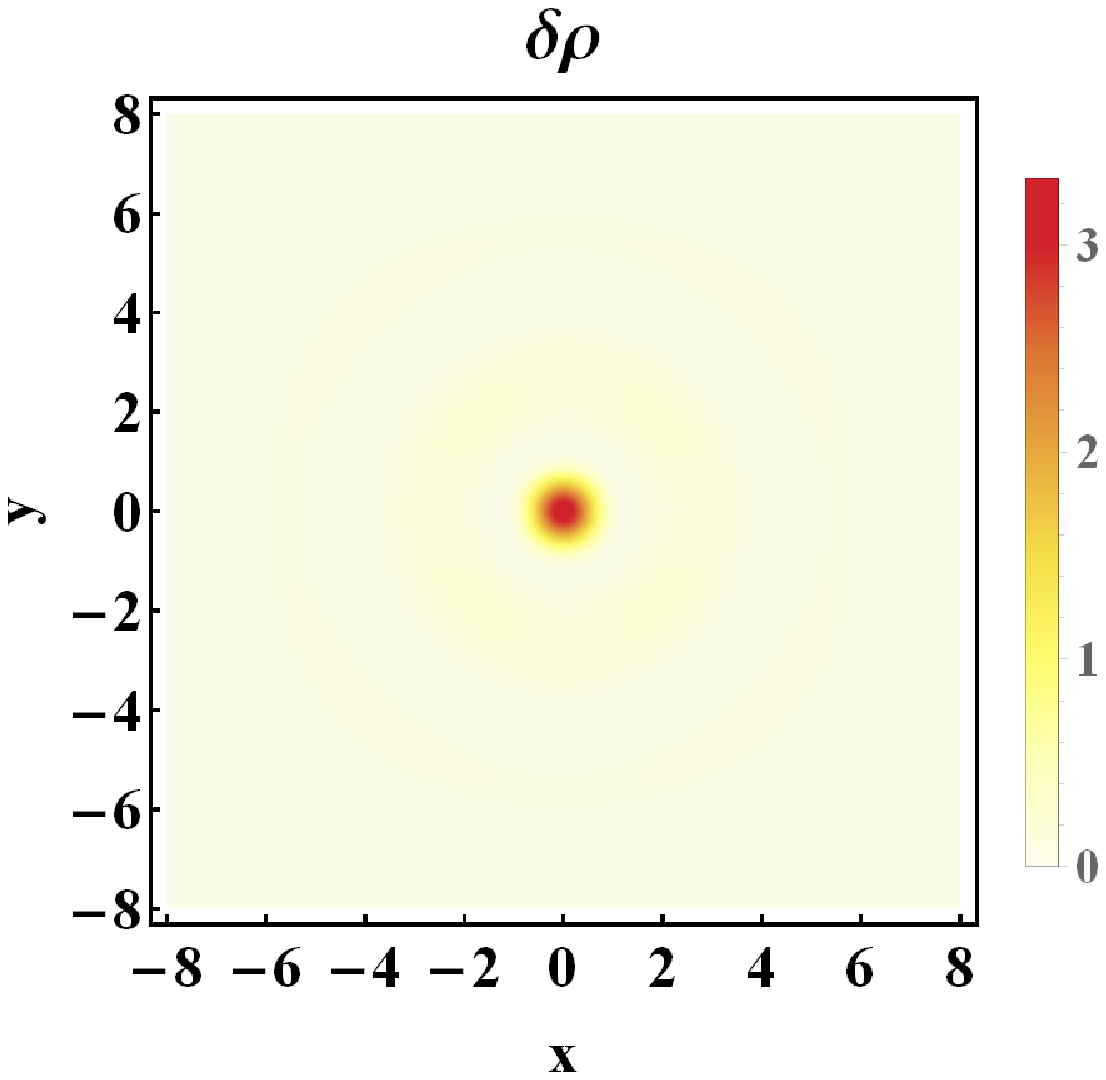} \\
  		$\phantom{a}J_z \;(J_x = J_y = 0)\phantom{a}$ & $\phantom{a}J_x \;(J_y = J_z = 0)\phantom{a}$
\end{tabular}
	\caption{(Color) The real-space dependence of the non-polarized as well as of the SP LDOS components for the positive energy Shiba state, for a magnetic impurity with $J_z=2$ (left column), and $J_x=2$ (right column). We take $t=1, \mu=3, \delta=0.01, \lambda=0.5, \Delta_s=0.2$.}
	\label{figCS}
\end{figure}

\begin{figure}[h!]
	\centering
	\begin{tabular}{cc}
		\textbf{z-impurity} & \textbf{x-impurity} \\
		\includegraphics*[width=0.48\columnwidth]{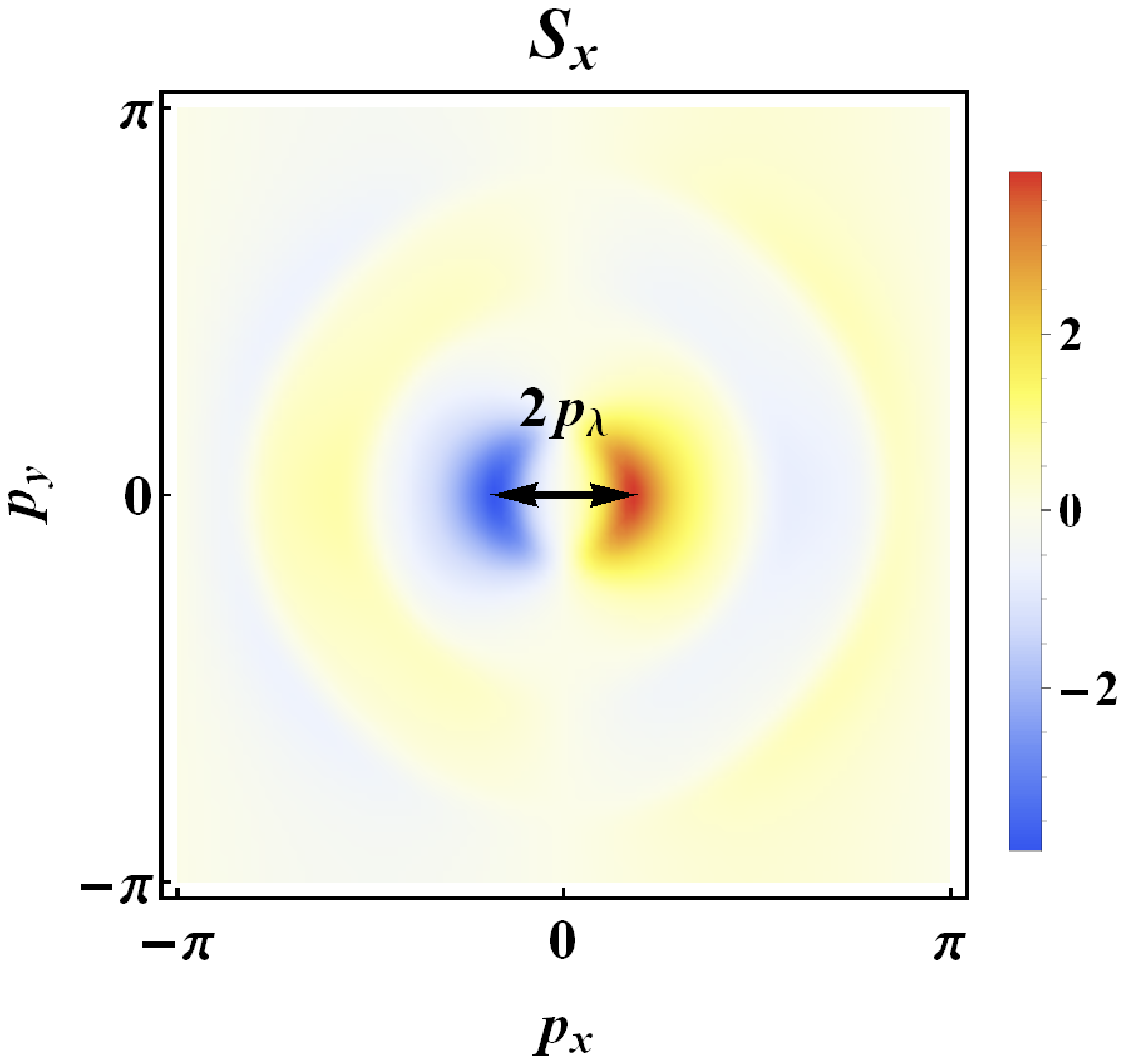} & 
		\includegraphics*[width=0.48\columnwidth]{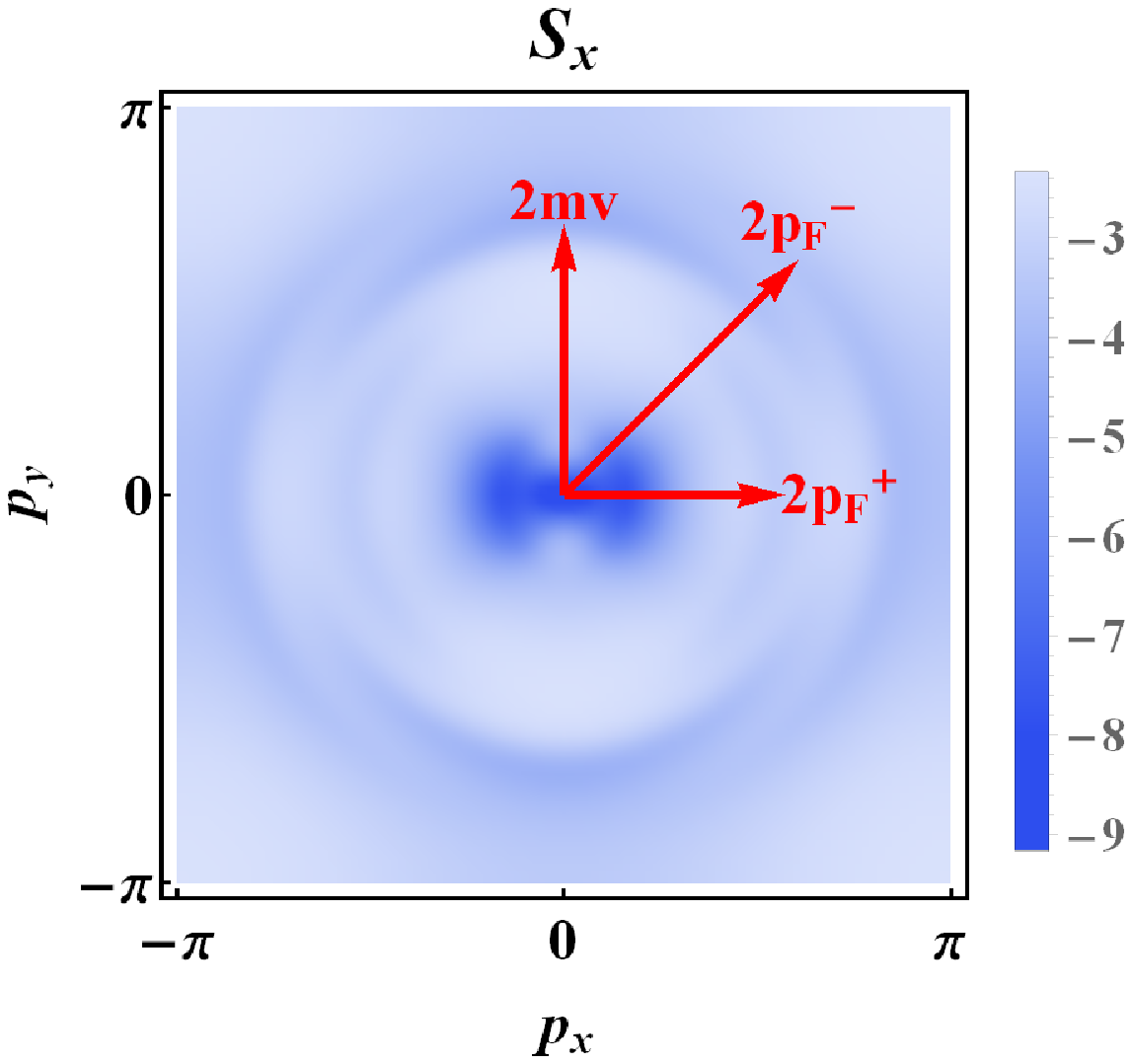}  \\
		\includegraphics*[width=0.48\columnwidth]{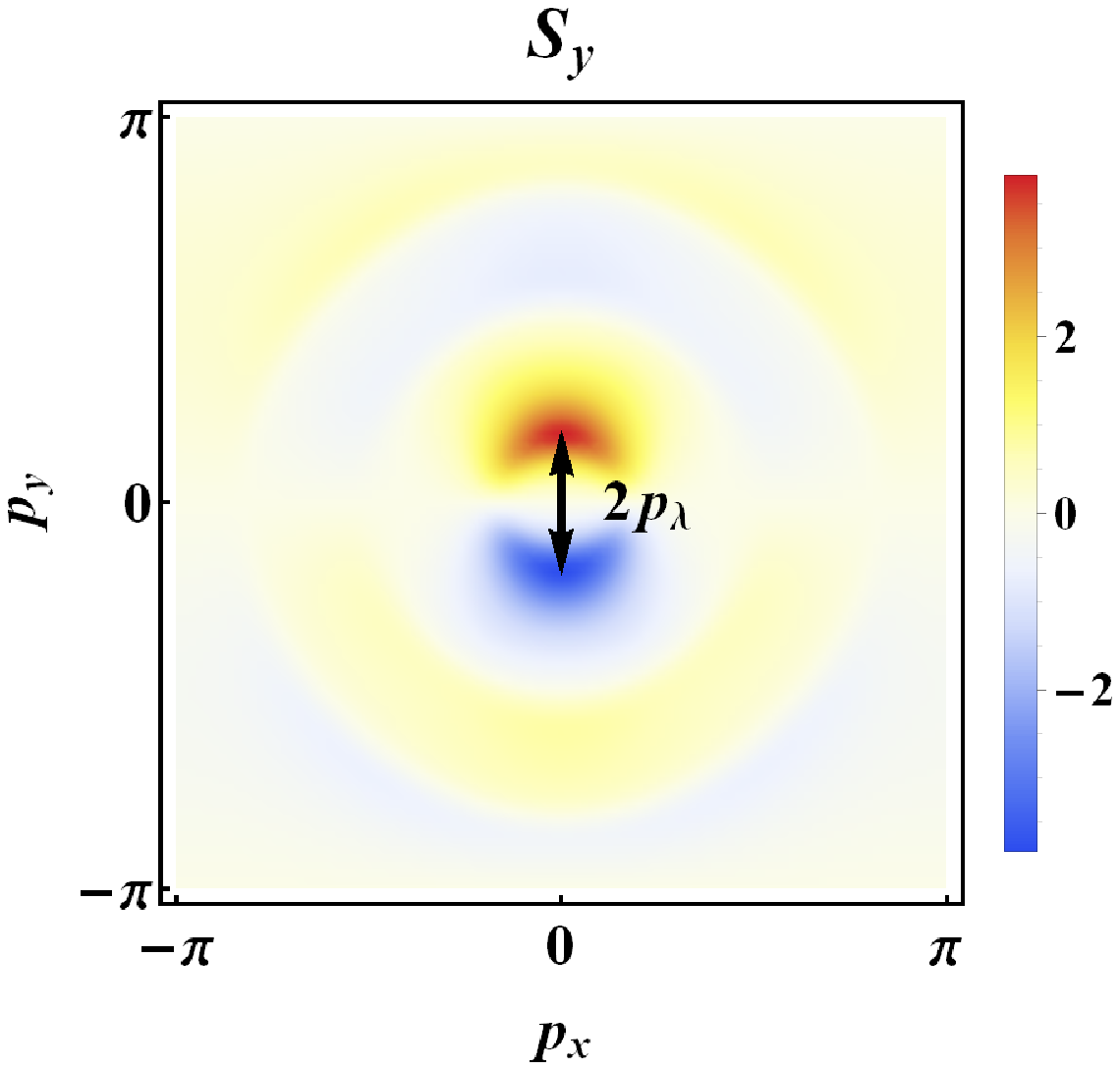} & 
		\includegraphics*[width=0.48\columnwidth]{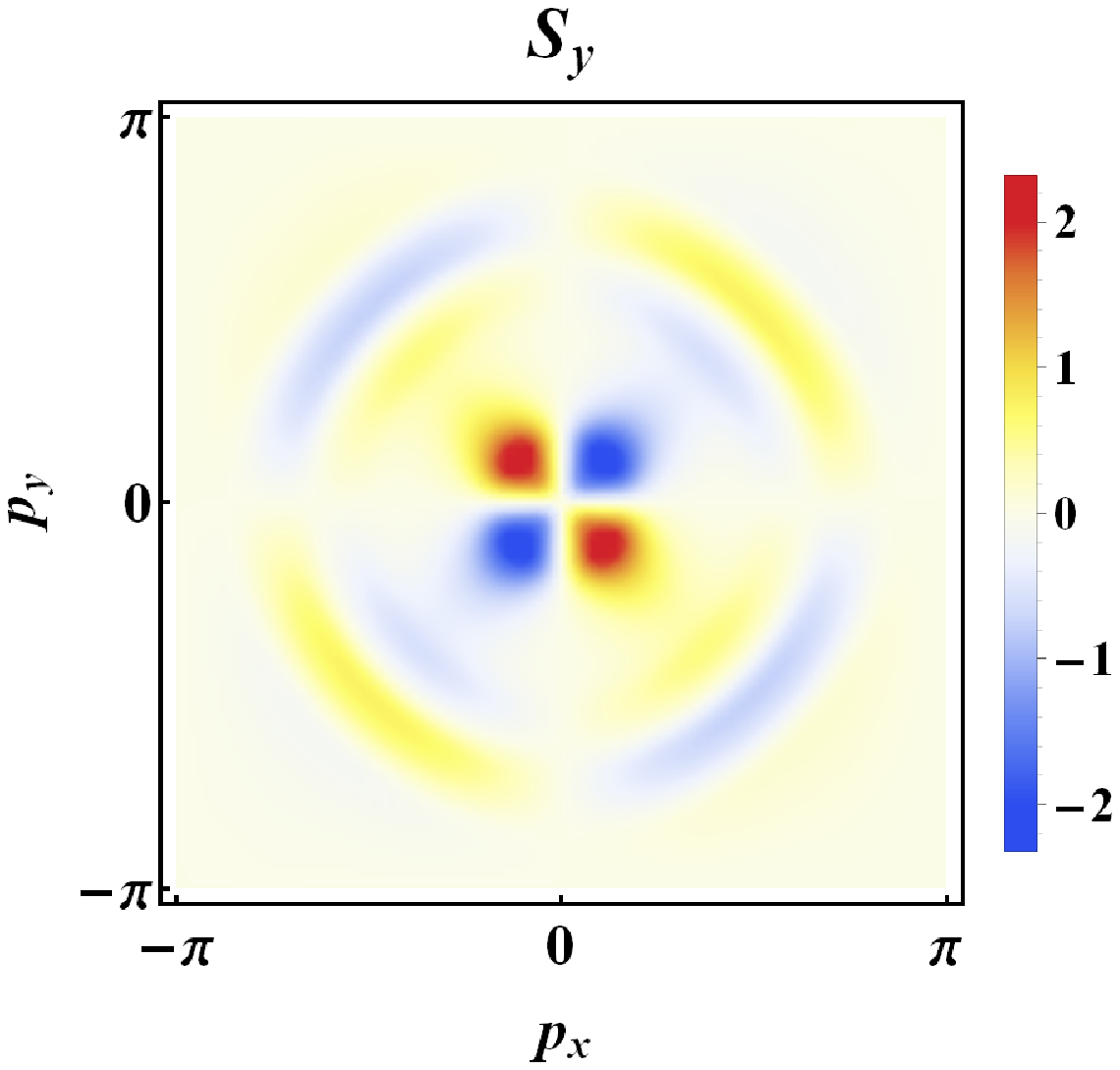}  \\
		\includegraphics*[width=0.48\columnwidth]{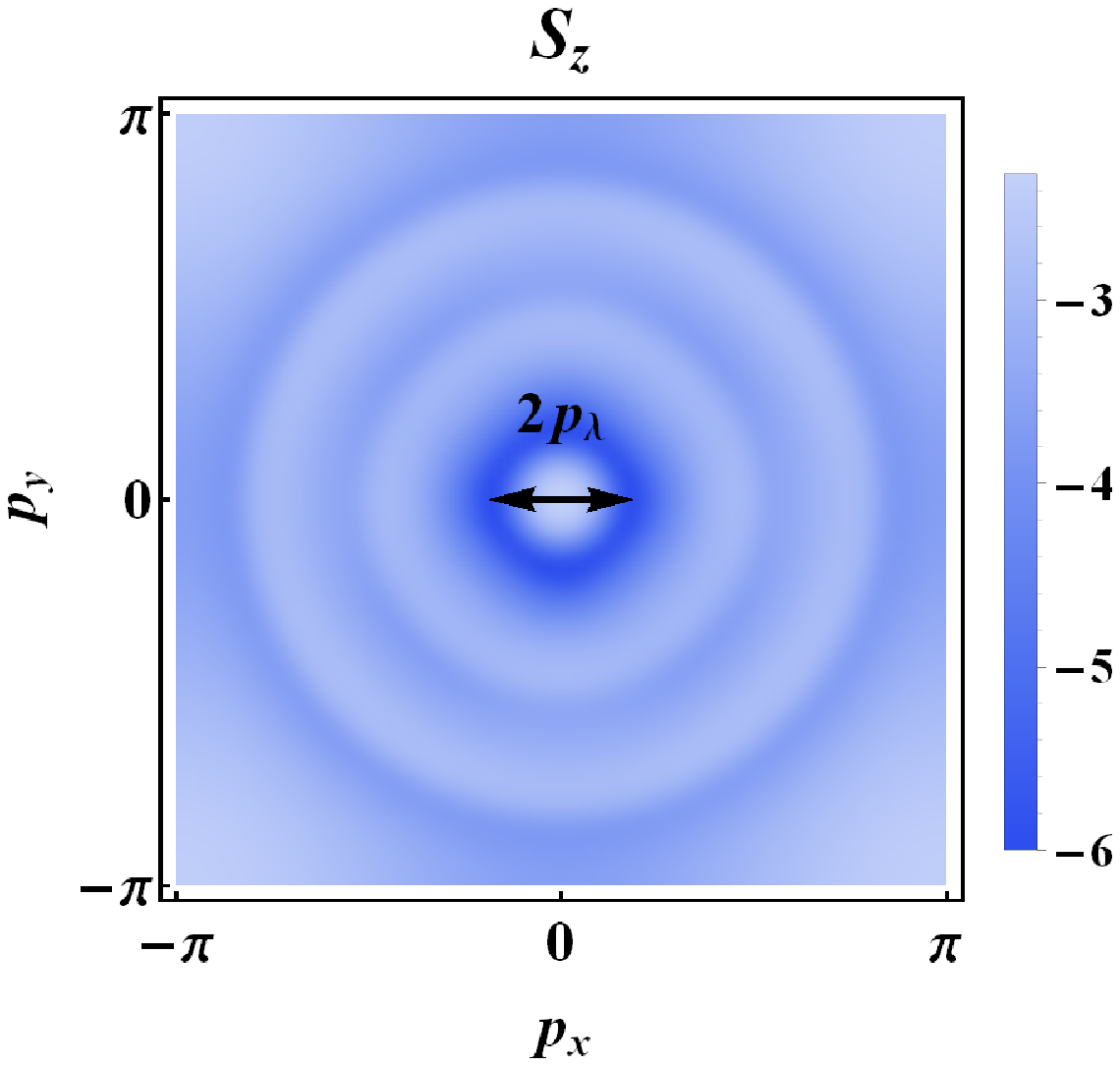} &
		\includegraphics*[width=0.48\columnwidth]{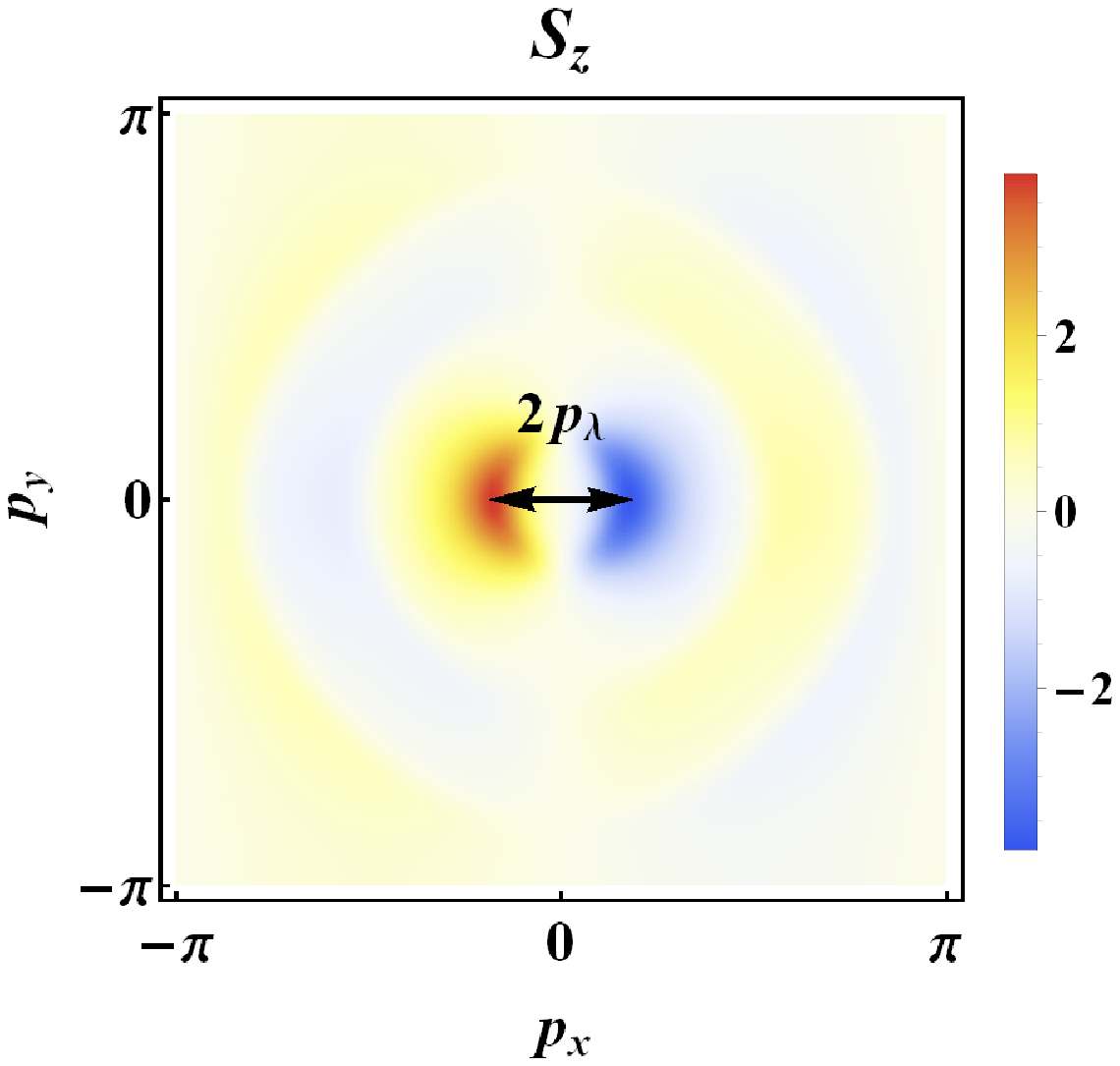}  \\
		\includegraphics*[width=0.48\columnwidth]{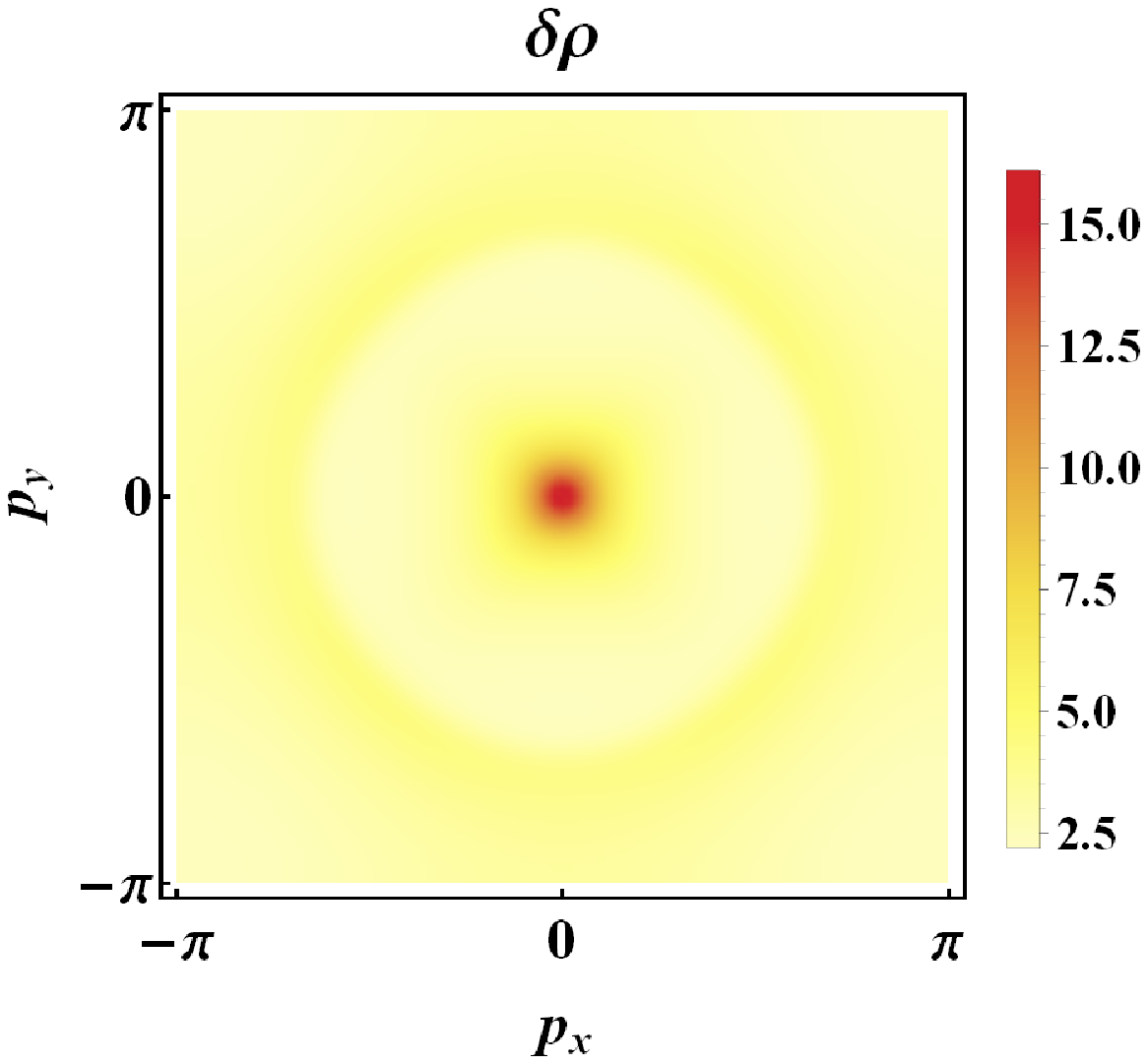} &
		\includegraphics*[width=0.48\columnwidth]{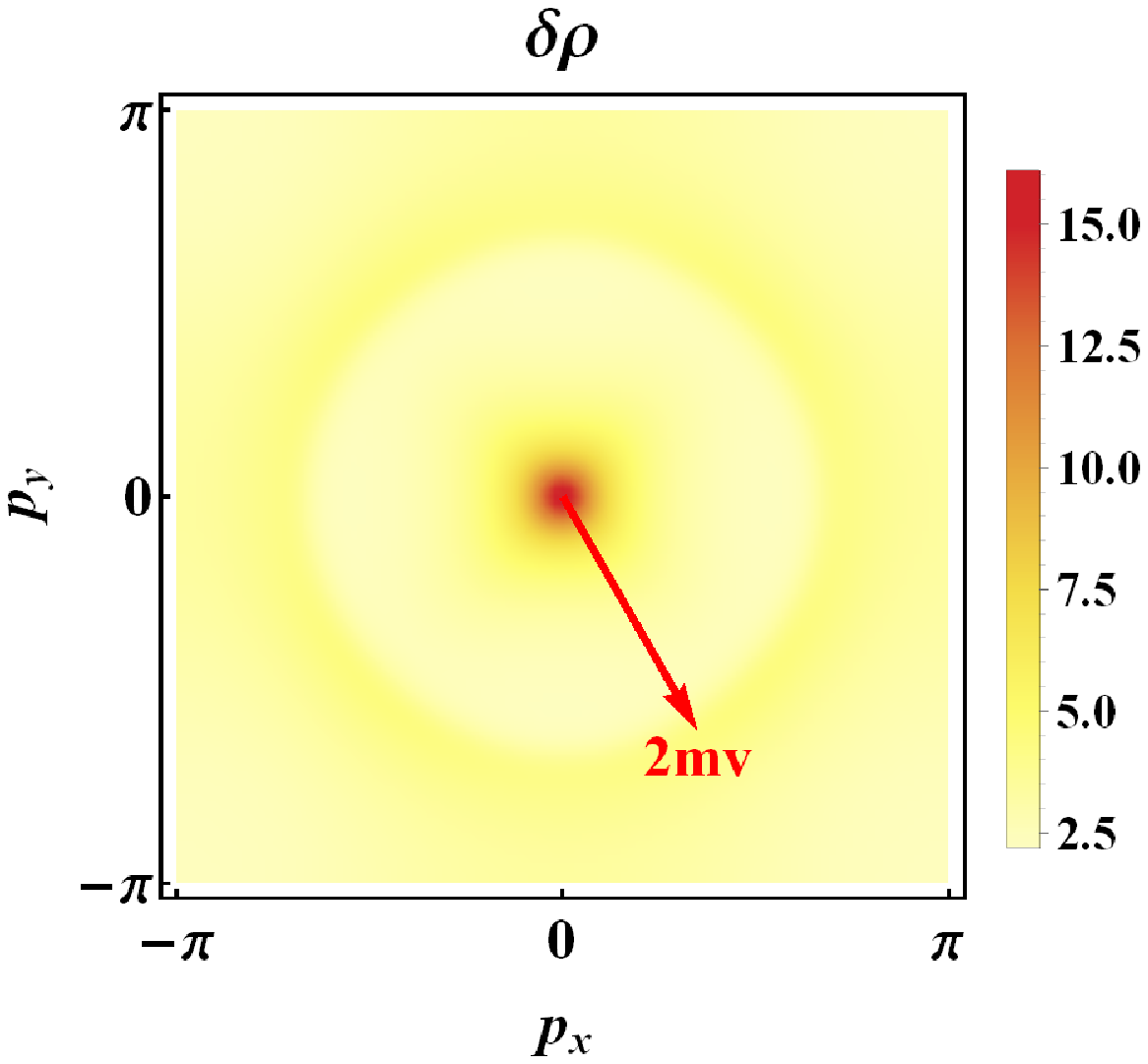} \\
  		$J_z \;(J_x = J_y = 0)\phantom{aa}$ & $J_x \;(J_y = J_z = 0)\phantom{aa}$
\end{tabular}
	\caption{(Color) The FT of the non-polarized as well as of the SP LDOS components for the positive energy Shiba state as a function of momentum, for a magnetic impurity with $J_z=2$ (left column), and $J_x=2$ (right column). We take $t=1, \mu=3, \delta=0.01, \lambda=0.5, \Delta_s=0.2$. For a $z$-impurity we depict the real part of the FT for $\delta \rho$ and for $S_z$, and the imaginary part for and $S_x$ and $S_y$, whereas for an $x$-impurity we take the imaginary part only for the $S_z$ component. Black two-headed arrows correspond to the value of $2p_\lambda \equiv 4m\lambda$ (see  the analytical results) and thus allow to extract the SO coupling constant directly from these strong features in momentum space. The other arrows correspond to the other important wavevectors that can be observed in these FTs, as identified with the help of the analytical results.}
	\label{figFTMS}
\end{figure}

We are interested in studying the spatial structure of the Shiba states in the presence of magnetic impurities oriented both perpendicular to the plane, and in plane. This can be done both in real space and momentum space by calculating the Fourier transform of the spin-polarized LDOS using the T-matrix technique detailed in the previous section.  We focus on the positive-energy Shiba state, noting that its negative energy counterpart exhibits a qualitatively similar behavior.
In Fig. \ref{figCS} we show the real-space dependence of the non-polarized and SP LDOS. Each of the panels corresponds to the interference patterns originating from different types of scattering. Note that the spin-orbit value cannot be accurately extracted from these type of measures, since the system contains oscillations with many different superposing wavevectors.
To overcome this problem we focus on the FT of these features, as it is oftentimes done in spatially resolved STM experiments, which allow for a more accurate separation of the different wavevectors\cite{Hasegawa1993,Crommie1993,Sprunger1997,Petersen2000,Hofmann1997,Vonau2004,Vonau2005,Simon2011}. Thus in Fig. \ref{figFTMS} we focus on the FT of the SP LDOS for two types of impurities with spin oriented along $z$ and $x$ axes respectively.

Note that the SO introduces non-zero spin components in the directions different from that of the impurity spin. These components exhibit either two-fold or four-fold symmetric patterns.
Also the SO is affecting strongly the spin component parallel to the impurity, in particular when the impurity is in-plane, in which case the structure of the SP LDOS around the impurity is no longer radially symmetric. However, as can be seen in the bottom panel of Fig. \ref{figFTMS}, the non-spin-polarized LDOS is not affected by the presence of SO, preserving a radially symmetric shape quasi-identical to that obtained in the absence of SO. Thus the SO coupling can be measured only via the spin-polarized components of the LDOS, and not the non-polarized LDOS.

These results, which are obtained using a numerical integration of the T-matrix equations,  are also supported  by analytical calculations  which help to understand the fine structure of the FT of the SP LDOS (see Appendices  for details). These calculations yield for the SP LDOS generated by a magnetic impurity perpendicular to the plane
\ba
S_x(\bs r) &=& +J_z^2 \left( 1+\frac{1}{\alpha^2} \right) \frac{e^{-2 p_s r}}{r} \cos\phi_{\bs r}\times
\nonumber\\&& \times \left\{ \sum\limits_\sigma \frac{\sigma \nu^2_\sigma}{p_F^\sigma} \cos \left(2p_F^\sigma r-\theta \right)  +  2 \nu^2 \frac{v_F^2}{v^2 p_F} \sin p_\lambda r \right\},\nonumber  \\
S_y(\bs r) &=& +J_z^2 \left( 1+\frac{1}{\alpha^2} \right) \frac{e^{-2 p_s r}}{r} \sin\phi_{\bs r}\times\nonumber\\&&\times \left\{ \sum\limits_\sigma \frac{\sigma \nu^2_\sigma}{p_F^\sigma} \cos \left(2p_F^\sigma r-\theta \right) +   2 \nu^2 \frac{v_F^2}{v^2 p_F} \sin p_\lambda r \right\},  \nonumber\\
S_z(r) &=& -J_z^2 \left( 1+\frac{1}{\alpha^2} \right)  \frac{e^{-2 p_s r}}{r}\times\nonumber \\&&\times \left\{ \sum\limits_\sigma \frac{\nu^2_\sigma}{p_F^\sigma} \sin \left(2p_F^\sigma r-\theta \right) -  2 \nu^2 \frac{v_F^2}{v^2 p_F} \cos p_\lambda r \right\}, \nonumber \\
\rho(r) &=& +J_z^2 \left( 1+\frac{1}{\alpha^2} \right) \frac{e^{-2 p_s r}}{r}\times\nonumber \\&&\times \left\{ 2\frac{\nu^2}{mv} +  2 \nu^2 \frac{v_F^2}{v^2 p_F} \sin \left(2mvr - \theta \right) \right\} ,
\ea
with
\begin{equation}
\tan\theta = \begin{cases} \frac{2\alpha}{1-\alpha^2}, \;\;\text{if}\; \alpha \neq 1 \\  +\infty, \;\;\text{if}\; \alpha = 1\end{cases}.
\end{equation}
We have introduced $e^{i\phi_{\bf r}}= \frac{x + i y}{r}$, and
\ba
p^\sigma_F &=&-\sigma m \lambda + m v,\\ 
p_\lambda &=& 2m\lambda,\\
p_s &=& \sqrt{\Delta_s^2-E_1^2}/v,
\ea
with $v = \sqrt{v_F^2+\lambda^2}$, and $v_F=\sqrt{2 \varepsilon_F/m}$. Here $p^\sigma_F,~p_\lambda$ and $p_s$ are the different momenta which can be read off from the SP LDOS.
For an in-plane magnetic impurity we have
\ba
&S_x^{s}(r) =& -J_x^2 \bigg( 1+\frac{1}{\alpha^2} \bigg) \bigg\{ \sum\limits_\sigma \frac{ \nu^2_\sigma}{p_F^\sigma} [1 +\sin (2p_F^\sigma r-2\beta)] ,
\nonumber \\&+2 \nu^2 \frac{v_F^2}{v^2 {p_F} }& [\cos p_\lambda r + \sin\left(2mvr -2\beta\right)]\bigg\}  \frac{e^{-2 p_s r}}{r}  \nonumber\\
&S_x^{a}(\bs r) =&+J_x^2 \left( 1+\frac{1}{\alpha^2} \right) \bigg\{ \sum\limits_\sigma  \frac{\nu^2_\sigma}{p_F^\sigma}[1 - \sin \left(2p_F^\sigma r-2\beta \right)] \nonumber
\\&-2 \nu^2 \frac{v_F^2}{v^2 p_F} &[\cos p_\lambda r + \sin\left(2mvr -2\beta\right)] \bigg\} \frac{e^{-2 p_s r}}{r} \cos 2\phi_{\bs r},\nonumber \\
&S_y(\bs r) = &+J_x^2 \left( 1+\frac{1}{\alpha^2} \right) \bigg\{ \sum\limits_\sigma  \frac{\nu^2_\sigma}{p_F^\sigma} [1 - \sin \left(2p_F^\sigma r-2\beta \right)]
\nonumber \\&-2 \nu^2 \frac{v_F^2}{v^2 p_F} &[\cos p_\lambda r - \sin\left(2mvr -\theta\right)] \bigg\}  \frac{e^{-2 p_s r}}{r} \sin 2\phi_{\bs r},\nonumber \\
&S_z(\bs r) =& -J_x^2 \left( 1+\frac{1}{\alpha^2} \right) \bigg\{ 2\sum\limits_\sigma \frac{\sigma \nu^2_\sigma}{p_F^\sigma} \cos \left(2p_F^\sigma r-\theta \right) \nonumber \\&&+ 4 \nu^2 \frac{v_F^2}{v^2 p_F} \sin p_\lambda r \bigg\}  \frac{e^{-2 p_s r}}{r} \cos\phi_{\bs r}, \nonumber \\
&\rho(r) =& +J_x^2 \left( 1+\frac{1}{\alpha^2} \right) \bigg\{ 4\frac{\nu^2}{mv} +  4 \nu^2 \frac{v_F^2}{v^2 p_F} \times \nonumber
\\&&\times \sin \left(2mvr - \theta \right) \bigg\} \frac{e^{-2 p_s r}}{r} ,
\ea
with $\tan\beta = \alpha$.

The $S_x$ component is the sum of symmetric part a $S_x^s$ and an asymmetric part $S_x^a$. Note that the features observed in the FT of the SP LDOS plots are well captured by the analytical calculations. In particular we note that the oscillations in the SP LDOS are dominated by the following four wavevectors:
 $$2p_F^\pm~,~p_F^+ + p_F^-=2mv,~{\rm~ and}~p_F^--p_F^+=p_\lambda \equiv 2m\lambda,$$ 
which should give rise in the FT to high-intensity features at these wavevectors (the red arrows in Fig.~\ref{figFTMS}). 
Indeed, we note in the numerical results for the FT of the SP LDOS the existence of four rings, corresponding to $2p_F^\pm$, $p_F^+ + p_F^-=2mv$ and $p_F^--p_F^+= p_\lambda$, having the proper two-fold or fold-fold symmetries, consistent with the $\cos/\sin \phi_r$ and $\cos/\sin 2\phi_r$ dependence of the SP LDOS obtained analytically. For example, in the $x$ component of the SP LDOS induced by an $x$ impurity, the $2p_F^+$, $2p_F^-$ and $p_\lambda$ rings have a maximum along $x$ and a minimum along $y$,  while the $2mv$ ring has a symmetry corresponding to a rotation by $90$ degrees. The $y$ component of the FT of the SP LDOS has a four-fold symmetry in which we can again identify the same wavevectors, while the $S_z$ component has a two-fold symmetry, and the $2mv$ vector is absent. Similarly, for the $S_x$ and the $S_y$ components of the SP LDOS induced by a $z$ impurity (these components should be zero in the absence of the SO coupling) only the $2p_F^\pm$ and $p_\lambda$ wave vectors are present, with similar symmetries, while the $S_z$ component is symmetric. Note also the central peak at $p_x=p_y=0$ which is due to the terms independent of position in the SP LDOS.

\begin{figure}
\includegraphics*[width=0.5\columnwidth]{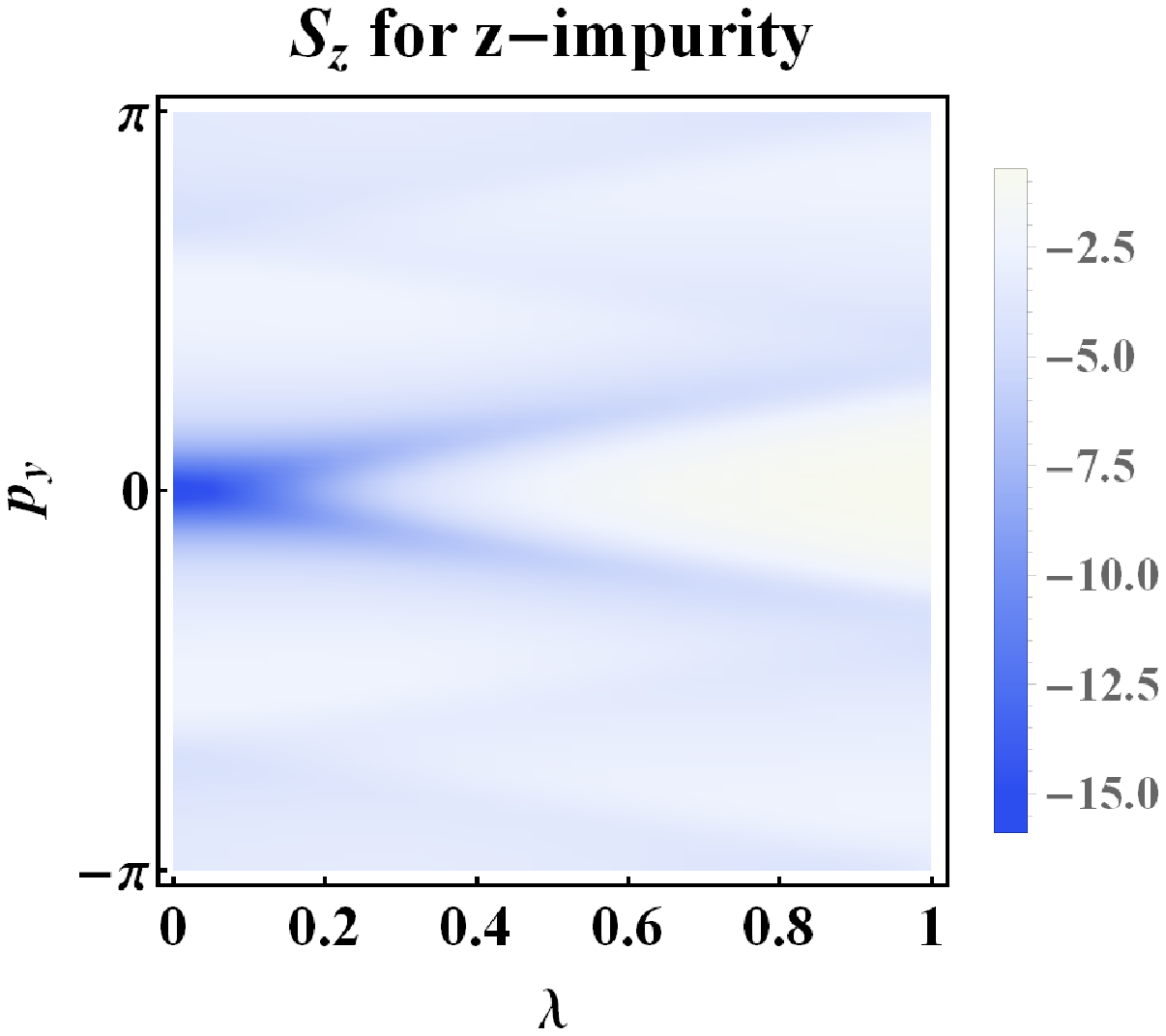}
\includegraphics*[width=0.48\columnwidth]{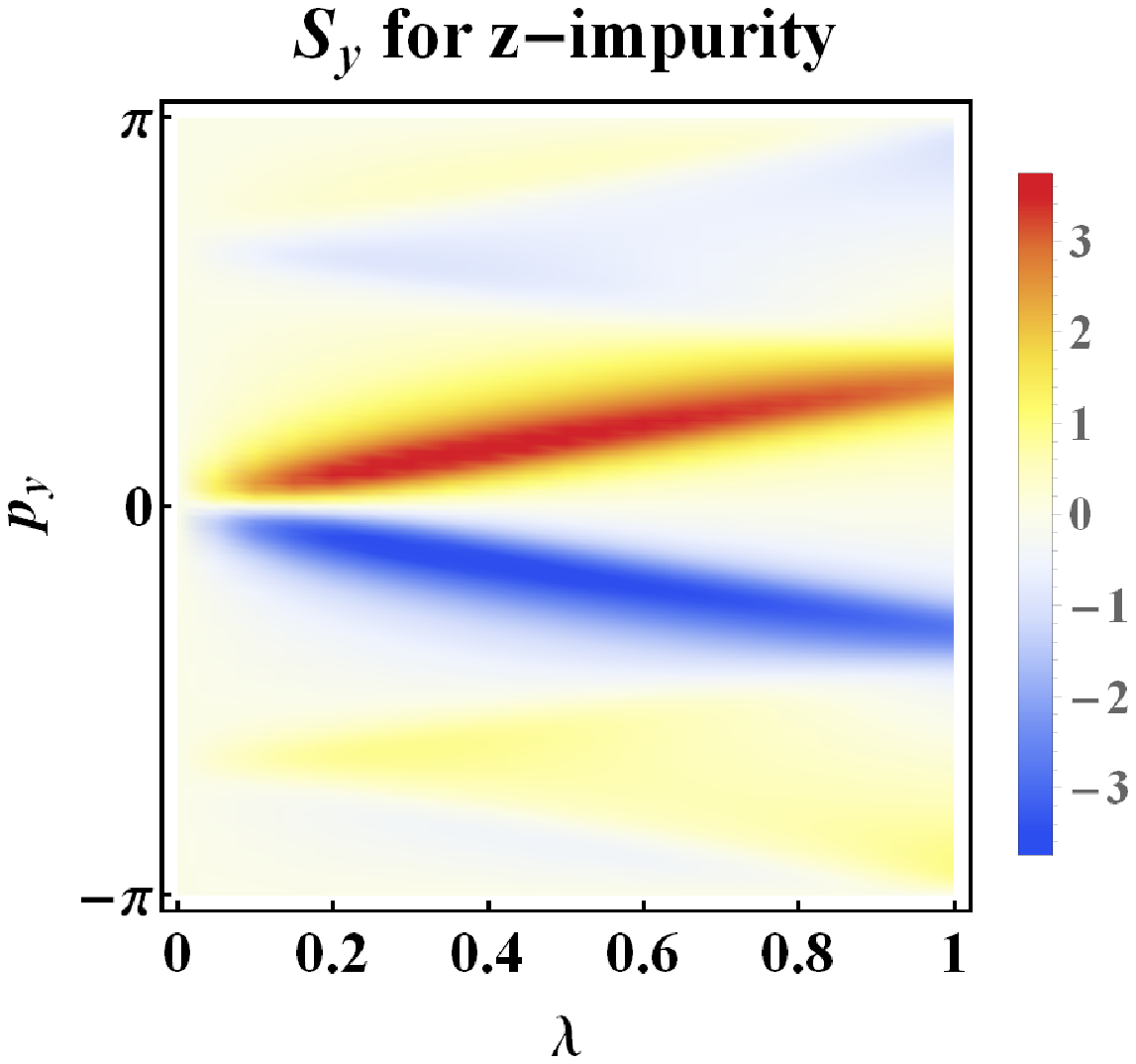}\\
\caption{(Color) The FT of various SP LDOS component for a Shiba state as a function of the SO coupling $\lambda$ and of $p_y$ (for $p_x=0$ - vertical cut). We take $t=1, \mu=3, \delta=0.01, \Delta_s=0.2, J_z=2$.}
\label{figHorCut}
\end{figure}

The most important observation is that all the components of the FT of the SP LDOS exhibit a strong feature at wave vector $p_\lambda$. Thus an experimental observation of this feature via spin-polarized STM would allow one to read-off directly the value of the SO coupling.  The spin orbit can also be read-off from the distance between the $2p_+$ and $2p_-$ peaks, though the intensity of these features is not as strong. This appears clearly  in Fig.~\ref{figHorCut}, in which we plot a horizontal cut though two of the FT -- SP LDOS above as a function of the SO coupling $\lambda$. 

Note that the only wave vector present in the non-polarized LDOS is $2 mv$, which has only a very weak dependence on $\lambda$ for not too large values of the SO with respect to the Fermi velocity, thus it is quasi-impossible to determine the SO coupling from a measurement without spin resolution. Note also the typical two-dimensional $1/r$ decay of the Friedel oscillations is overlapping in this case with an exponential decay with wave vector $p_s$.

\subsection{Comparison to the metallic phase} 
A similar analysis can be performed for impurity states forming in the vicinity of a magnetic impurity in a metallic system. Here the classical magnetic impurity does not lead to any localized bound states  at a specific energy, and the intensity of the impurity contribution is roughly independent of energy. 
\begin{figure}[h!]
	\centering
	\begin{tabular}{cc}
		\textbf{z-impurity} & \textbf{x-impurity} \\
		\includegraphics*[width=0.48\columnwidth]{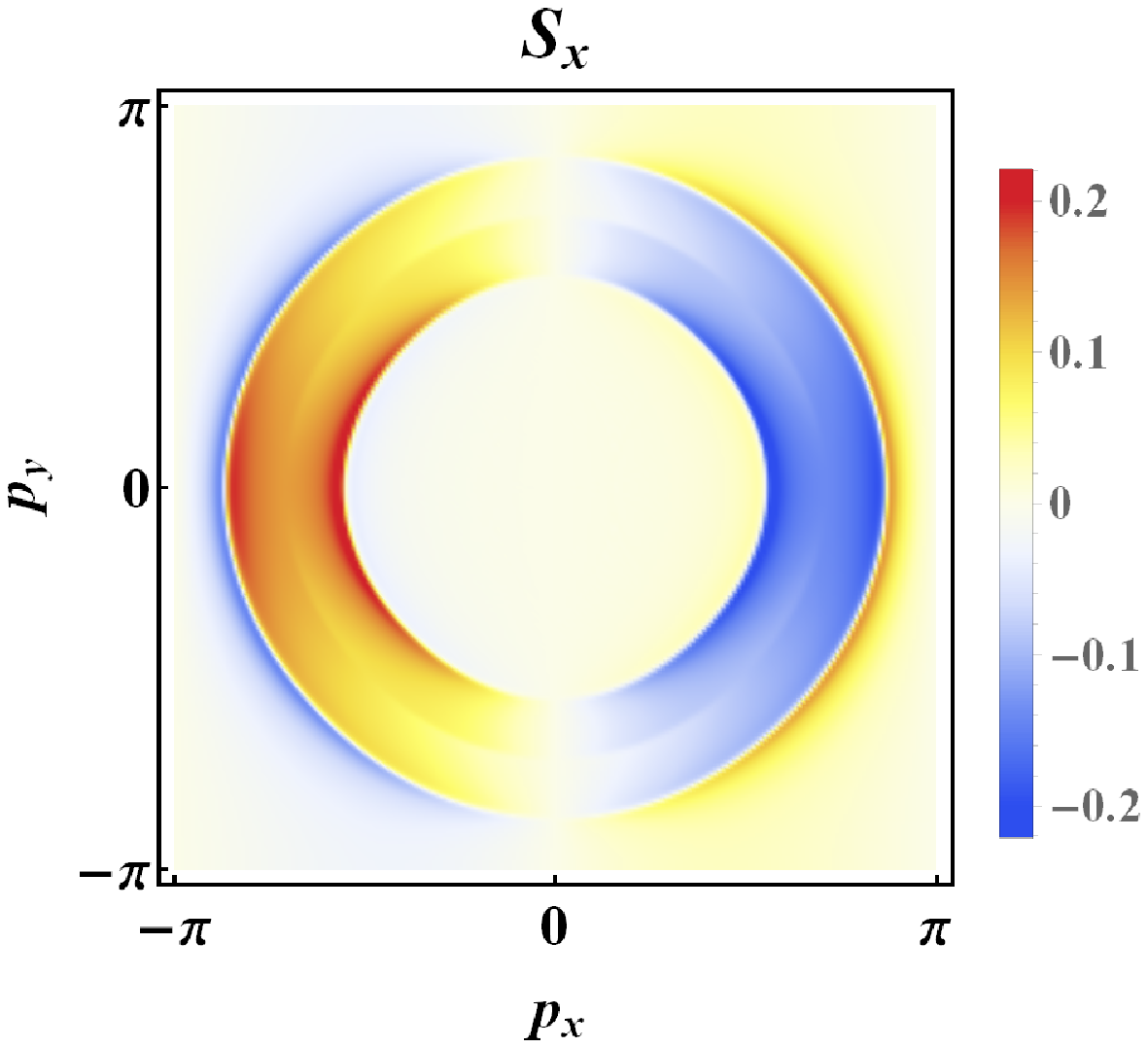} &
		\includegraphics*[width=0.48\columnwidth]{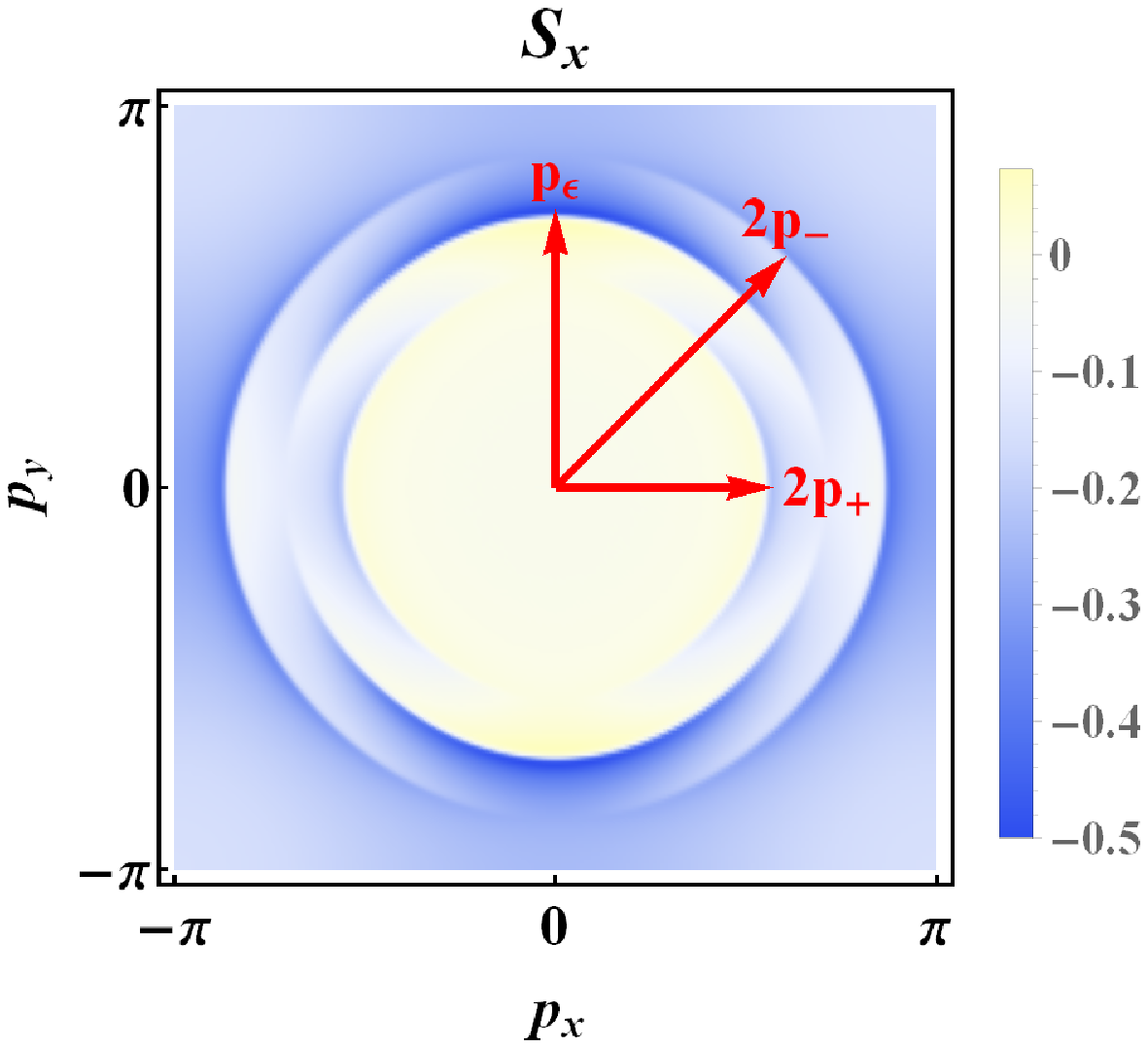} \\
		\includegraphics*[width=0.48\columnwidth]{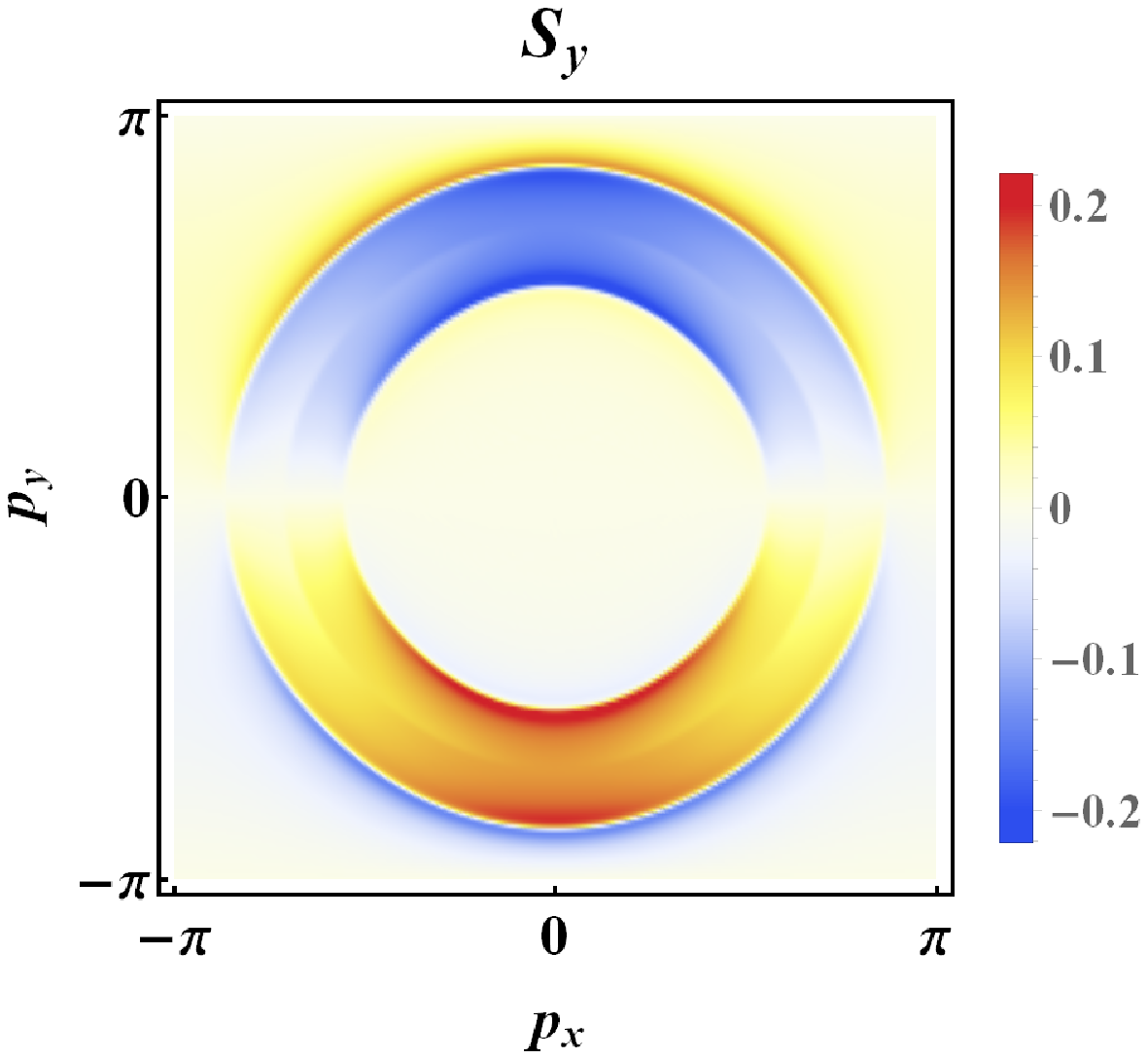} &
		\includegraphics*[width=0.48\columnwidth]{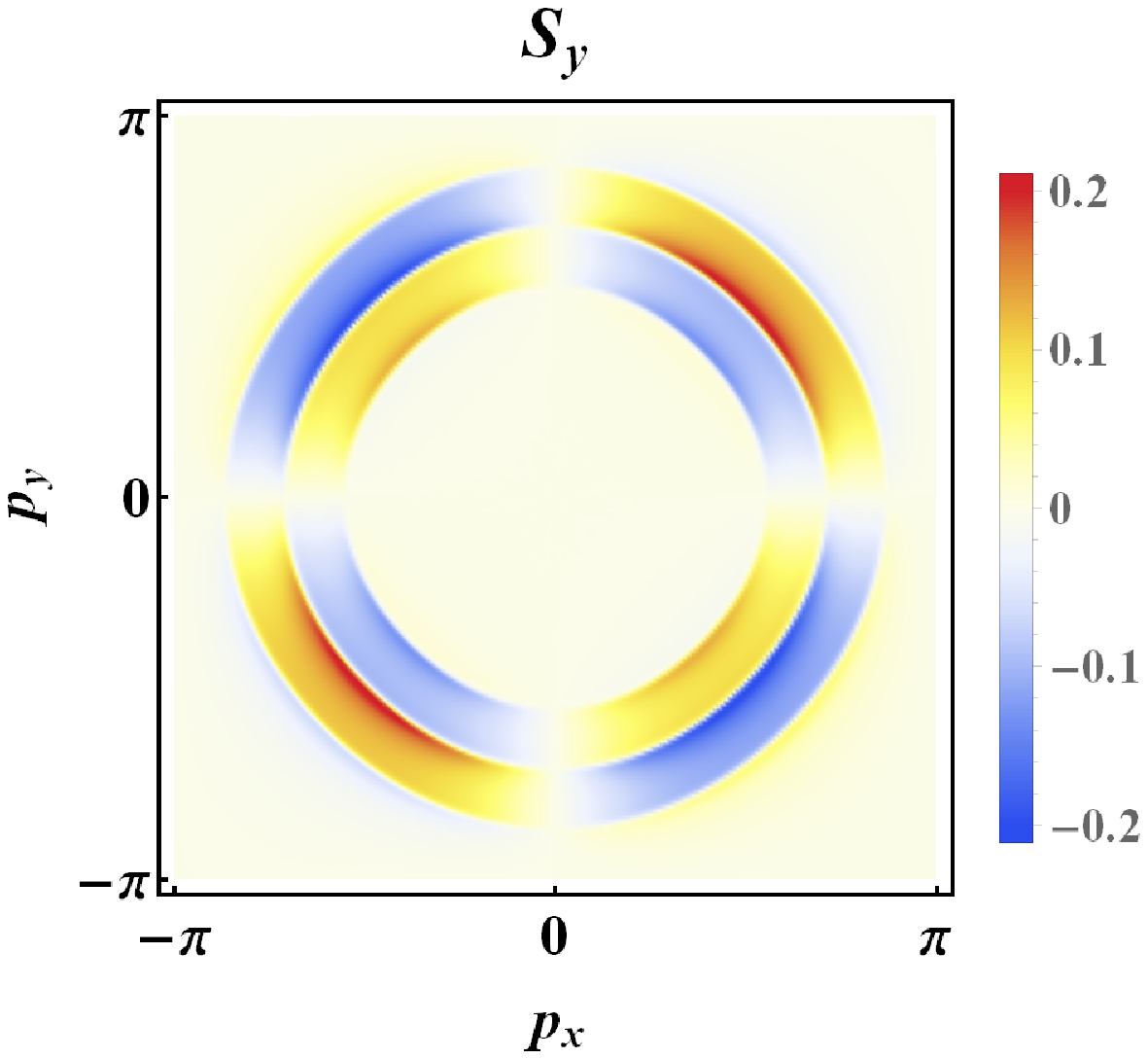} \\
		\includegraphics*[width=0.48\columnwidth]{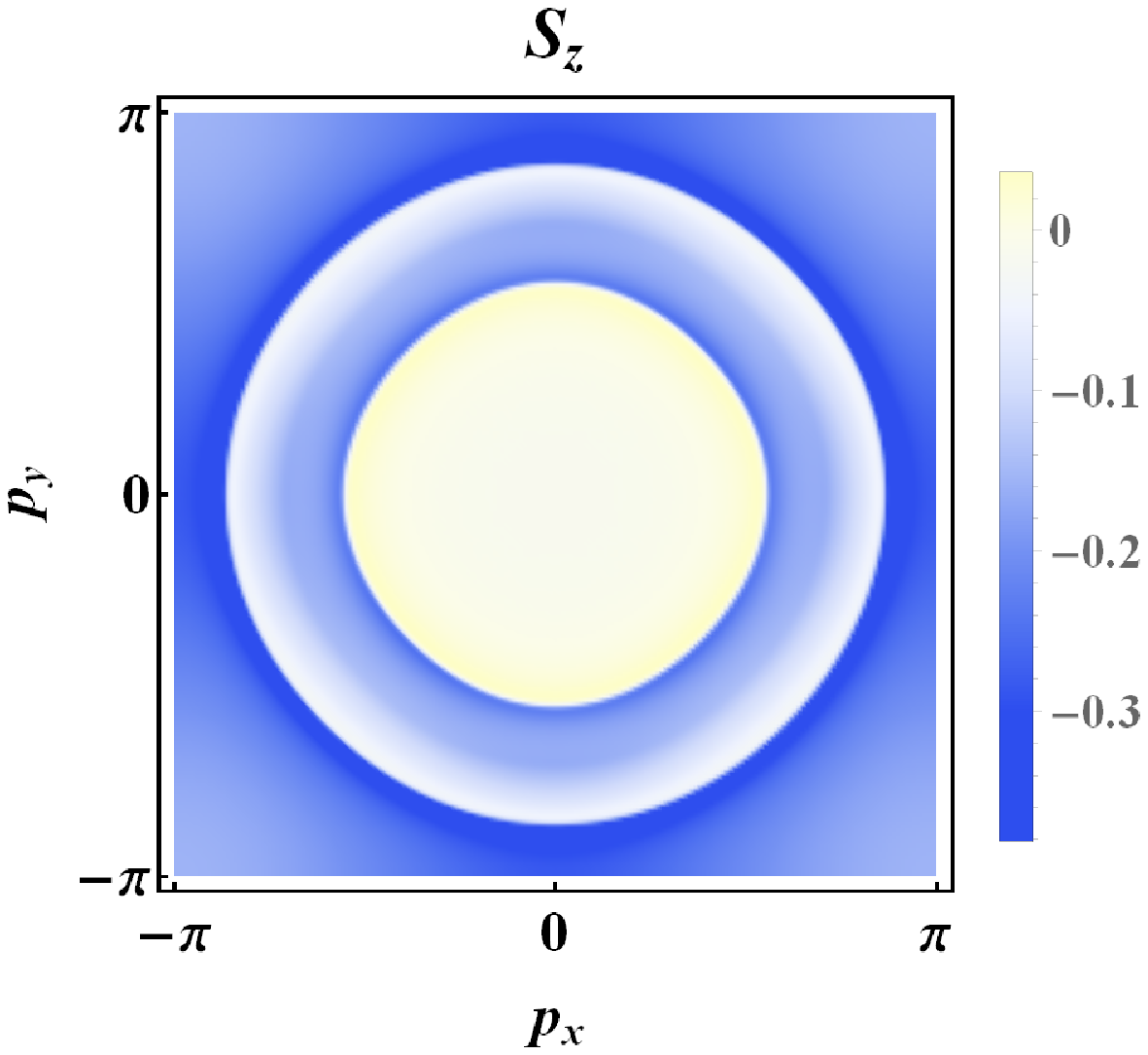} &
		\includegraphics*[width=0.48\columnwidth]{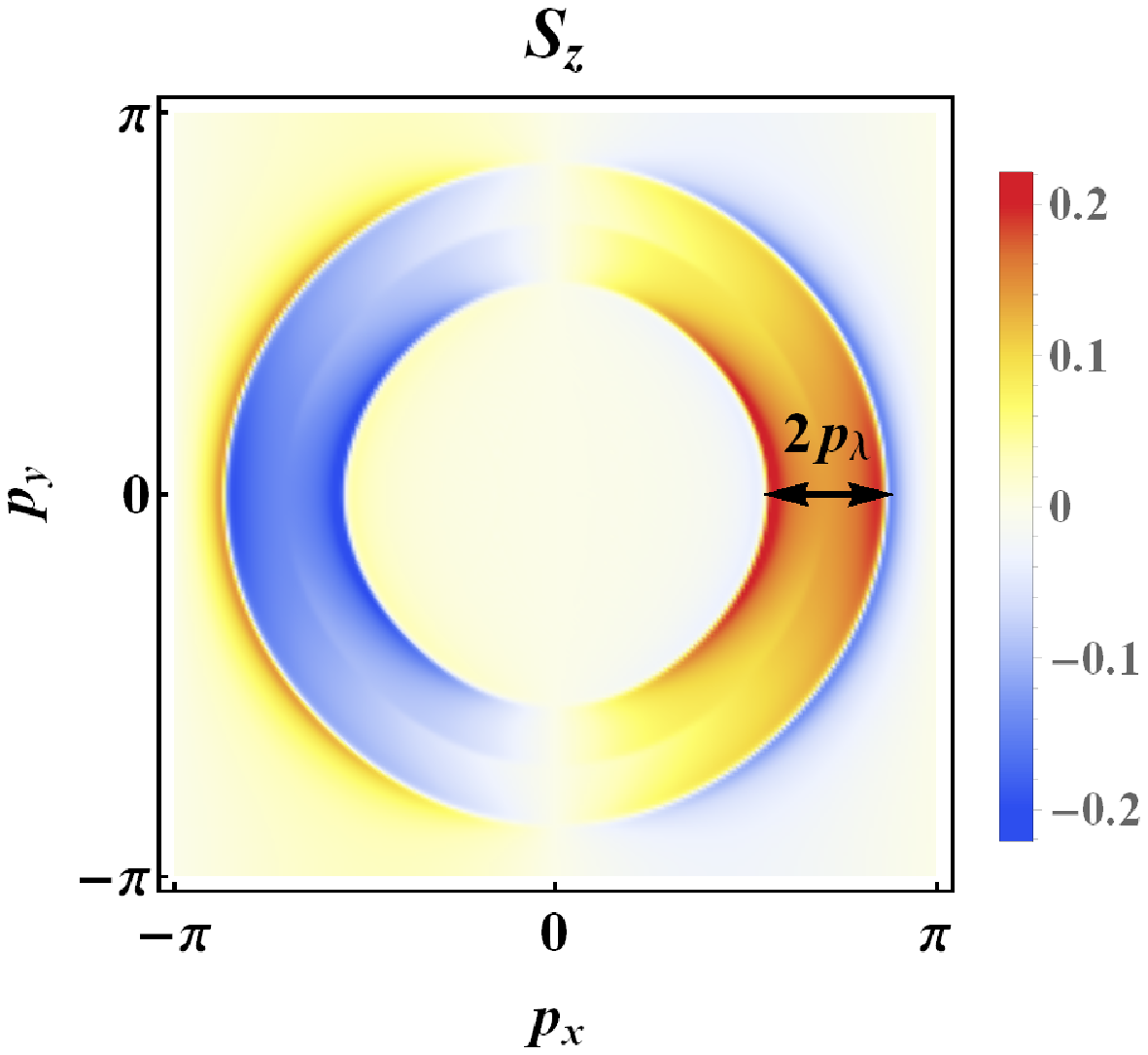} \\
		\includegraphics*[width=0.48\columnwidth]{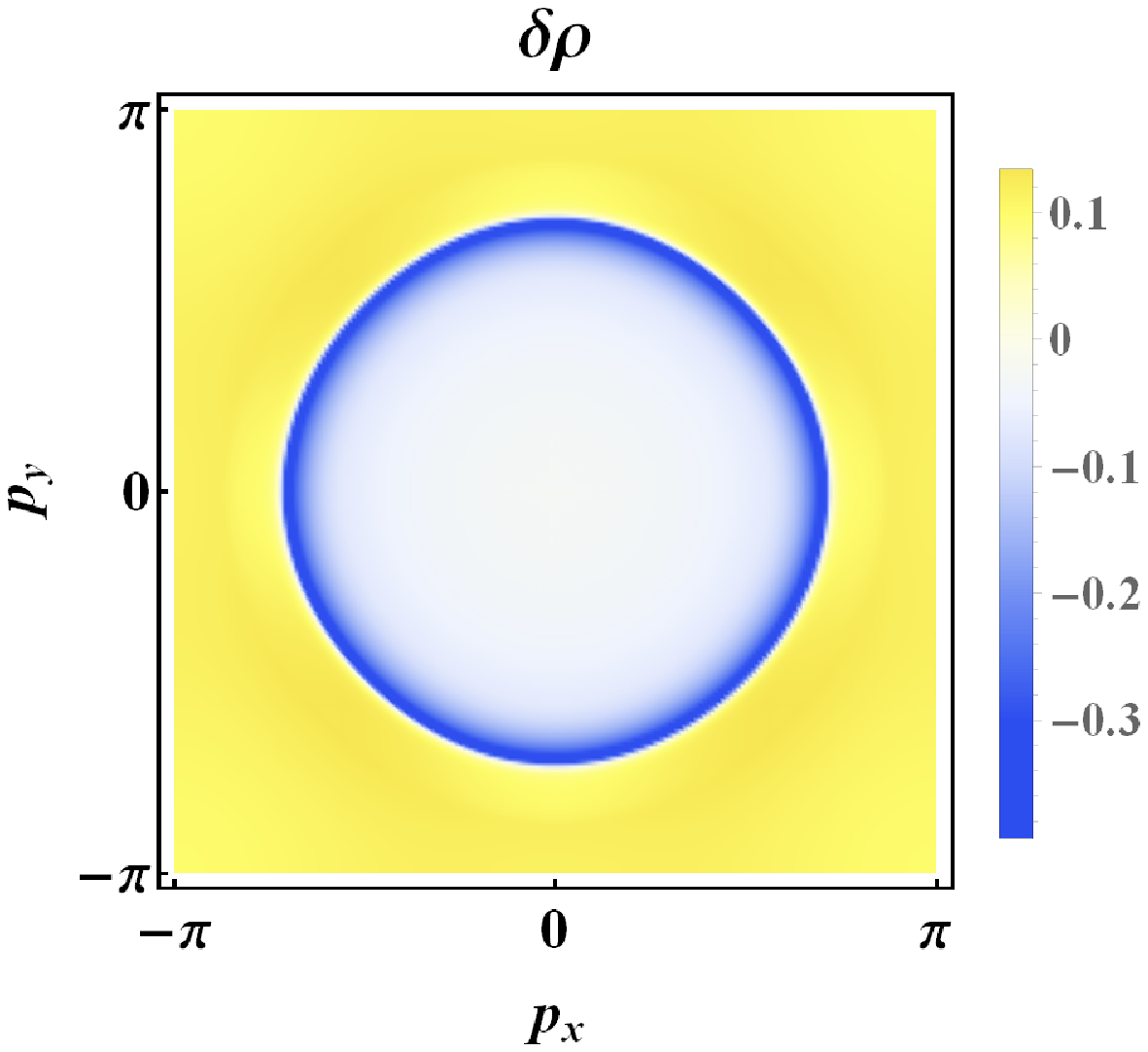} &
		\includegraphics*[width=0.48\columnwidth]{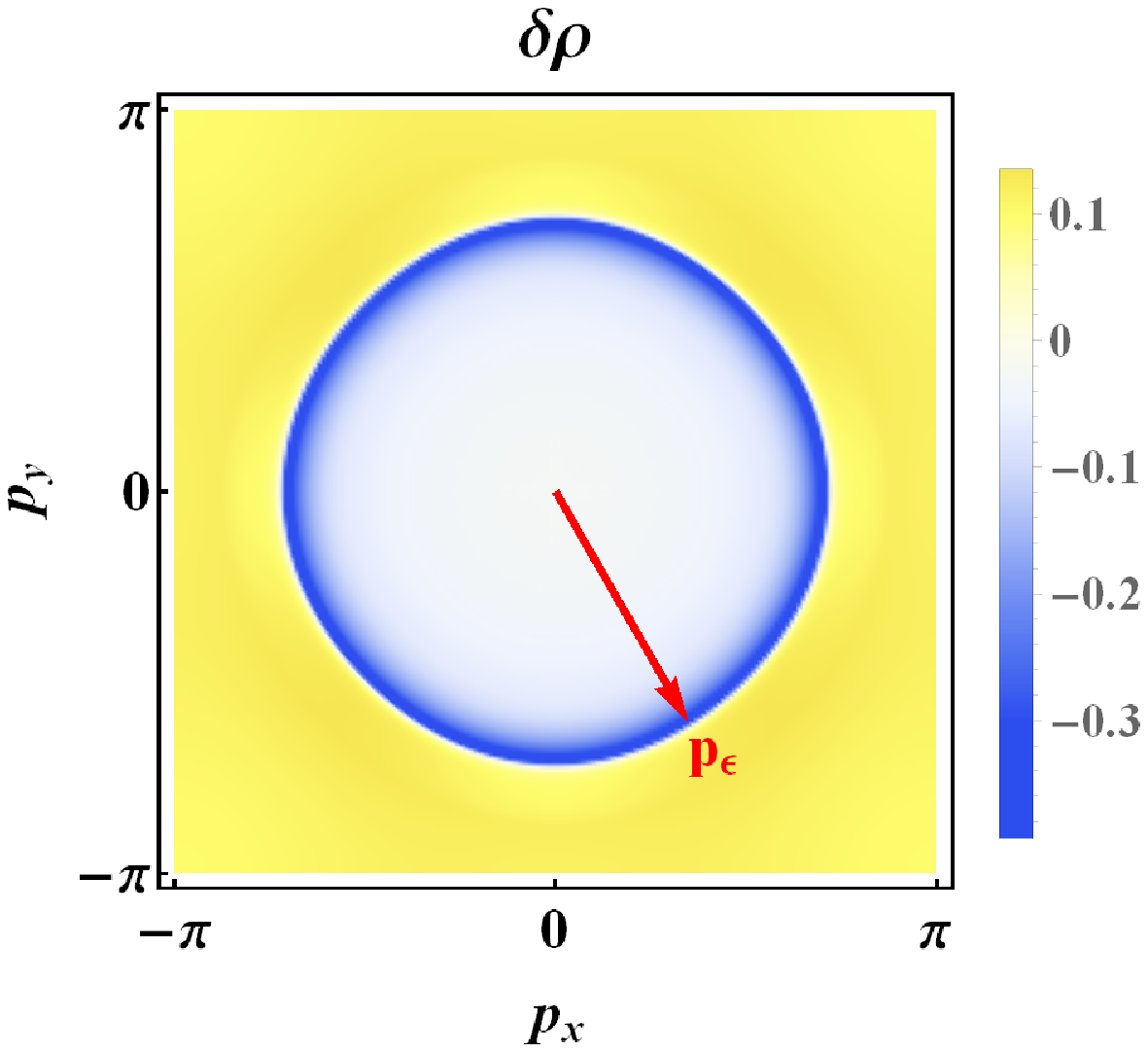} \\
  		$J_z \;(J_x = J_y = 0)\phantom{aa}$ & $J_x \;(J_y = J_z = 0)\phantom{aa}$
\end{tabular}
	\caption{(Color) The FT of the impurity contributions to the non-polarized and SP LDOS for an energy $E=0.1$ and for a magnetic impurity with $J_z=2$ (left column), and $J_x=2$ (right column). We take the inverse quasiparticle lifetime $\delta=0.03$ and we set $t=1, \mu=3, \lambda=0.5, \Delta_s=0$. For a $z$-impurity we depict the real part of the FT for $\delta \rho$ and for $S_z$, and the imaginary part for $S_x$ and $S_y$, whereas for an $x$-impurity we take the imaginary part only for the $S_z$ component. Unlike in the SC case, the strong peaks appearing in the center and at $p_\lambda$ are absent here. The arrows denote the wavevectors of the observed features as identified from the analytical calculations.}
	\label{figFTMetal}
\end{figure}

Thus in Fig.~\ref{figFTMetal} we plot the FT of the impurity contribution to the LDOS and SP LDOS at a fixed energy $E=0.1$. We note that we have similar features to those observed in the SC regime, with the main differences being that the long-wavelength central features are now absent, and that the FT peaks are much sharper than in the SC regime. This behavior can be explained from the analytical expressions of the non-polarized and SP LDOS, whose derivation is presented in Appendix B. The results are presented below for an out-of-plane spin impurity:
\begin{eqnarray}
S_x(\bs r) &\sim& \frac{J}{1+\alpha^2} \frac{\cos \phi_{\bs r}}{r} \sum\limits_\sigma \sigma \frac{\nu^2_\sigma}{p_\sigma}\sin 2p_\sigma r ,\nonumber \\
S_y(\bs r) &\sim& \frac{J}{1+\alpha^2} \frac{\sin \phi_{\bs r}}{r} \sum\limits_\sigma \sigma \frac{\nu^2_\sigma}{p_\sigma}\sin 2p_\sigma r ,\nonumber  \\
S_z(r) &\sim& -\frac{J}{1+\alpha^2} \frac{2}{r} \sum\limits_\sigma  \frac{\nu^2_\sigma}{p_\sigma}\cos 2p_\sigma r, \nonumber  \\
\rho(r) &\sim& -\frac{J}{1+\alpha^2} 4\alpha\nu^2 \frac{v_F^2}{v^2} \frac{1}{\sqrt{p_F^2 + 2mE + E^2/v^2}}\times  \nonumber\\ &&\times\frac{\sin p_\varepsilon r}{r} ,
\end{eqnarray}
while for an $x$ directed impurity (in-plane):
\begin{eqnarray}
S_x(\bs r) &\sim& -\frac{J}{1+\alpha^2} \bigg\{2\nu^2 \frac{v_F^2}{v^2}\frac{1-\cos 2\phi_{\bs r}}{r}  \frac{\cos p_\varepsilon r}{\sqrt{p_F^2 + 2mE + E^2/v^2}} 
\nonumber \\&&
+\sum\limits_\sigma \frac{\nu^2_\sigma}{p_\sigma}\frac{1+\cos 2\phi_{\bs r}}{r}\cos 2p_\sigma r  , 
\nonumber \\
S_y(\bs r) &\sim& -\frac{J}{1+\alpha^2} \frac{\sin 2\phi_{\bs r}}{r} \bigg[ -2\nu^2 \frac{v_F^2}{v^2} \frac{\cos p_\varepsilon r}{\sqrt{p_F^2 + 2mE + E^2/v^2}} 
\nonumber \\&&
+ \sum\limits_\sigma \frac{\nu^2_\sigma}{p_\sigma}\cos 2p_\sigma r\bigg], \nonumber\\
S_z(\bs r) &\sim& -\frac{J}{1+\alpha^2} \frac{\cos \phi_{\bs r}}{r} \sum\limits_\sigma \sigma \frac{\nu^2_\sigma}{p_\sigma}\sin 2p_\sigma r ,\nonumber \\
\rho(r) &\sim& -\frac{J}{1+\alpha^2} \cdot \frac{\alpha}{r} \cdot 4\nu^2 \frac{v_F^2}{v^2} \frac{\sin p_\varepsilon r}{\sqrt{p_F^2 + 2mE + E^2/v^2}} ,
\end{eqnarray}
with $p_F=m v_F$, $p_\sigma = p_F^\sigma + E/v \neq 0$, $p_\varepsilon \equiv p_++p_- = 2 \left( mv + E/v \right)$ and $\nu_\sigma = \nu \left[1-\sigma \frac{\lambda}{v}\right]$.

Note that these expressions are very similar to those obtained in the SC regime, except that the wave vectors of the oscillations now do not include $p_\lambda$. However, this could still be read-off experimentally from the difference between $p_-$ and $p_+$. 
Another important difference between the SC and non-SC regimes is the presence of the exponentially decaying term in the expressions describing the LDOS dependence for the Shiba states in the SC regime. The Shiba states have an exponential decay for distances larger than the superconducting coherence length, while the impurity states in the non-SC regime only decay algebraically as $1/r$. 
In the Fourier space this is translated into a much larger broadening of the features corresponding to the Shiba states in the SC regime with respect to that of the features corresponding to the impurity contributions in metals.  The width of the peaks in the latter is solely controlled by the inverse quasiparticle lifetime $\delta$ and is generally quite small. 

Note also that in both regimes one needs to use the spin-polarized LDOS and magnetic impurities to be able to extract the value of the SO coupling, while the non-polarized LDOS is not sensitive to this wavevector. Last but not least, as described in Appendix B, the LDOS perturbations induced by a non-magnetic impurity do not show any direct signature of the SO coupling (the only contributing wavevector is $2mv$ in the metallic regime, while in the SC regime no Shiba state form for a non-magnetic impurity), thus the only manner to have access to the SO coupling is via spin-polarized STM in the presence of magnetic impurities.

\section{One-dimensional systems}

While in one-dimensional systems superconductivity is not intrinsic, a superconducting gap can be opened via proximitizing them with a superconducting substrate. For such systems it is thus particularly interesting  to study the FT of the SP LDOS for both the superconducting and non-superconducting regimes, as both these regimes can be achieved experimentally at low temperature for the same materials.

We consider the Hamiltonian given by Eqs.(\ref{H0}-\ref{hm}), where we set $p_y \to 0$, and we
perform a T-matrix analysis similar to that described in the previous section for both the SC and non-SC phases, for different directions of the magnetic impurity. The wire is considered to be oriented along the $x$ direction, and the SO coupling is oriented along $y$ \cite{Oreg2010,Lutchyn2010}. We thus expect a similar and more exotic behavior for impurities directed along $x$ and $z$, and a more classical behavior for impurities with the spin parallel to the direction of the SO, thus oriented along $y$.

The energies and wave functions of the Shiba states can be found using the same procedure as for the two-dimensional systems (see Appendix C). This yields for the energies of the states:
\begin{align*}
	E_{1,\bar{1}} = \pm \frac{1-\alpha^2}{1+\alpha^2} \Delta_s, \; \text{where}\; \alpha = J/v.
\end{align*}

\begin{figure}[h]
	\centering
	\begin{tabular}{cc}
		\textbf{SC case} & \textbf{Non-SC case} \\
		\includegraphics*[width=0.48\columnwidth]{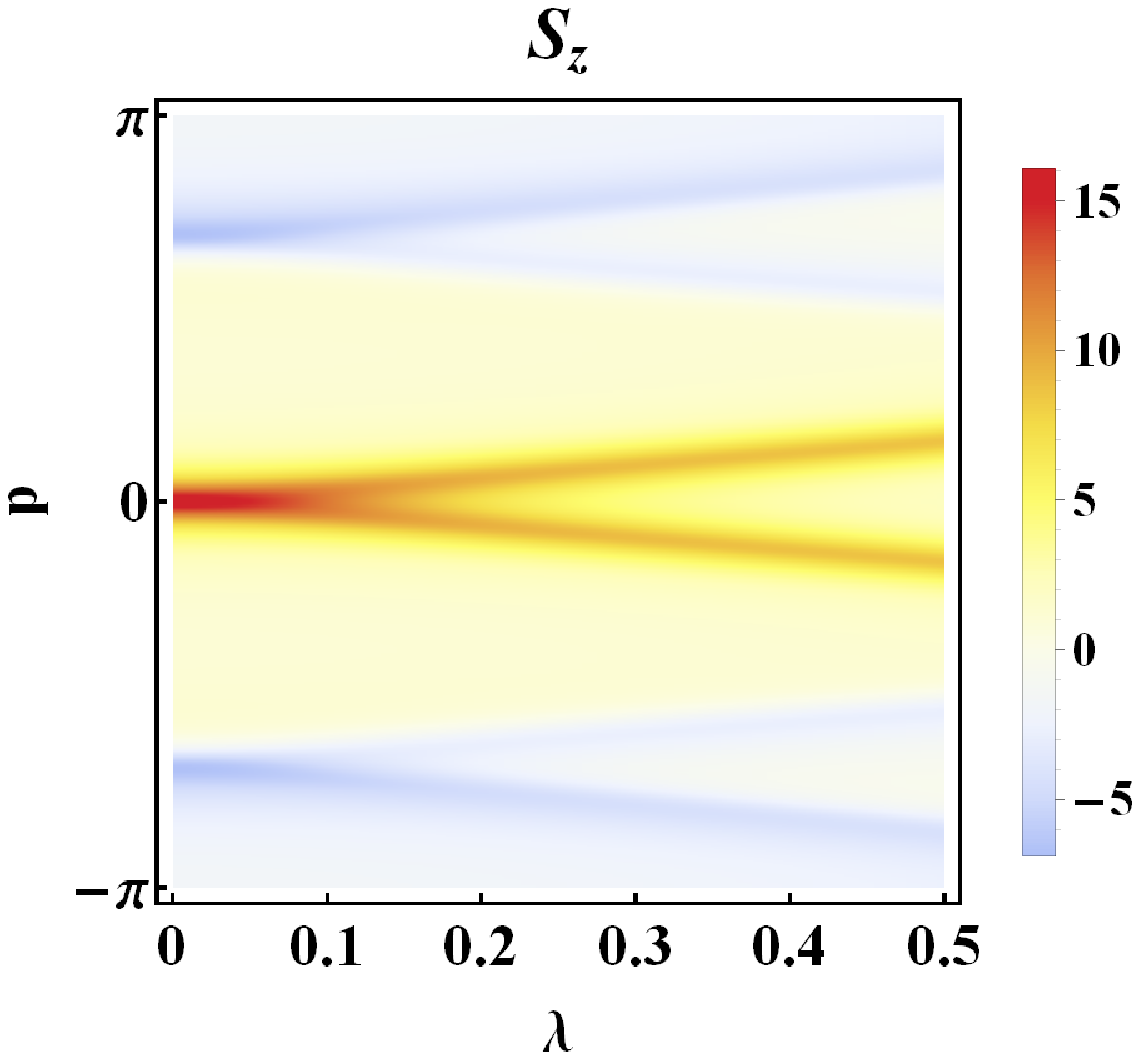} &
		\includegraphics*[width=0.48\columnwidth]{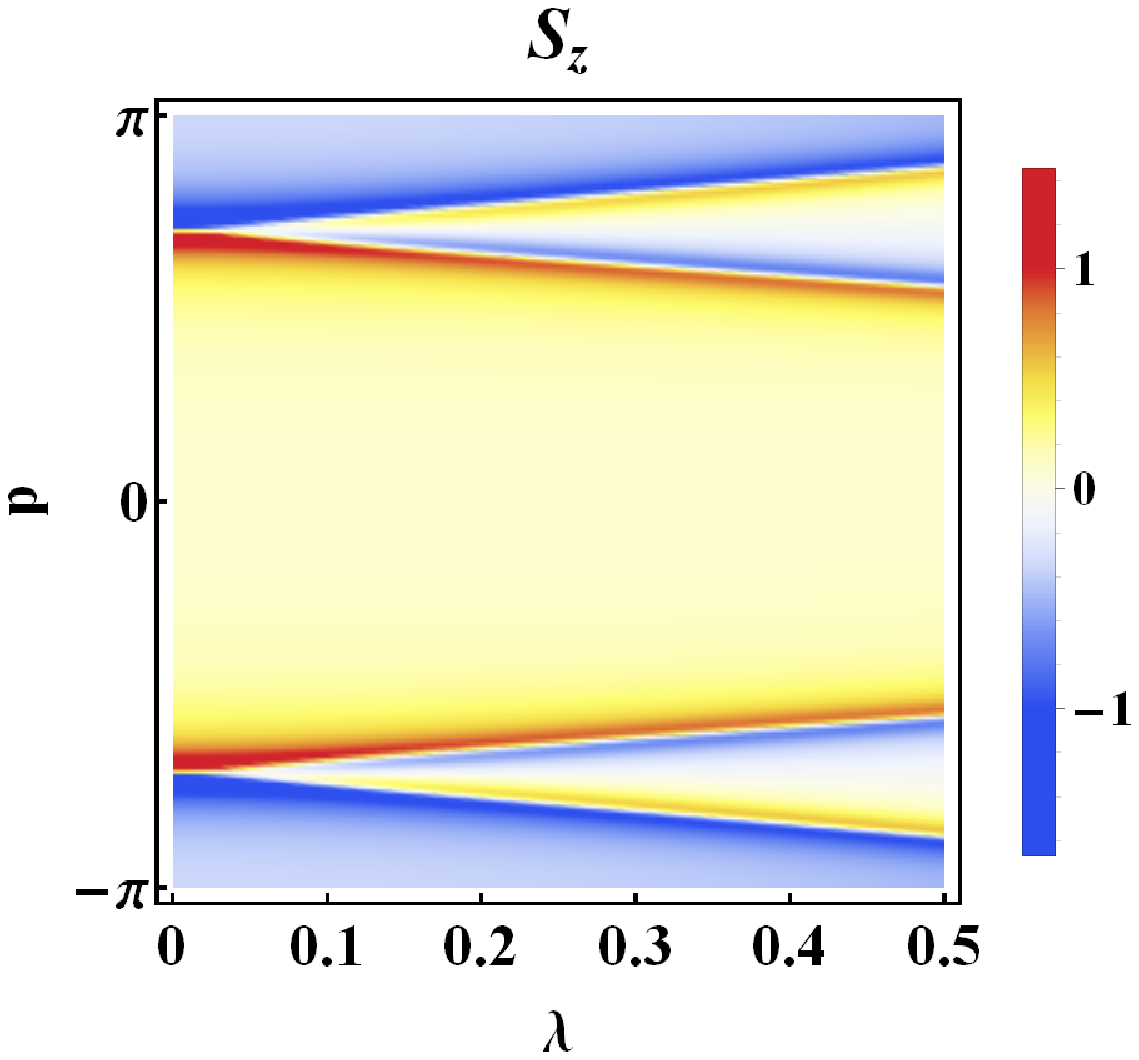}\\
		\includegraphics*[width=0.48\columnwidth]{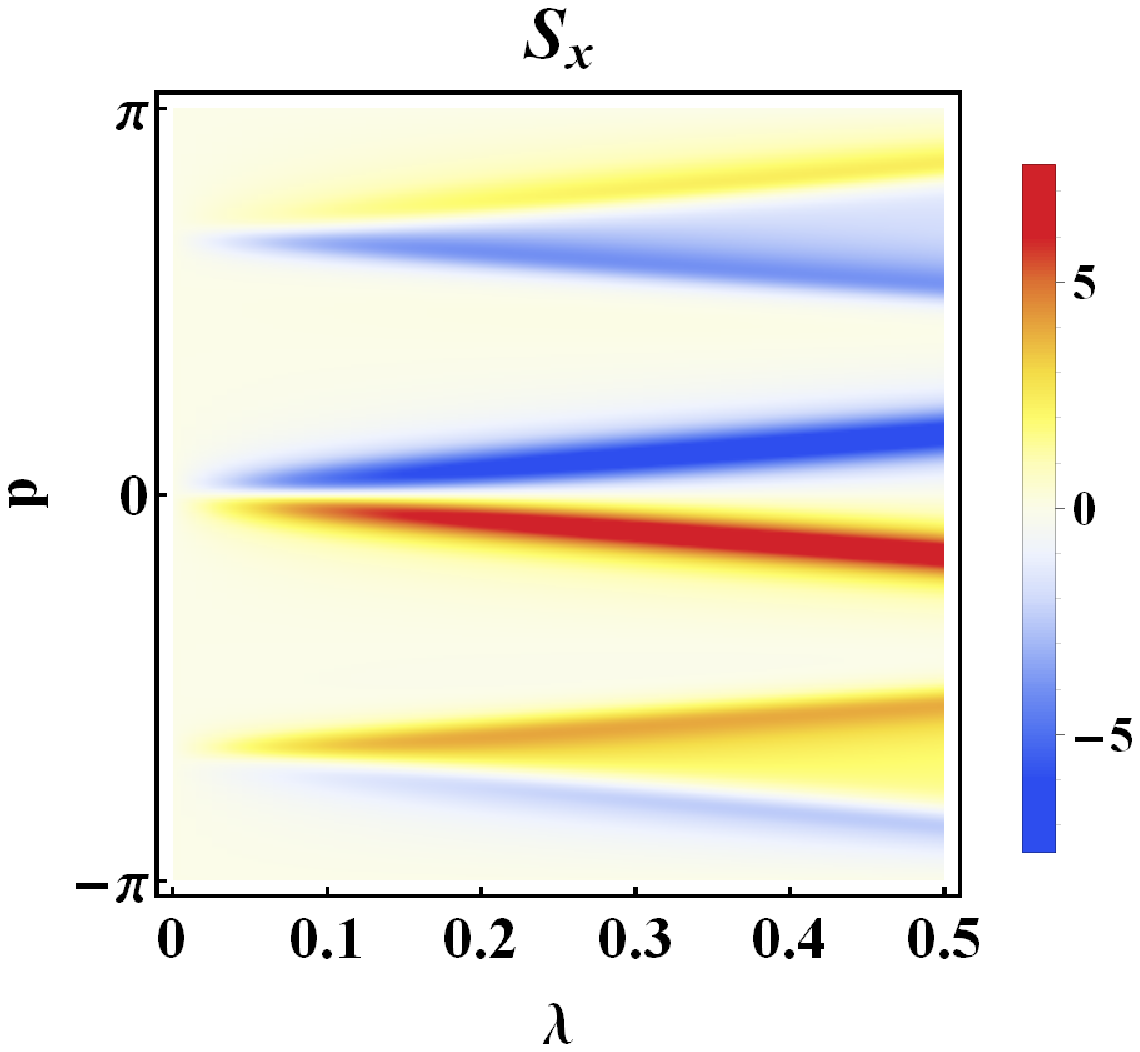} &
		\includegraphics*[width=0.48\columnwidth]{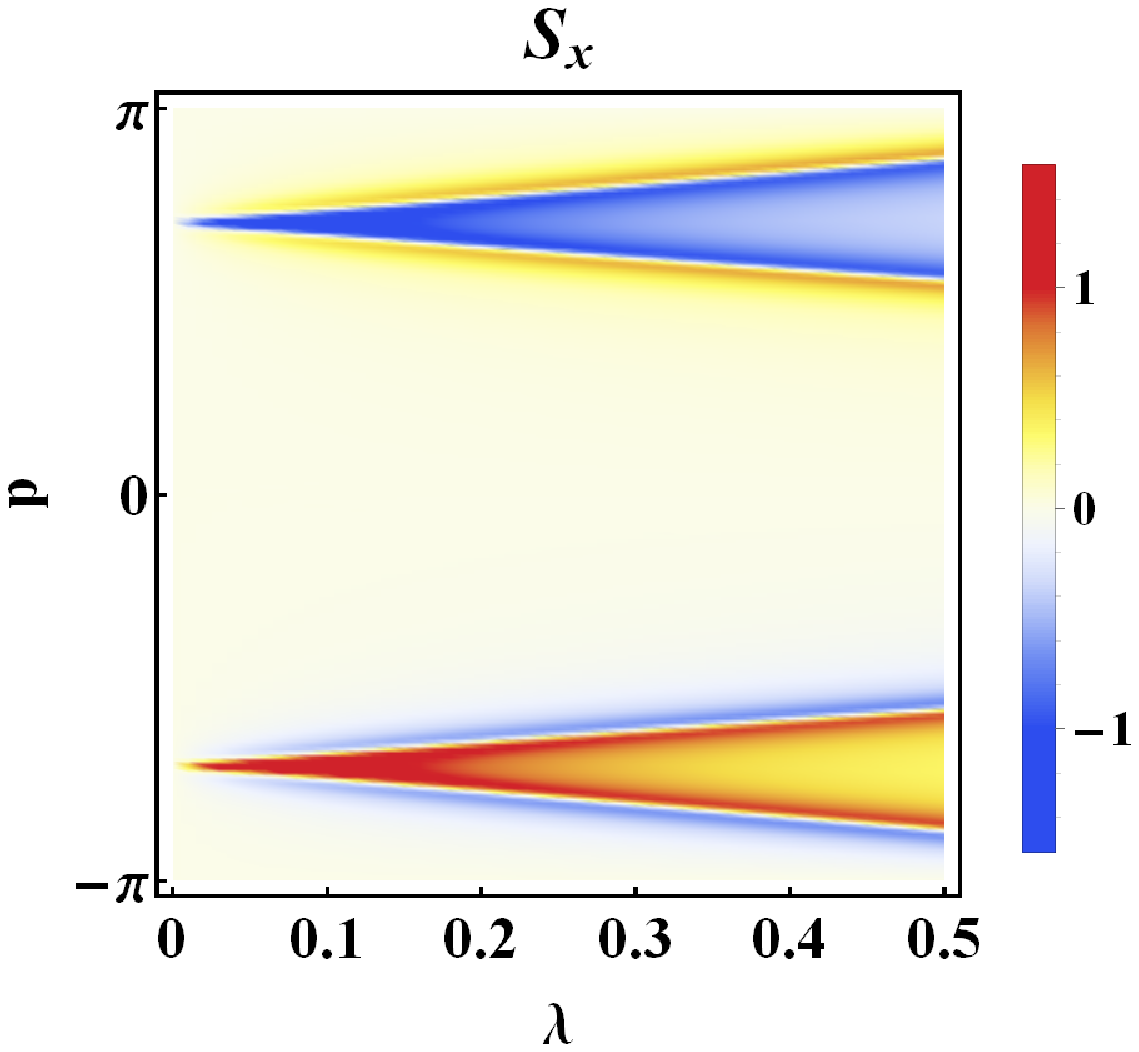}\\
		\includegraphics*[width=0.48\columnwidth]{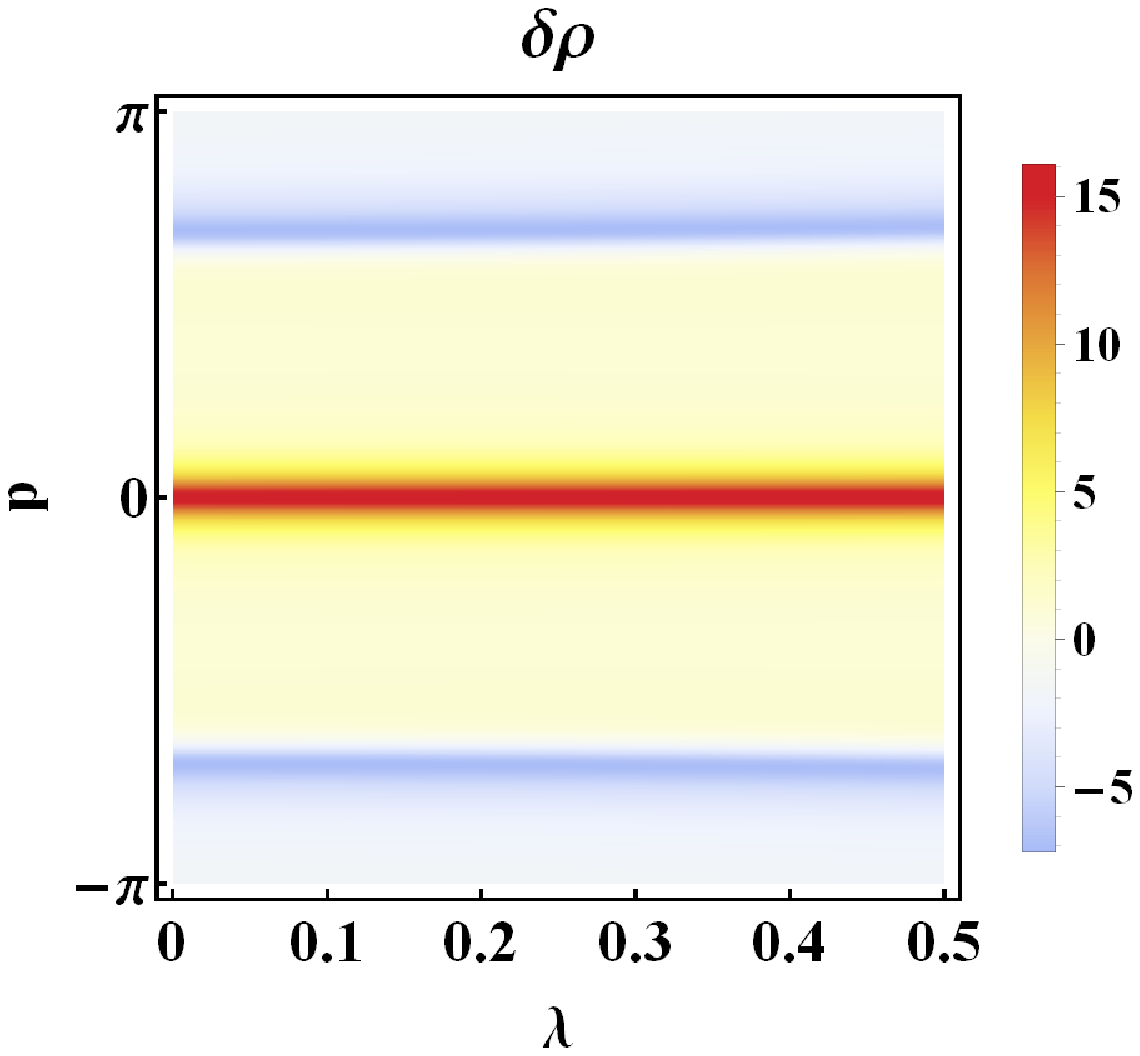} &
		\includegraphics*[width=0.48\columnwidth]{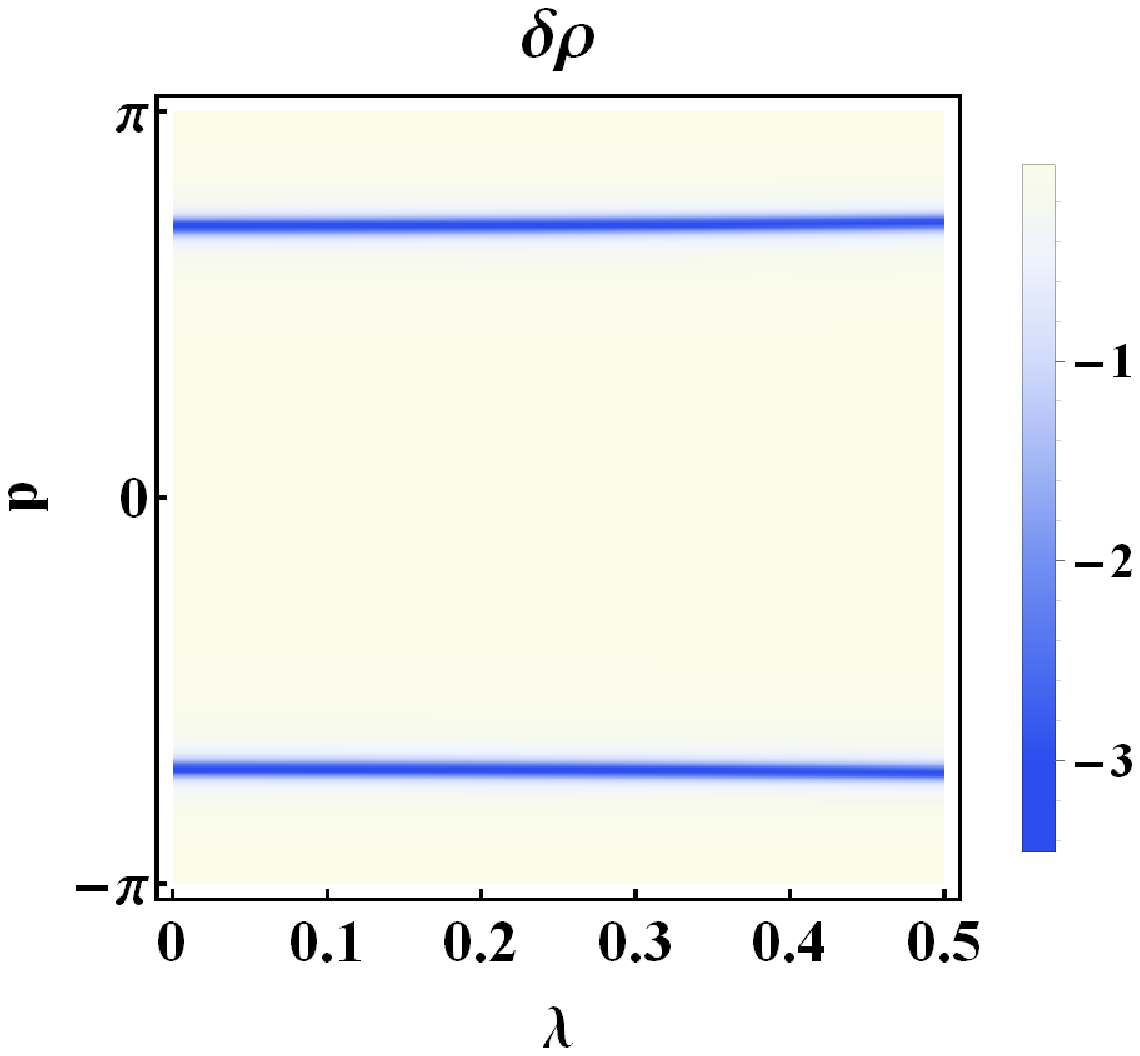}
\end{tabular}
	\caption{(Color) The FT of various SP LDOS component for a Shiba state (left column), and for an impurity state at $E=0.1$ (right column), as a function of the SO coupling $\lambda$ and of momentum $p$, for an impurity perpendicular to the wire and directed along $z$. We set $t=1, \mu=1$. We take $\Delta_s=0.2, J_z=4, \delta=0.01$ in the SC case and $\Delta_s=0, J_z=2, \delta=0.05$ in the non-SC case.}
	\label{figFT1D}
\end{figure}

The FT of the positive energy state as a function of momentum and the SO coupling is presented in Fig.~\ref{figFT1D} for a SC (left column) and non-SC state (right column), for an impurity directed along $z$. For this situation the spin of the Shiba state has two non-zero components, one parallel to the wire, and one parallel to the impurity spin, and these two components are depicted in 
Fig.~\ref{figFT1D}. 
Note that, similar to the two-dimensional case, there is a split of the FT features increasing linearly with the SO coupling strength. Also note that in the non-SC phase the central feature, whose wave vector is given by $p_\lambda$, is absent, and that the FT features are broadened in the SC regime with respect to the non-SC one. Also, same as in the two-dimensional case, the SO affects the spin-polarized components but almost do not change the non-polarized LDOS, as it can be seen in Fig.~\ref{figFT1D} where it appears that the non-polarized LDOS FT features do not evolve with the SO coupling. 

These results are confirmed by analytical calculations. Below we give the spin components and the LDOS in the SC state for an impurity directed along $z$ obtained analytically (see Appendix C), for the positive energy Shiba state:

\ba
S_x(x)& =& \frac{1+\alpha^2}{4} [2\sin p_\lambda x+\sin (2mv |x|+p_\lambda x-2\theta)
\nonumber \\&&-\sin (2mv |x|-p_\lambda x-2\theta)] \cdot e^{-2\omega |x|/v} \nonumber\\
S_y(x) &= &0 \nonumber  \\
S_z(x) &= &-\frac{1+\alpha^2}{4} [2\cos p_\lambda x+\cos (2mv |x|+p_\lambda x-2\theta)
\nonumber\\&&+\cos (2mv |x|-p_\lambda x-2\theta)] \cdot e^{-2\omega |x|/v} \nonumber\\
\rho(x) &= &\frac{1+\alpha^2}{2} [1+\cos(2mv |x|-2\theta)] \cdot e^{-2\omega |x|/v}
\ea
where $\tan \theta = \alpha$.
We also present the FT of the SP LDOS for the non-SC phase for the impurity contribution corresponding to the energy $E$ (see Appendix D):
\begin{align*}
S_x(x) &= +\frac{\alpha}{1+\alpha^2} \cdot \frac{1}{\pi v} \left[ \cos (p_\varepsilon |x| - p_\lambda x) -\cos (p_\varepsilon |x| + p_\lambda x) \right] \\
S_y(x) &= 0 \\
S_z(x) &= +\frac{\alpha}{1+\alpha^2} \cdot \frac{1}{\pi v}  \left[ \sin (p_\varepsilon |x| - p_\lambda x ) +\sin (p_\varepsilon |x| + p_\lambda x ) \right] \\
\rho(x) &= -\frac{2\alpha^2}{1+\alpha^2} \cdot \frac{1}{\pi v} \cos p_\varepsilon x 
\end{align*}
As before, in the expressions above $p_\varepsilon = 2(mv+E/v), p_\lambda = 2m\lambda$.

Indeed these calculations confirm our observations, in the SC state the dominant wave vectors are $2p_F^\pm=2mv\pm p_\lambda$, $2mv$ and $p_\lambda$, while in the non-SC phase only $p_\epsilon \pm p_\lambda$, and $2mv$. 

Similar results are obtained if the impurity is oriented along $x$, with the only difference that the $x$ and $z$ components will be interchanged, up to on overall sign change (see Appendices C and D). 
For impurities parallel to $y$, and thus to the SO vector, we expect the SP LDOS to be less exotic, and indeed in this case the only non-zero component of the impurity SP LDOS is $S_y$.  In the SC regime we thus find
\begin{align*}
S_x(x) &= 0 \\
S_y(x) &= -(1+\alpha^2)[1+ \cos(2 mv |x|-2\theta) ]\cdot e^{-2\omega |x|/v} \\
S_z(x) &= 0 \\
\rho(x) &= +(1+\alpha^2) [1+ \cos(2 mv |x|-2\theta) ]\cdot e^{-2\omega |x|/v}
\end{align*}
while in the non-SC regime we have
\begin{align*}
S_x(x) &= 0 \\
S_y(x) &= +\frac{2\alpha}{1+\alpha^2} \cdot \frac{1}{\pi v} \sin p_\varepsilon |x| \\
S_z(x) &= 0 \\
\rho(x) &= -\frac{2\alpha^2}{1+\alpha^2} \cdot \frac{1}{\pi v} \cos p_\varepsilon x
\end{align*}
We see that $S_y$ exhibits features only at the $2 mv$ and correspondingly at the $p_\epsilon$ wave vectors, same as the non-polarized LDOS, thus not allowing for the detection of the SO coupling. 

For intermediate directions of the impurity spin, all three components will be present, with the $x$ and $z$ exhibiting all the wave vectors, while the $y$ component solely the $2mv$, and with relative intensities given by the relative components of the impurity spin. 

Thus, we conclude that, same as in the 2D case, the SO can be measured using spin-polarized STM and magnetic impurities; moreover, in the 1D case one needs to consider impurities that have a non-zero component perpendicular to the direction of the SO. 

\section{Conclusions}
We have analyzed the formation of Shiba states and impurity states in 1D and 2D superconducting and metallic systems with Rashba SO coupling. In particular we have studied the Fourier transform of the local density of states of Shiba states in SCs and of the impurity states in metals, both non-polarized and spin-polarized. We have shown that the spin-polarized density of states contains information that allows one to extract experimentally the strength of the SO coupling. In particular the features observed in the FT of the SP LDOS split with a magnitude proportional to the SO coupling strength. Moreover, the Friedel oscillations in the SP LDOS in the SC regime show a combination of wavelengths, out of which the SO length can be read off directly and non-ambiguously. We note that these signatures are only visible in the spin-polarized quantities and in the presence of magnetic impurities.
For  non-spin-polarized measurements, no such splitting is present and the wave vectors observed in the FT of the SP LDOS basically do not depend on the SO coupling. 
When comparing the results for the SC Shiba states to the impurity contribution in the metallic state and we find a few interesting differences, such as a broadening of the FT features corresponding to a spatial exponential decay of the Shiba states compared to the non-SC case. 
Moreover, the FT of the SP LDOS in the SC regime exhibits extra features with a wavelength equal to the SO length which are not present in the non-SC phase. It would be interesting to generalize our results to more realistic calculations which may include some specific lattice characteristics, more realistic material-dependent tight-binding parameters for the band structure and the SO coupling values. However, we should note that our results have a fully general characteristic, independent of the band structure or other material characteristics, and that the features in the FT of the non-polarized LDOS will correspond to split features in the spin-polarized LDOS, and thus the spin-orbit can be measured unequivocally from the split obtained from the comparison between the non-polarized and spin-polarized measurements. We have checked that up to a rotation in the spin space our results hold also for other types of SO coupling such as Dresselhaus.

According to our knowledge, the FT-STM is a well-established experimental technique which does not deal with large systematic errors \cite{Hasegawa1993,Crommie1993,Sprunger1997,Petersen2000,Hofmann1997,Vonau2004,Vonau2005,Simon2011}. The experimental data presented e.g. in Ref. \onlinecite{Simon2011} shows that the resolution in the Fourier space (momentum space) reaches $0.05 \;\mathrm{\AA}^{-1}$, whereas a typical value of spin-orbit coupling wave vector $p_\lambda \sim 0.15 \; \mathrm{\AA}^{-1}$ (see e.g. Ref. \onlinecite{Manchon2015}), and thus it is sufficient to resolve the features originating from the spin-orbit coupling. Moreover, we would like to point that the exponent $e^{-2p_s r}$ defines in the real space how far the impurity-induced states are extended, and it manifests in the momentum space as the widening of the ring-like features appearing at particular momenta. The condition of resolving the spin-orbit is thus $2p_s < p_\lambda$, otherwise the widening is large enough to blur the spin-orbit feature. This condition can be rewritten in a more explicit way, namely
	$$
		\frac{1}{\sqrt{1+(\lambda/v_F)^2}}\cdot \frac{\alpha}{1+\alpha^2} \cdot \frac{\Delta_s}{\varepsilon_F} < \frac{\lambda}{v_F}
	$$
	For any realistic parameters the first two factors on the left side are of the order of unity, and $\Delta_s/\varepsilon_F \sim 10^{-3}$ for superconductors. However, for realistic values of the spin-orbit coupling $\lambda$, this inequality holds and therefore there should not be any technical problem with resolving those features.

 Our results can be tested using for example materials such as Pb, Bi, NbSe$_2$ or InAs and InSb wires, which are known to have a strong SO coupling, using spin-polarized STM which is nowadays becoming more and more available \cite{Wiesendanger2009}.

While finalizing this manuscript we became aware of a recent work \cite{Ruitenbeek2016} focusing on issues similar to some of the subjects (in particular the real space Friedel oscillations in the metallic regime) addressed in our work.

\acknowledgements

This work is supported by the ERC Starting Independent Researcher Grant NANOGRAPHENE 256965. PS  would like to acknowledge interesting discussions with T. Cren and financial support from the  French Agence Nationale de la Recherche through the contract Mistral.

\bibliography{biblio_SO}

\widetext
\appendix

\section{Analytical calculation of the Shiba states wave functions for a 2D system}
We can calculate analytically the non-polarized and the SP LDOS for the Shiba states exploiting the model described by the Hamiltonians in Eqs.~(\ref{H0}-\ref{hm}). All the integrations below are performed using a linearization  around the Fermi energy. The energies of the Shiba states can be found by solving the 
corresponding eigenvalue equation\cite{Pientka2013} 
\begin{equation}
	\left[\mathbb{I}_4-V G_0(E,{\bm r}={\bm 0}) \right]\Phi(\bm 0) = 0
\label{eigenvap}
\end{equation}
where $G_0(E,{\bm r})$ is the retarded Green's function in real space obtained by a Fourier transform from the retarded Green's function in momentum space $G_0(E,{\bm p})=\left[(E+i\delta)\mathbb{I}_4-\mathcal{H}_0({\bm p})\right]^{-1}$, where $\delta$ is the inverse quasiparticle lifetime. In all the calculations below we take the limit of $\delta \to +0$, and we specify $+i0$ only in the cases when it affects the results. The wave functions of the Shiba states at ${\bm r}=0$ are given by the eigenfunctions obtained from the equation above. Their spatial dependence is determined using
\begin{equation}
	\Phi (\bm r) = G_0(E, \bm r) V \Phi(\bm 0)
\label{eigenfap}
\end{equation}
Consequently, the non-polarized and the SP LDOS are given by
\begin{equation}
	\rho(E, \bm r) = \Phi^\dag(\bm r) 
		\begin{pmatrix} 	
			0 & 0 \\ 
			0 & \sigma_0
		\end{pmatrix} \Phi (\bm r),
\label{LDOS_ap}
\end{equation}
\begin{equation}
	\bm S(E, \bm r) = \Phi^\dag(\bm r) 
		\begin{pmatrix} 	
			0 & 0 \\ 
			0 & \bm \sigma
		\end{pmatrix} \Phi (\bm r).
\label{SPLDOS_ap}
\end{equation}
Thus, in order to find the energies and the wave functions corresponding to the Shiba states we need to find the real-space Green's function. This is obtained simply by a Fourier transform of  the unperturbed Green's function in momentum space, $G_0(E,{\bm p})$. We start by writing down the unperturbed Green's function in momentum space, which is given by $G_0(E,\bm p) = \frac{1}{2}\sum\limits_{\sigma =\pm} G^\sigma_0(E,\bm p)$, where
\begin{align}
G^\sigma_0(E,\mathbf{p}) = -\frac{1}{\xi^2_\sigma+\omega^2}
	\bpm
		1 & i\sigma e^{-i\phi_{\mathbf{p}}}  \\
		-i\sigma e^{i\phi_{\mathbf{p}}} & 1 \\
	\epm
	\otimes
	\bpm
		E+\xi_\sigma & \Delta_s \\
		\Delta_s & E-\xi_\sigma
	\epm,
\end{align}
where $\omega = \sqrt{\Delta_s^2-E^2},\, \xi_\sigma = \xi_{\mathbf{p}} + \sigma \lambda p$. To obtain its real-space dependence one needs to perform the Fourier transform:
$$
G^\sigma_0(E,\mathbf{r}) = \int \frac{d\mathbf{p}}{(2\pi)^2} G^\sigma_0(E,\mathbf{p}) e^{i\mathbf{p r}}
$$
We will have four types of integrals:
\begin{align}
X^\sigma_0(\mathbf{r}) &= -\int \frac{d\mathbf{p}}{(2\pi)^2} \frac{e^{i\mathbf{p r}}}{\xi_\sigma^2+\omega^2} \\
X^\sigma_1(\mathbf{r}) &= -\int \frac{d\mathbf{p}}{(2\pi)^2} \frac{\xi_\sigma\, e^{i\mathbf{p r}}}{\xi_\sigma^2+\omega^2}\\
X^\sigma_2(s,\mathbf{r}) &= -\int \frac{d\mathbf{p}}{(2\pi)^2} \frac{-is \sigma e^{is\phi_{\mathbf{p}}}\,e^{i\mathbf{p r}}}{\xi_\sigma^2+\omega^2} \\
X^\sigma_3(s,\mathbf{r}) &= -\int \frac{d\mathbf{p}}{(2\pi)^2} \frac{-is \sigma e^{is\phi_{\mathbf{p}}}\,\xi_\sigma\, e^{i\mathbf{p r}}}{\xi_\sigma^2+\omega^2}
\end{align}
Since the spectrum is split by SO coupling, there will be two Fermi momenta which can be found the following way:
$$
\frac{p^2}{2m} + \sigma \lambda p - \varepsilon_F = 0, \quad p^\sigma_F = \frac{-\sigma \lambda + \sqrt{\lambda^2+2\varepsilon_F/m}}{1/m}
$$
For $p>0$ we linearize the spectrum around the Fermi momenta, thus:
$$
\xi_\sigma \approx \left(\frac{p_F^\sigma}{m}+\sigma \lambda \right)(p-p_F^\sigma) = \sqrt{\lambda^2+2\varepsilon_F/m}\;(p-p_F^\sigma) \equiv v (p-p_F^\sigma),
$$ 
therefore $p = p_F + \xi_\sigma/v$, where $v = \sqrt{v_F^2+\lambda^2}$. We rewrite:
$$
\frac{d\mathbf{p}}{(2\pi)^2} = \frac{m}{2\pi} \left[1-\sigma \frac{\lambda}{v}\right] d\xi_\sigma \frac{d\phi}{2\pi} = \nu_\sigma d\xi_\sigma \frac{d\phi}{2\pi},
$$ 
where $\nu_\sigma = \nu \left[1-\sigma \frac{\lambda}{v}\right]$, with $\nu = m/2\pi$. Due to the symmetry all the integrals are zero at $\mathbf{r=0}$ except for the first one, namely,
\begin{align}
X^\sigma_0(\mathbf{0}) = -\nu_\sigma \frac{\pi}{\omega}.
\end{align}
All the coordinate dependences can be calculated using the formalism introduced in Ref. \onlinecite{Kaladzhyan2016}. Finally we get:
\begin{align}
X^\sigma_0(\mathbf{r}) &= -2\nu_\sigma \cdot \frac{1}{\omega} \cdot \Im \mathrm{K}_0 \left[-i(1+i \Omega_\sigma)p_F^\sigma r \right] \\
X^\sigma_1(\mathbf{r}) &= -2\nu_\sigma \cdot \Re \mathrm{K}_0 \left[-i(1+i \Omega_\sigma)p_F^\sigma r \right]  \\
X^\sigma_2(s,\mathbf{r}) &= 2s\sigma\nu_\sigma \cdot \frac{1}{\omega} \cdot e^{i s \phi_{\mathbf{r}}}\cdot \Re \mathrm{K}_1 \left[-i(1+i \Omega_\sigma)p_F^\sigma r \right]\\
X^\sigma_3(s,\mathbf{r}) &= -2s\sigma\nu_\sigma \cdot e^{i s \phi_{\mathbf{r}}}\cdot \Im \mathrm{K}_1 \left[-i(1+i \Omega_\sigma)p_F^\sigma r \right],
\end{align}
where $\Omega_\sigma = \omega/p_F^\sigma v$ defines the inverse superconducting decay length, and $p_S = \omega/v$. Therefore, the Green's function can be written as
\begin{align}
G^\sigma_0(E,\mathbf{r}) =
	\bpm
			E X^\sigma_0(r)+X^\sigma_1(r) & E X^\sigma_2(-,\mathbf{r})+X^\sigma_3(-,\mathbf{r}) & \Delta_s X^\sigma_0(r) & \Delta_s X^\sigma_2(-,\mathbf{r}) \\ 
			E X^\sigma_2(+,\mathbf{r})+X^\sigma_3(+,\mathbf{r}) & E X^\sigma_0(r)+X^\sigma_1(r) & \Delta_s X^\sigma_2(+,\mathbf{r}) & \Delta_s X^\sigma_0(r) \\
		 	\Delta_s X^\sigma_0(r) & \Delta_s X^\sigma_2(-,\mathbf{r}) & E X^\sigma_0(r)-X^\sigma_1(r) & E X^\sigma_2(-,\mathbf{r})-X^\sigma_3(-,\mathbf{r})\\
		 	\Delta_s X^\sigma_2(+,\mathbf{r}) & \Delta_s X^\sigma_0(r) & E X^\sigma_2(+,\mathbf{r})-X^\sigma_3(+,\mathbf{r}) & E X^\sigma_0(r)-X^\sigma_1(r) 
	\epm.
\end{align}
Thus we have:
\begin{align}
G_0(E,\mathbf{r}=\mathbf{0}) = - \frac{\pi \nu}{\sqrt{\Delta_s^2-E^2}}
		\bpm 	E \sigma_0 & \Delta_s \sigma_0 \\ 
				\Delta_s \sigma_0 & E \sigma_0  \\
		\epm.
\end{align}
\subsection{z-impurity}
The coordinate dependence of the eigenfunctions is given by
\begin{align}
\Phi_{\bar 1}(\mathbf{r}) = +\frac{J_z}{2}\sum\limits_{\sigma =\pm}
	\bpm
		(E_{\bar 1}-\Delta_s)X_0^\sigma(r) + X_1^\sigma(r) \\
		(E_{\bar 1}-\Delta_s) X^\sigma_2(+,\mathbf{r})+X^\sigma_3(+,\mathbf{r})\\
		-(E_{\bar 1}-\Delta_s)X_0^\sigma(r) + X_1^\sigma(r) \\
		-(E_{\bar 1}-\Delta_s) X^\sigma_2(+,\mathbf{r})+X^\sigma_3(+,\mathbf{r})
	\epm, \;
\Phi_1(\mathbf{r}) = -\frac{J_z}{2}\sum\limits_{\sigma =\pm}
	\bpm
		(E_1+\Delta_s) X^\sigma_2(-,\mathbf{r})+X^\sigma_3(-,\mathbf{r})\\
		(E_1+\Delta_s)X_0^\sigma(r) + X_1^\sigma(r) \\
		(E_1+\Delta_s) X^\sigma_2(-,\mathbf{r})-X^\sigma_3(-,\mathbf{r})\\
		(E_1+\Delta_s)X_0^\sigma(r) - X_1^\sigma(r) 
	\epm.
\end{align}
Using these expressions we can compute the asymptotic behavior  of the non-polarized and SP LDOS  in coordinate space for the state with positive energy (thus we omit index $1$ below):
\begin{align}
S_x(\bs r) &= +J_z^2 \left( 1+\frac{1}{\alpha^2} \right) \left\{ \sum\limits_\sigma \sigma \nu^2_\sigma \frac{\cos \left(2p_F^\sigma r-\theta \right)}{p_F^\sigma} +   2 \nu^2 \frac{v_F^2}{v^2} \cdot\frac{\sin p_\lambda r}{p_F} \right\} \cdot \frac{e^{-2 p_s r}}{r} \cos\phi_{\bs r} \\
S_y(\bs r) &= +J_z^2 \left( 1+\frac{1}{\alpha^2} \right) \left\{ \sum\limits_\sigma \sigma \nu^2_\sigma \frac{\cos \left(2p_F^\sigma r-\theta \right)}{p_F^\sigma} +   2 \nu^2 \frac{v_F^2}{v^2} \cdot\frac{\sin p_\lambda r}{p_F} \right\} \cdot \frac{e^{-2 p_s r}}{r} \sin\phi_{\bs r} \\
S_z(r) &= -J_z^2 \left( 1+\frac{1}{\alpha^2} \right) \left\{ \sum\limits_\sigma \nu^2_\sigma \frac{\sin \left(2p_F^\sigma r-\theta \right)}{p_F^\sigma} -  2 \nu^2 \frac{v_F^2}{v^2} \cdot\frac{\cos p_\lambda r}{p_F} \right\} \cdot \frac{e^{-2 p_s r}}{r} \\
\rho(r) &= +J_z^2 \left( 1+\frac{1}{\alpha^2} \right) \left\{ 2\frac{\nu^2}{mv} +  2 \nu^2 \frac{v_F^2}{v^2} \cdot\frac{\sin \left(2mvr - \theta \right)}{p_F} \right\} \cdot \frac{e^{-2 p_s r}}{r}
\end{align}
with
$
\tan\theta = \begin{cases} \frac{2\alpha}{1-\alpha^2}, \;\;\text{if}\; \alpha \neq 1 \\  +\infty, \;\;\text{if}\; \alpha = 1\end{cases},
$
and $p_\lambda = 2m\lambda$. Performing the Fourier transforms of these expressions we can obtain information about the main features and symmetries that we observe in momentum space:
\begin{align}
S_x(\bs p) &= +2\pi i J_z^2 \left( 1+\frac{1}{\alpha^2} \right) \cos\phi_{\bs p} \int\limits_0^{+\infty} dr J_1\left(p r \right) \left\{ \sum\limits_\sigma \sigma \nu^2_\sigma \frac{\cos \left(2p_F^\sigma r-\theta \right)}{p_F^\sigma} +   2 \nu^2 \frac{v_F^2}{v^2} \cdot\frac{\sin p_\lambda r}{p_F} \right\} \cdot e^{-2 p_s r}  \\
S_y(\bs p) &= +2\pi i J_z^2 \left( 1+\frac{1}{\alpha^2} \right) \sin\phi_{\bs p} \int\limits_0^{+\infty} dr J_1\left(p r \right) \left\{ \sum\limits_\sigma \sigma \nu^2_\sigma \frac{\cos \left(2p_F^\sigma r-\theta \right)}{p_F^\sigma} +   2 \nu^2 \frac{v_F^2}{v^2} \cdot\frac{\sin p_\lambda r}{p_F} \right\} \cdot e^{-2 p_s r}  \\
S_z(p) &= -2\pi J_z^2 \left( 1+\frac{1}{\alpha^2} \right) \int\limits_0^{+\infty} dr J_0\left(p r \right) \left\{ \sum\limits_\sigma \nu^2_\sigma \frac{\sin \left(2p_F^\sigma r-\theta \right)}{p_F^\sigma} -  2 \nu^2 \frac{v_F^2}{v^2} \cdot\frac{\cos p_\lambda r}{p_F} \right\} \cdot e^{-2 p_s r} \\
\rho(p) &= +2\pi J_z^2 \left( 1+\frac{1}{\alpha^2} \right)\int\limits_0^{+\infty} dr J_0\left(p r \right) \left\{ 2\frac{\nu^2}{mv} +  2 \nu^2 \frac{v_F^2}{v^2} \cdot\frac{\sin \left(2mvr - \theta \right)}{p_F} \right\} \cdot e^{-2 p_s r}
\end{align}
\subsection{x-impurity}
The coordinate dependence of the eigenfunctions is given by
\begin{align}
\Phi_{\bar 1}(\mathbf{r}) &= +\frac{J_x}{2}\sum\limits_{\sigma =\pm}
	\bpm
	+(E_{\bar 1}-\Delta_s)\left[ X_0^\sigma(r)+X^\sigma_2(-,\mathbf{r})\right] + X_1^\sigma(r) + X^\sigma_3(-,\mathbf{r}) \\
	+(E_{\bar 1}-\Delta_s)\left[ X_0^\sigma(r)+X^\sigma_2(+,\mathbf{r})\right] + X_1^\sigma(r) + X^\sigma_3(+,\mathbf{r}) \\
	-(E_{\bar 1}-\Delta_s)\left[ X_0^\sigma(r)+X^\sigma_2(-,\mathbf{r})\right] + X_1^\sigma(r) + X^\sigma_3(-,\mathbf{r}) \\
	-(E_{\bar 1}-\Delta_s)\left[ X_0^\sigma(r)+X^\sigma_2(+,\mathbf{r})\right] + X_1^\sigma(r) + X^\sigma_3(+,\mathbf{r})	
	\epm, \\
\Phi_1(\mathbf{r}) &= -\frac{J_x}{2}\sum\limits_{\sigma =\pm}
	\bpm
	+(E_1+\Delta_s)\left[ X_0^\sigma(r)-X^\sigma_2(-,\mathbf{r})\right] + X_1^\sigma(r) - X^\sigma_3(-,\mathbf{r}) \\
	-(E_1+\Delta_s)\left[ X_0^\sigma(r)-X^\sigma_2(+,\mathbf{r})\right] - X_1^\sigma(r) + X^\sigma_3(+,\mathbf{r}) \\
	+(E_1+\Delta_s)\left[ X_0^\sigma(r)-X^\sigma_2(-,\mathbf{r})\right] - X_1^\sigma(r) + X^\sigma_3(-,\mathbf{r}) \\
	-(E_1+\Delta_s)\left[ X_0^\sigma(r)-X^\sigma_2(+,\mathbf{r})\right] + X_1^\sigma(r) - X^\sigma_3(+,\mathbf{r})
	\epm.
\end{align}
For the positive energy state we compute the asymptotic behavior of the non-polarized and SP LDOS in coordinate space. We write $S_x(\bs r) = S_x^{s}(r) + S_x^{a}(\bs r)$:
\begin{align}
S_x^{s}(r)& = -J_x^2 \left( 1+\frac{1}{\alpha^2} \right) \left\{ \sum\limits_\sigma  \nu^2_\sigma \frac{1 + \sin \left(2p_F^\sigma r-2\beta \right)}{p_F^\sigma} 
+2 \nu^2 \frac{v_F^2}{v^2} \cdot\frac{\cos p_\lambda r + \sin\left(2mvr -2\beta\right)}{p_F} \right\} \cdot \frac{e^{-2 p_s r}}{r} \\
S_x^{a}(\bs r) &= +J_x^2 \left( 1+\frac{1}{\alpha^2} \right) \left\{ \sum\limits_\sigma  \nu^2_\sigma \frac{1 - \sin \left(2p_F^\sigma r-2\beta \right)}{p_F^\sigma} 
-2 \nu^2 \frac{v_F^2}{v^2} \cdot\frac{\cos p_\lambda r + \sin\left(2mvr -2\beta\right)}{p_F} \right\} \cdot \frac{e^{-2 p_s r}}{r} \cos 2\phi_{\bs r}\\
S_y(\bs r) &= +J_x^2 \left( 1+\frac{1}{\alpha^2} \right) \left\{ \sum\limits_\sigma  \nu^2_\sigma \frac{1 - \sin \left(2p_F^\sigma r-2\beta \right)}{p_F^\sigma} 
-2 \nu^2 \frac{v_F^2}{v^2} \cdot\frac{\cos p_\lambda r - \sin\left(2mvr -\theta\right)}{p_F} \right\} \cdot \frac{e^{-2 p_s r}}{r} \sin 2\phi_{\bs r}\\
S_z(\bs r)& = -J_x^2 \left( 1+\frac{1}{\alpha^2} \right) \left\{ 2\sum\limits_\sigma \sigma \nu^2_\sigma \frac{\cos \left(2p_F^\sigma r-\theta \right)}{p_F^\sigma} + 4 \nu^2 \frac{v_F^2}{v^2} \cdot\frac{\sin p_\lambda r}{p_F} \right\} \cdot \frac{e^{-2 p_s r}}{r} \cos\phi_{\bs r} \phantom{aaaaaaaa} \\
\rho(r) &= +J_x^2 \left( 1+\frac{1}{\alpha^2} \right) \left\{ 4\frac{\nu^2}{mv} +  4 \nu^2 \frac{v_F^2}{v^2} \cdot\frac{\sin \left(2mvr - \theta \right)}{p_F} \right\} \cdot \frac{e^{-2 p_s r}}{r} \phantom{aaaaaaaaaaaaaaaaaaaaaaaa}
\end{align}
with $\tan\beta = \alpha$. Same as before, performing the Fourier transforms of these expressions allows us to obtain information about the most important features and symmetries we observe in momentum space:
\begin{align}
S_x^{s}(p) = -2\pi J_x^2 \left( 1+\frac{1}{\alpha^2} \right) \int\limits_0^{+\infty} dr J_0\left( pr \right) \left\{ \sum\limits_\sigma  \nu^2_\sigma \frac{1 + \sin \left(2p_F^\sigma r-2\beta \right)}{p_F^\sigma} + \right. \phantom{aaaaaaaaaaaaaaaaaaaaaaa} \\
\left. +2 \nu^2 \frac{v_F^2}{v^2} \cdot\frac{\cos p_\lambda r + \sin\left(2mvr -2\beta\right)}{p_F} \right\} \cdot e^{-2 p_s r}\\
S_x^{a}(\bs p) = -2\pi J_x^2 \left( 1+\frac{1}{\alpha^2} \right) \cos 2\phi_{\bs p} \int\limits_0^{+\infty} dr J_2\left(p r \right) \left\{ \sum\limits_\sigma  \nu^2_\sigma \frac{1 - \sin \left(2p_F^\sigma r-2\beta \right)}{p_F^\sigma} - \right. \phantom{aaaaaaaaaaaaaaaaa} \\
\left. -2 \nu^2 \frac{v_F^2}{v^2} \cdot\frac{\cos p_\lambda r + \sin\left(2mvr -2\beta\right)}{p_F} \right\} \cdot e^{-2 p_s r}\\
S_y(\bs p) = -2\pi J_x^2 \left( 1+\frac{1}{\alpha^2} \right) \sin 2\phi_{\bs p} \int\limits_0^{+\infty} dr J_2\left(p r \right) \left\{ \sum\limits_\sigma  \nu^2_\sigma \frac{1 - \sin \left(2p_F^\sigma r-2\beta \right)}{p_F^\sigma} - \right. \phantom{aaaaaaaaaaaaaaaaa} \\
\left. -2 \nu^2 \frac{v_F^2}{v^2} \cdot\frac{\cos p_\lambda r - \sin\left(2mvr -\theta\right)}{p_F} \right\} \cdot e^{-2 p_s r}\\
S_z(\bs p) = -2\pi i J_x^2 \left( 1+\frac{1}{\alpha^2} \right) \cos\phi_{\bs p} \int\limits_0^{+\infty} dr J_1\left(p r \right) \left\{ 2\sum\limits_\sigma \sigma \nu^2_\sigma \frac{\cos \left(2p_F^\sigma r-\theta \right)}{p_F^\sigma} + 4 \nu^2 \frac{v_F^2}{v^2} \cdot\frac{\sin p_\lambda r}{p_F} \right\} \cdot e^{-2 p_s r} \\
\rho(p) = +2\pi J_x^2 \left( 1+\frac{1}{\alpha^2} \right)\int\limits_0^{+\infty} dr J_0\left(p r \right) \left\{ 4\frac{\nu^2}{mv} +  4 \nu^2 \frac{v_F^2}{v^2} \cdot\frac{\sin \left(2mvr - \theta \right)}{p_F} \right\} \cdot e^{-2 p_s r} \phantom{aaaaaaaaaaaaaaaaa}
\end{align}

\section{The SPDOS for a 2D metallic system in the presence of a magnetic impurity}
The low-energy Hamiltonian can be written as
\begin{align}
H_0 = \xi_p \sigma_0 + \lambda (p_y \sigma_x - p_x \sigma_y) = \bpm \xi_p & i \lambda p_- \\ -i \lambda p_+ & \xi_p \epm,
\end{align}
where $\xi_p = \frac{p^2}{2m}-\varepsilon_F$.
The corresponding spectrum is given by $\mathcal{E} = \xi_p \pm \lambda p$. The retarded Green's function reads
\begin{align}
G_0(E,\bs p) = \frac{1}{(E-\xi_p+i0)^2-\lambda^2p^2} \bpm E-\xi_p+i0 & i \lambda p_- \\ -i \lambda p_+ & E-\xi_p+i0 \epm
\end{align}
To compute the eigenvalues for a single localized impurity we calculate
$$
G_0(E,\bs r = \bs 0) = \int \frac{d\bs p}{(2\pi)^2}\frac{E-\xi_p+i0}{(E-\xi_p+i0)^2-\lambda^2p^2} \bpm 1 & 0 \\ 0 & 1 \epm = \frac{1}{2} \sum\limits_\sigma \int \frac{d\bs p}{(2\pi)^2} \frac{1}{E-\xi_\sigma+i0} \bpm 1 & 0 \\ 0 & 1 \epm,
$$
where $\xi_\sigma = \xi_p + \sigma \lambda p$. For $p>0$ we linearize the spectrum around Fermi momenta, thus:
$$
\xi_\sigma \approx \left(\frac{p_F^\sigma}{m}+\sigma \lambda \right)(p-p_F^\sigma) = \sqrt{\lambda^2+2\varepsilon_F/m}\;(p-p_F^\sigma) \equiv v (p-p_F^\sigma),
$$
with $p_F^\sigma = m \left[-\sigma \lambda + v \right]$, and thus we rewrite:
$$
\frac{d\mathbf{p}}{(2\pi)^2} = \frac{m}{2\pi} \left[1-\sigma \frac{\lambda}{v}\right] d\xi_\sigma \frac{d\phi}{2\pi} = \nu_\sigma d\xi_\sigma \frac{d\phi}{2\pi},
$$ 
where $\nu_\sigma = \nu \left[1-\sigma \frac{\lambda}{v}\right]$, with $\nu = m/2\pi$. Thus we get:
$$
\int \frac{d\bs p}{(2\pi)^2} \frac{1}{E-\xi_\sigma+i0} = \nu_\sigma \int d\xi_\sigma \frac{1}{E-\xi_\sigma+i0} = -i\pi \nu_\sigma,
$$
and therefore:
\begin{align}
G_0(E,\bs r = \bs 0) = \frac{1}{2} \sum\limits_\sigma \left( -i\pi \nu_\sigma \right) \bpm 1 & 0 \\ 0 & 1 \epm = -i\pi\nu \bpm 1 & 0 \\ 0 & 1 \epm
\end{align}
Since there is no energy dependence, there will be no impurity-induced states. To find the coordinate dependence of the Green's function we calculate:
\begin{align}
X_0^\sigma(r) &= \int \frac{d\bs p}{(2\pi)^2} \frac{e^{i\bs{pr}}}{E-\xi_\sigma+i0} \\
X_1^\sigma(s,\bs r) &= \int \frac{d\bs p}{(2\pi)^2} \frac{-is e^{i s \phi_{\bs p}}\,e^{i\bs{pr}}}{E-\xi_\sigma+i0}
\end{align}
Below we use the Sokhotsky formula:
\begin{align*}
\frac{1}{x+i0} = \mathcal{P}\frac{1}{x} - i\pi \delta(x)
\end{align*}
\begin{align*}
X_0^\sigma(r) &= \int \frac{d\bs p}{(2\pi)^2} \frac{e^{i\bs{pr}}}{E-\xi_\sigma+i0} = \nu_\sigma \int d\xi_\sigma \int \frac{d\phi_{\bs p}}{2\pi}\frac{e^{i p r \cos \left(\phi_{\bs p} - \phi_{\bs r}\right)}}{E-\xi_\sigma + i0} = \nu_\sigma \int d\xi_\sigma \frac{J_0\left[ \left(p_F^\sigma +\xi_\sigma/v \right) r\right]}{E-\xi_\sigma + i0} = \\
&= \nu_\sigma \left\{ \mathcal{P}\negthickspace\int d\xi_\sigma \frac{J_0\left[ \left(p_F^\sigma +\xi_\sigma/v \right) r\right]}{E-\xi_\sigma} -i\pi \int d\xi_\sigma\, \delta\left(E-\xi_\sigma \right) J_0\left[ \left(p_F^\sigma +\xi_\sigma/v \right) r\right] \right\} = \spadesuit
\end{align*}
We calculate separately the first integral:
\begin{align*}
&\mathcal{P}\negthickspace\int d\xi_\sigma \frac{J_0\left[ \left(p_F^\sigma +\xi_\sigma/v \right) r\right]}{E-\xi_\sigma} = \frac{2}{\pi} \int\limits_1^{+\infty} \frac{du}{\sqrt{u^2-1}} \mathcal{P}\negthickspace\int d\xi_\sigma \frac{\sin\left[ \left(p_F^\sigma +\xi_\sigma/v \right) r\right]}{E-\xi_\sigma} = \\
&= \frac{2}{\pi} \Im \int\limits_1^{+\infty} \frac{du}{\sqrt{u^2-1}} \mathcal{P}\negthickspace\int d\xi_\sigma \frac{e^{i \left(p_F^\sigma +\xi_\sigma/v \right) r} }{E-\xi_\sigma} = \frac{2}{\pi} \Im \int\limits_1^{+\infty} \frac{du}{\sqrt{u^2-1}} e^{i p_\sigma r u} \cdot \mathcal{P}\negthickspace\int dx \frac{e^{-i \frac{r}{v}x}}{x} = \clubsuit
\end{align*}
\begin{align*}
\mathcal{P}\negthickspace\int dx \frac{e^{-i \frac{r}{v}x}}{x} = \mathcal{P}\negthickspace\int \frac{\cos \frac{r}{v} x}{x} dx - i \mathcal{P}\negthickspace\int \frac{\sin \frac{r}{v} x}{x} dx = 0-i\pi = -i\pi
\end{align*}
Therefore:
\begin{align*}
\clubsuit = -2 \Im \int\limits_1^{+\infty} \frac{i e^{i p_\sigma r u}}{\sqrt{u^2-1}} du = -2 \int\limits_1^{+\infty} \frac{\cos p_\sigma r u }{\sqrt{u^2-1}} du = \pi Y_0\left( p_\sigma r \right), \; p_\sigma \neq 0 \phantom{aaaaaaaaaa}
\end{align*}
\begin{align*}
\spadesuit = \pi \nu_\sigma \left[Y_0\left(p_\sigma r\right) - i J_0\left(p_\sigma r\right) \right]. \phantom{aaaaaaaaaaaaaaaaaaaaaaaaaaaaaaaaaaaaaaaaaaaaaa}
\end{align*}
The second integral is
\begin{align*}
X_1^\sigma(s,\bs r) &= \negthickspace\int\negthickspace \frac{d\bs p}{(2\pi)^2} \frac{-is e^{i s \phi_{\bs p}}\,e^{i\bs{pr}}}{E-\xi_\sigma+i0} = \nu_\sigma \negthickspace\int\negthickspace d\xi_\sigma \negthickspace\int \negthickspace \frac{d\phi_{\bs p}}{2\pi}\frac{-is e^{i s \phi_{\bs p}}\,e^{i p r \cos \left(\phi_{\bs p} - \phi_{\bs r}\right)}}{E-\xi_\sigma + i0} = se^{i s \phi_{\bs r}}\, \nu_\sigma \negthickspace \int \negthickspace d\xi_\sigma \frac{J_1\left[ \left(p_F^\sigma +\xi_\sigma/v \right) r\right]}{E-\xi_\sigma + i0} = \\
&= se^{i s \phi_{\bs r}}\cdot \nu_\sigma \left\{ \mathcal{P}\negthickspace\int d\xi_\sigma \frac{J_1\left[ \left(p_F^\sigma +\xi_\sigma/v \right) r\right]}{E-\xi_\sigma} -i\pi \int d\xi_\sigma \, \delta\left(E-\xi_\sigma \right) J_1\left[ \left(p_F^\sigma +\xi_\sigma/v \right) r\right] \right\} = \heartsuit
\end{align*}
We calculate separately the first integral:
\begin{align*}
&\mathcal{P}\negthickspace\int \negthickspace d\xi_\sigma \frac{J_1\left[ \left(p_F^\sigma +\xi_\sigma/v \right) r\right]}{E-\xi_\sigma} = \mathcal{P}\negthickspace\int\negthickspace dx \frac{J_1\left[ \left(p_\sigma - x/v \right) r\right]}{x} = -\frac{\partial}{\partial r} \mathcal{P}\negthickspace\int\negthickspace dx \frac{J_0\left[ \left(p_\sigma - x/v \right) r\right]}{x\left(p_\sigma - x/v \right)} = \\
&= -\frac{\partial}{\partial r}\mathcal{P}\negthickspace\int\negthickspace dy \frac{J_0\left[ \left(p_\sigma - y \right) r\right]}{y\left(p_\sigma - y \right)} = -\frac{\partial}{\partial (p_\sigma r)} \left[ \mathcal{P}\negthickspace\int\negthickspace dy \frac{J_0\left[ \left(p_\sigma - y \right) r\right]}{y} + \mathcal{P}\negthickspace\int\negthickspace dy \frac{J_0\left[ \left(p_\sigma - y \right) r\right]}{p_\sigma - y}\right] = \\
&= -\frac{\partial}{\partial (p_\sigma r)} \frac{2}{\pi} \Im\negthickspace \int\limits_1^{+\infty} \negthickspace\frac{du}{\sqrt{u^2-1}}\left[\mathcal{P}\negthickspace\int\negthickspace \frac{e^{i (p_\sigma-y) r u}}{y
} dy + \mathcal{P}\negthickspace\int\negthickspace \frac{e^{i (p_\sigma-y) r u}}{p_\sigma-y} dy \right] = -2 \frac{\partial}{\partial (p_\sigma r)} \Im \negthickspace\int\limits_1^{+\infty} \negthickspace\frac{i du}{\sqrt{u^2-1}}\left[1-e^{i p_\sigma r u}\right] = \\
&= -2 \int\limits_1^{+\infty} \frac{u \sin p_\sigma r u }{\sqrt{u^2-1}} du = 2 \frac{\partial}{\partial (p_\sigma r)} \int\limits_1^{+\infty} \frac{\cos p_\sigma r u }{\sqrt{u^2-1}} du = -\pi \frac{\partial}{\partial (p_\sigma r)} Y_0\left(p_\sigma r\right) = \pi Y_1\left(p_\sigma r\right), \; p_\sigma \neq 0
\end{align*}
Therefore:
\begin{align*}
\heartsuit = \pi \nu_\sigma \left[Y_1\left(p_\sigma r\right) - i J_1\left(p_\sigma r\right) \right]. \phantom{aaaaaaaaaaaaaaaaaaaaaaaaaaaaaaaaaaaaaaaaaaaaaa}
\end{align*}
Finally:
\begin{align}
X_0^\sigma(r) &= \pi \nu_\sigma \left[Y_0\left(p_\sigma r\right) - i J_0\left(p_\sigma r\right) \right]\\
X_1^\sigma(s,\bs r) &= s e^{is\phi_{\bs r}} \Big\{ \pi \nu_\sigma \left[Y_1\left(p_\sigma r\right) - i J_1\left(p_\sigma r\right) \right]\Big\} \equiv s e^{is\phi_{\bs r}} \tilde{X}_1^\sigma(r),
\end{align}
where $p_\sigma = p_F^\sigma + E/v \neq 0$. Thus the Green's function for $\bs r \neq \bs 0$ can be written as:
\begin{align}
G_0(E,\bs r) = \frac{1}{2} \sum\limits_\sigma 
	\begin{pmatrix}
		X_0^\sigma(r) & -\sigma e^{-i\phi_{\bs r}} \tilde{X}_1^\sigma(r) \\
		\sigma e^{i\phi_{\bs r}} \tilde{X}_1^\sigma(r) & X_0^\sigma(r)
	\end{pmatrix}
\end{align}
Below we compute the T-matrix for different types of impurities. Impurity potentials take the following forms:
\begin{align}
V_{sc} = U \bpm 1 & 0 \\ 0 & 1 \epm, \quad V_z = J_z \bpm 1 & 0 \\ 0 & -1 \epm, \quad V_x = J_x \bpm 0 & 1 \\ 1 & 0 \epm
\end{align}
The corresponding T-matrices are
\begin{align}
T_{sc} = \frac{U}{1+i\pi\nu U} \bpm 1 & 0 \\ 0 & 1 \epm, \quad T_z = \bpm \frac{J}{1+i\pi\nu J} & 0 \\ 0 & -\frac{J}{1-i\pi\nu J}\epm, \quad \quad T_x = \frac{J}{1+\pi^2\nu^2 J^2} \bpm -i \pi \nu J & 1 \\ 1 & -i \pi \nu J \epm
\end{align}
For each type of impurity we can compute the SP and non-polarized LDOS using
\begin{align}
\Delta G(E,\bs r) = G_0(E,-\bs r) T(E) G_0(E,\bs r)
\label{DGapp}
\end{align}
\begin{align}
S_x(E,\mathbf{r}) &= -\frac{1}{\pi} \left[ \Im \Delta G_{12} + \Im \Delta G_{21}\right]\\
S_y(E,\mathbf{r}) &= -\frac{1}{\pi} \left[ \Re \Delta G_{12} - \Re \Delta G_{21}\right]\\
S_z(E,\mathbf{r}) &= -\frac{1}{\pi} \left[ \Im \Delta G_{11} - \Im \Delta G_{22}\right]\\
\Delta\rho(E,\mathbf{r}) &= -\frac{1}{\pi} \left[ \Im \Delta G_{11} + \Im \Delta G_{22}\right]
\label{SPLDOSapp}
\end{align}
\subsection*{Asymptotic expansions of Bessel functions}
Since the integrals are expressed in terms of Neumann function and Bessel function of the first kind, we  give their asymptotic behavior for $x \to +\infty$:
\begin{align*}
J_0\left(x \right) \sim +\sqrt{\frac{2}{\pi x}} \cos \left( x - \frac{\pi}{4}\right), \quad\quad
J_1\left(x \right) \sim -\sqrt{\frac{2}{\pi x}} \cos \left( x + \frac{\pi}{4}\right) \\
Y_0\left(x \right) \sim -\sqrt{\frac{2}{\pi x}} \cos \left( x + \frac{\pi}{4}\right), \quad\quad
Y_1\left(x \right) \sim -\sqrt{\frac{2}{\pi x}} \cos \left( x - \frac{\pi}{4}\right)
\end{align*}
\subsection*{Fourier transforms in 2D}
\begin{align}
\mathcal{F}\left[ f(r) \right] = 2\pi \int\limits_0^{+\infty} r J_0\left(pr \right) f(r) dr \phantom{aaaaaaaaaaaaaaaaaaaaaaaaaaaaa}\\
\mathcal{F}\left[\cos\phi_{\bs r} f(r) \right] = 2\pi i\, \cos\phi_{\bs p} \cdot \int\limits_0^{+\infty} r J_1\left(pr \right) f(r) dr, \quad\quad
\mathcal{F}\left[\sin\phi_{\bs r} f(r) \right] = 2\pi i\, \sin\phi_{\bs p} \cdot \int\limits_0^{+\infty} r J_1\left(pr \right) f(r) dr \\
\mathcal{F}\left[\cos 2\phi_{\bs r} f(r) \right] = -2\pi\, \cos 2\phi_{\bs p} \int\limits_0^{+\infty} r J_2\left(pr \right) f(r) dr, \quad\quad
\mathcal{F}\left[\sin 2\phi_{\bs r} f(r) \right] = -2\pi\, \sin 2\phi_{\bs p} \int\limits_0^{+\infty} r J_2\left(pr \right) f(r) dr 
\end{align}
\subsection{z-impurity}
We denote $\alpha = \pi \nu J$ and write the asymptotic expansions of the non-polarized and SP LDOS components in coordinate space:
\begin{align}
S_x(\bs r) &\sim \frac{J}{1+\alpha^2} \frac{\cos \phi_{\bs r}}{r} \sum\limits_\sigma \sigma \frac{\nu^2_\sigma}{p_\sigma}\sin 2p_\sigma r \\
S_y(\bs r) &\sim \frac{J}{1+\alpha^2} \frac{\sin \phi_{\bs r}}{r} \sum\limits_\sigma \sigma \frac{\nu^2_\sigma}{p_\sigma}\sin 2p_\sigma r \\
S_z(r) &\sim -\frac{J}{1+\alpha^2} \frac{2}{r} \sum\limits_\sigma  \frac{\nu^2_\sigma}{p_\sigma}\cos 2p_\sigma r \\
\rho(r) &\sim -\frac{J}{1+\alpha^2} 4\alpha\nu^2 \frac{v_F^2}{v^2} \frac{1}{\sqrt{p_F^2 + 2mE + E^2/v^2}}  \cdot \frac{\sin p_\varepsilon r}{r}, 
\end{align}
where $p_\varepsilon = 2\left(mv+E/v\right)$. 
and we get for $p_\sigma > 0$:
\begin{align}
S_x(\bs p) &\sim +\frac{J}{1+\alpha^2} \cdot 2\pi i\, \cos \phi_{\bs p} \int\limits_0^{+\infty} dr J_1\left( pr \right) \sum\limits_\sigma \sigma \frac{\nu^2_\sigma}{p_\sigma}\sin 2p_\sigma r \\
S_y(\bs p) &\sim +\frac{J}{1+\alpha^2} \cdot 2\pi i\, \sin \phi_{\bs p} \int\limits_0^{+\infty} dr J_1\left( pr \right) \sum\limits_\sigma \sigma \frac{\nu^2_\sigma}{p_\sigma}\sin 2p_\sigma r \\
S_z(p) &\sim -\frac{J}{1+\alpha^2} \cdot 4\pi \int\limits_0^{+\infty} dr J_0\left( pr \right) \sum\limits_\sigma  \frac{\nu^2_\sigma}{p_\sigma}\cos 2p_\sigma r \\
\rho(p) &\sim -\frac{J}{1+\alpha^2} \cdot 8\pi \alpha \nu^2 \frac{v_F^2}{v^2} \frac{1}{\sqrt{p_F^2 + 2mE + E^2/v^2}} \int\limits_0^{+\infty} dr J_0\left( pr \right) \sin p_\varepsilon r 
\end{align}
\subsection{x-impurity}
\begin{align}
S_x(\bs r) &\sim -\frac{J}{1+\alpha^2} \frac{1}{r} \left\{2\nu^2 \frac{v_F^2}{v^2} \frac{\cos p_\varepsilon r}{\sqrt{p_F^2 + 2mE + E^2/v^2}} + \sum\limits_\sigma \frac{\nu^2_\sigma}{p_\sigma}\cos 2p_\sigma r \; + \right. \phantom{aaaaaaaaaaaaaaaaaaaaaaaaaaa}\\
& \phantom{aaaaaaaaaaaaaaaaaaaaaaaaaaa}\left. + \cos 2\phi_{\bs r} \left[-2\nu^2 \frac{v_F^2}{v^2} \frac{\cos p_\varepsilon r}{\sqrt{p_F^2 + 2mE + E^2/v^2}} + \sum\limits_\sigma \frac{\nu^2_\sigma}{p_\sigma}\cos 2p_\sigma r \right] \right\} \\
S_y(\bs r) &\sim -\frac{J}{1+\alpha^2} \frac{\sin 2\phi_{\bs r}}{r} \left[ -2\nu^2 \frac{v_F^2}{v^2} \frac{\cos p_\varepsilon r}{\sqrt{p_F^2 + 2mE + E^2/v^2}} + \sum\limits_\sigma \frac{\nu^2_\sigma}{p_\sigma}\cos 2p_\sigma r\right] \\
S_z(\bs r) &\sim -\frac{J}{1+\alpha^2} \frac{\cos \phi_{\bs r}}{r} \sum\limits_\sigma \sigma \frac{\nu^2_\sigma}{p_\sigma}\sin 2p_\sigma r \\
\rho(r) &\sim -\frac{J}{1+\alpha^2} \cdot \frac{\alpha}{r} \cdot 4\nu^2 \frac{v_F^2}{v^2} \frac{\sin p_\varepsilon r}{\sqrt{p_F^2 + 2mE + E^2/v^2}} 
\end{align}
With the corresponding Fourier transforms:
\begin{align}
S_x(\bs p) &= S^{sym}_x(p) + S^{asym}_x(\bs p) = -\frac{J}{1+\alpha^2} \cdot 2\pi \negthickspace\int \limits_0^{+\infty} \negthickspace dr J_0\left( pr \right) \left[ 2\nu^2 \frac{v_F^2}{v^2} \frac{\cos p_\varepsilon r}{\sqrt{p_F^2 + 2mE + E^2/v^2}} + \sum\limits_\sigma \frac{\nu^2_\sigma}{p_\sigma}\cos 2p_\sigma r \right] - \\
&-\frac{J}{1+\alpha^2} \cdot 2\pi \cos 2\phi_{\bs p} \negthickspace\int \limits_0^{+\infty} \negthickspace dr J_2\left( pr \right) \left[ 2\nu^2 \frac{v_F^2}{v^2} \frac{\cos p_\varepsilon r}{\sqrt{p_F^2 + 2mE + E^2/v^2}} - \sum\limits_\sigma \frac{\nu^2_\sigma}{p_\sigma}\cos 2p_\sigma r \right] \\
S_y(\bs p) &= -\frac{J}{1+\alpha^2} \cdot 2\pi \sin 2\phi_{\bs p} \negthickspace\int \limits_0^{+\infty} \negthickspace dr J_2\left( pr \right) \left[ 2\nu^2 \frac{v_F^2}{v^2} \frac{\cos p_\varepsilon r}{\sqrt{p_F^2 + 2mE + E^2/v^2}} - \sum\limits_\sigma \frac{\nu^2_\sigma}{p_\sigma}\cos 2p_\sigma r \right] \\
S_z(\bs p) &\sim -\frac{J}{1+\alpha^2} \cdot 2\pi i\, \cos \phi_{\bs p} \int\limits_0^{+\infty} dr J_1\left( pr \right) \sum\limits_\sigma \sigma \frac{\nu^2_\sigma}{p_\sigma}\sin 2p_\sigma r \\
\rho(p) &\sim -\frac{J}{1+\alpha^2} \cdot 8\pi \alpha \nu^2 \frac{v_F^2}{v^2} \frac{1}{\sqrt{p_F^2 + 2mE + E^2/v^2}} \int\limits_0^{+\infty} dr J_0\left( pr \right) \sin p_\varepsilon r 
\end{align}

\section{Analytical calculation of the Shiba states wave functions for a 1D system}
The unperturbed Green's function in momentum space is $G_0(E,p) = \frac{1}{2}\sum\limits_{\sigma =\pm} G^\sigma_0(E,p)$, where
\begin{align}
G^\sigma_0(E,p) = -\frac{1}{\xi^2_\sigma+\Delta_s^2-E^2}
	\bpm
		1 & i\sigma  \\
		-i\sigma & 1 \\
	\epm
	\otimes
	\bpm
		E+\xi_\sigma & \Delta_s \\
		\Delta_s & E-\xi_\sigma
	\epm,
\end{align}
where $\xi_\sigma = \xi_p + \sigma \lambda p$. To get the coordinate value one needs to perform the Fourier transform:
$$
G^\sigma_0(E,x) = \int \frac{dp}{2\pi} G^\sigma_0(E,p) e^{i p x}
$$
We will have two types of integrals:
\begin{align}
X^\sigma_0(x) &= -\int \frac{dp}{2\pi} \frac{e^{ipx}}{\xi_\sigma^2+\omega^2}, \\
X^\sigma_1(x) &= -\int \frac{dp}{2\pi} \frac{\xi_\sigma e^{ipx}}{\xi_\sigma^2+\omega^2},
\end{align}
where $\omega^2 = \Delta_s^2 - E^2$. Since the spectrum is split by SO coupling, there will be two Fermi momenta which can be found the following way:
$$
\frac{p^2}{2m} + \sigma \lambda p - \varepsilon_F = 0, \quad p^\sigma_F = \frac{-\sigma \lambda + \sqrt{\lambda^2+2\varepsilon_F/m}}{1/m} \equiv m \left[-\sigma \lambda + v \right]
$$
For $p>0$ we linearize the spectrum around Fermi momenta, thus:
$$
\xi_\sigma \approx \left(\frac{p_F^\sigma}{m}+\sigma \lambda \right)(p-p_F^\sigma) = \sqrt{\lambda^2+2\varepsilon_F/m}\;(p-p_F^\sigma) \equiv v (p-p_F^\sigma),
$$ 
therefore $p = p^{\sigma}_F + \xi_\sigma/v$ and we get:
\begin{align*}
X^\sigma_0(x) &= -\int \frac{dp}{2\pi} \frac{e^{ipx}}{\xi_\sigma^2+\omega^2} = - \left[ \int\limits_0^{+\infty} \frac{dp}{2\pi} \frac{e^{ipx}}{\xi_\sigma^2+\omega^2} + \int\limits_0^{+\infty} \frac{dp}{2\pi} \frac{e^{-ipx}}{\xi_{-\sigma}^2+\omega^2} \right] = \clubsuit \phantom{aaaaaaaaaaaaaaaaaaaaaaa}
\end{align*}
\begin{align*}
&\int\limits_0^{+\infty} \frac{dp}{2\pi} \frac{e^{ipx}}{\xi_\sigma^2+\omega^2} \approx \frac{1}{2\pi v} e^{i p_F^\sigma x }\int d\xi_\sigma  \frac{ e^{i \xi_\sigma x/v}}{\xi_\sigma^2+\omega^2} = \frac{1}{2v\omega} e^{i p_F^\sigma x } e^{-\omega |x|/v} \\
&\int\limits_0^{+\infty} \frac{dp}{2\pi} \frac{e^{-ipx}}{\xi_{-\sigma}^2+\omega^2} \approx \frac{1}{2\pi v} e^{-i p_F^{-\sigma} x} \int d\xi_{-\sigma}  \frac{ e^{-i \xi_{-\sigma} x/v}}{\xi_{-\sigma}^2+\omega^2} = \frac{1}{2v\omega} e^{-i p_F^{-\sigma} x} e^{-\omega |x|/v} 
\end{align*}
\begin{align*}
\clubsuit = - \frac{1}{2v\omega} \left[e^{i m \left[-\sigma \lambda + v \right] x} + e^{-i m \left[\sigma \lambda + v \right] x}\right] e^{-\omega |x|/v}  = -\frac{1}{v} \cdot \frac{1}{\omega} \cos mvx \; e^{-i\sigma m \lambda x}\, e^{-\omega |x|/v} \phantom{aaaaaaa}
\end{align*}
\begin{align*}
X^\sigma_1(x) &= -\int \frac{dp}{2\pi} \frac{\xi_\sigma\,e^{ipx}}{\xi_\sigma^2+\omega^2} = - \left[ \int\limits_0^{+\infty} \frac{dp}{2\pi} \frac{\xi_\sigma\,e^{ipx}}{\xi_\sigma^2+\omega^2} + \int\limits_0^{+\infty} \frac{dp}{2\pi} \frac{\xi_{-\sigma}\,e^{-ipx}}{\xi_{-\sigma}^2+\omega^2} \right] = \spadesuit \phantom{aaaaaaaaaaaaaaaaaaaaaaa}
\end{align*}
\begin{align*}
&\int\limits_0^{+\infty} \frac{dp}{2\pi} \frac{\xi_\sigma\, e^{ipx}}{\xi_\sigma^2+\omega^2} \approx \frac{1}{2\pi v} e^{i p_F^\sigma x }\int d\xi_\sigma  \frac{\xi_\sigma\, e^{i \xi_\sigma x/v}}{\xi_\sigma^2+\omega^2} = \frac{i}{2v} \sgn x \, e^{i p_F^\sigma x } e^{-\omega |x|/v} \\
&\int\limits_0^{+\infty} \frac{dp}{2\pi} \frac{\xi_{-\sigma}\,e^{-ipx}}{\xi_{-\sigma}^2+\omega^2} \approx \frac{1}{2\pi v} e^{-i p_F^{-\sigma} x} \int d\xi_{-\sigma}  \frac{\xi_{-\sigma}\, e^{-i \xi_{-\sigma} x/v}}{\xi_{-\sigma}^2+\omega^2} = -\frac{i}{2v} \sgn x\, e^{-i p_F^{-\sigma} x} e^{-\omega |x|/v} 
\end{align*}
\begin{align*}
\spadesuit = - \frac{i}{2v}  \sgn x \left[e^{i m \left[-\sigma \lambda + v \right] x} - e^{-i m \left[\sigma \lambda + v \right] x}\right] e^{-\omega |x|/v}  = \frac{1}{v} \cdot \sin mv\left|x \right| \; e^{-i\sigma m \lambda x}\, e^{-\omega |x|/v} \phantom{aaaaaaa} \\
\end{align*}
Finally:
\begin{align}
X^\sigma_0(x) &= -\frac{1}{v} \cdot \frac{1}{\omega} \cos mvx \; e^{-i\sigma m \lambda x}\, e^{-\omega |x|/v} \\
X^\sigma_1(x) &= +\frac{1}{v} \cdot \sin mv\left|x \right| \; e^{-i\sigma m \lambda x}\, e^{-\omega |x|/v}
\end{align}
and
\begin{align}
G_0(E,x) = \frac{1}{2}\sum\limits_{\sigma=\pm} 
	\bpm
		1 & i\sigma  \\
		-i\sigma & 1 \\
	\epm
	\otimes
	\bpm
		E X^\sigma_0(x) + X^\sigma_1(x) & \Delta_s X^\sigma_0(x) \\
		\Delta_s X^\sigma_0(x) & E X^\sigma_0(x) - X^\sigma_1(x)
	\epm
\end{align}
\begin{align}
G_0(\epsilon,x=0) = -\frac{1}{v} \frac{1}{\sqrt{1-\epsilon^2}}
		\bpm 	\epsilon \sigma_0 & \sigma_0 \\
				\sigma_0 & \epsilon \sigma_0  \\
		\epm, \quad\text{where}\; \epsilon=\frac{E}{\Delta_s}
\end{align}
The eigenvalues and eigenfunctions at ${\bm r}={\bm 0}$ can be obtained using Eq. (\ref{eigenv})
The energy levels are
\begin{align}
	E_{1,\bar{1}} = \pm \frac{1-\alpha^2}{1+\alpha^2} \Delta_s, \; \text{where}\; \alpha = J/v.
\end{align}
In case of an impurity along the $z$-axis the corresponding eigenvectors are
\begin{align}
	\Phi_{\bar{1}}( 0) = \begin{pmatrix} 1 & 0 & -1 & 0 \end{pmatrix}^T, \; \Phi_1(0) = \begin{pmatrix} 0 & 1 & 0 & 1 \end{pmatrix}^T
\end{align}
and in case of an impurity along the $x$-axis:
\begin{align}
	\Phi_{\bar{1}}(0) = \begin{pmatrix} 1 & 1 & -1 & -1 \end{pmatrix}^T, \; \Phi_1(0) = \begin{pmatrix} 1 & -1 & 1 & -1 \end{pmatrix}^T.
\end{align}

\subsection{z-impurity}
\begin{align}
\Phi_{\bar 1} (x) = +\frac{J_z}{2}\sum\limits_{\sigma} 
	\bpm
		+\left(E_{\bar 1}-\Delta_s\right)X_0^\sigma(x) + X_1^\sigma(x) \\
		-i\sigma \left[ \left(E_{\bar 1}-\Delta_s\right)X_0^\sigma(x) + X_1^\sigma(x) \right] \\
		-\left(E_{\bar 1}-\Delta_s\right)X_0^\sigma(x) + X_1^\sigma(x) \\
		+i\sigma \left[ \left(E_{\bar 1}-\Delta_s\right)X_0^\sigma(x) - X_1^\sigma(x) \right]
	\epm, \;
 \Phi_{1} (x) = -\frac{J_z}{2}\sum\limits_{\sigma} 
	\bpm
		+i\sigma \left[ \left(E_{1}+\Delta_s\right)X_0^\sigma(x) + X_1^\sigma(x) \right] \\
		 \left(E_{1}+\Delta_s\right)X_0^\sigma(x) + X_1^\sigma(x) \\
		+i\sigma \left[ \left(E_{1}+\Delta_s\right)X_0^\sigma(x) - X_1^\sigma(x) \right] \\
		 \left(E_{1}+\Delta_s\right)X_0^\sigma(x) - X_1^\sigma(x) 		
	\epm.
\end{align}
Using these expressions we can compute the non-polarized and SP LDOS in both coordinate and momentum space for the positive energy state (omitting the index $1$):
\begin{align}
S_x(x) &= \frac{1+\alpha^2}{4} [2\sin p_\lambda x+\sin (2mv |x|+p_\lambda x-2\theta)-\sin (2mv |x|-p_\lambda x-2\theta)] \cdot e^{-2\omega |x|/v} \\
S_y(x) &= 0 \\
S_z(x) &= -\frac{1+\alpha^2}{4} [2\cos p_\lambda x+\cos (2mv |x|+p_\lambda x-2\theta)+\cos (2mv |x|-p_\lambda x-2\theta)] \cdot e^{-2\omega |x|/v} \\
\rho(x) &= \frac{1+\alpha^2}{2} [1+\cos(2mv |x|-2\theta)] \cdot e^{-2\omega |x|/v}
\end{align}
where $\tan \theta = \alpha$. We perform the Fourier transform to get the momentum space behavior, exploiting the following 'standard' integrals:
\begin{align}
&\int e^{-2\omega |x|/v} e^{-ipx}dx = 2\frac{2\omega/v}{p^2+(2\omega/v)^2} \\
&\int \cos p_\lambda x \cdot e^{-2\omega |x|/v} e^{-ipx}dx = \frac{2\omega}{v} \left[\frac{1}{(p+p_\lambda)^2+(2\omega/v)^2} + \frac{1}{(p-p_\lambda)^2+(2\omega/v)^2}\right] \\
&\int \sin p_\lambda x \cdot e^{-2\omega |x|/v} e^{-ipx}dx = i\frac{2\omega}{v} \left[\frac{1}{(p+p_\lambda)^2+(2\omega/v)^2} - \frac{1}{(p-p_\lambda)^2+(2\omega/v)^2}\right] \\
&\int \sin 2mv |x| \cdot e^{-2\omega |x|/v} e^{-ipx}dx = \frac{p+2mv}{(p+2mv)^2+(2\omega/v)^2} - \frac{p-2mv}{(p-2mv)^2+(2\omega/v)^2}
\end{align}
We rewrite these expressions using $p_F^\pm$, thus we get:
\begin{align}
&\int \cos p_\lambda x \cdot e^{-2\omega |x|/v} e^{-ipx}dx = \frac{2\omega}{v} \left\{\frac{1}{\left[p+(p_F^--p_F^+)\right]^2+(2\omega/v)^2} + \frac{1}{(\left[p-(p_F^--p_F^+)\right]^2+(2\omega/v)^2}\right\} \\
&\int \sin p_\lambda x \cdot e^{-2\omega |x|/v} e^{-ipx}dx = i\frac{2\omega}{v} \left\{\frac{1}{\left[p+(p_F^--p_F^+)\right]^2+(2\omega/v)^2} - \frac{1}{\left[p-(p_F^--p_F^+)\right]^2+(2\omega/v)^2}\right\} \\
&\int \sin 2mv |x| \cdot e^{-2\omega |x|/v} e^{-ipx}dx = \frac{p+(p_F^-+p_F^+)}{\left[ p+(p_F^-+p_F^+)\right]^2+(2\omega/v)^2} - \frac{p-(p_F^-+p_F^+)}{\left[p-(p_F^-+p_F^+)\right]^2+(2\omega/v)^2}
\end{align}
For the last two integrals we introduce symbols $\sum\limits_{p'}$ and $\widetilde{\sum\limits_{p'}}$ (wide tilde signify that we take the difference, not sum), where $p' \in \left\{p-p_\lambda,\, p+p_\lambda\right\}$. Thus we have
\begin{align}
\int \cos(2mv |x|-2\theta)\cos p_\lambda x \cdot e^{-2\omega |x|/v} e^{-ipx}dx = \phantom{aaaaaaaaaaaaaaaaaaaaaaaaaaaaaaaaaaaaaaaaaaa} \\
= \frac{1}{2}\sum\limits_{p'}\left\{\frac{1-\alpha^2}{1+\alpha^2} \cdot \frac{2\omega}{v} \left[\frac{1}{(p'+2mv)^2+(2\omega/v)^2} + \frac{1}{(p'-2mv)^2+(2\omega/v)^2}\right] + \right. \phantom{aaaaaaaaaaaaaa}\\
\left. + \frac{2\alpha}{1+\alpha^2} \cdot \left[\frac{p'+2mv}{(p'+2mv)^2+(2\omega/v)^2} + \frac{p'-2mv}{(p'-2mv)^2+(2\omega/v)^2}\right]\right\}
\end{align}
\begin{align}
\int \cos(2mv |x|-2\theta)\sin p_\lambda x \cdot e^{-2\omega |x|/v} e^{-ipx}dx = \phantom{aaaaaaaaaaaaaaaaaaaaaaaaaaaaaaaaaaaaaaaaaaa} \\
= \frac{1}{2i}\widetilde{\sum\limits_{p'}}\left\{\frac{1-\alpha^2}{1+\alpha^2} \cdot \frac{2\omega}{v} \left[\frac{1}{(p'+2mv)^2+(2\omega/v)^2} + \frac{1}{(p'-2mv)^2+(2\omega/v)^2}\right] + \right. \phantom{aaaaaaaaaaaaaa}\\
\left. + \frac{2\alpha}{1+\alpha^2} \cdot \left[\frac{p'+2mv}{(p'+2mv)^2+(2\omega/v)^2} + \frac{p'-2mv}{(p'-2mv)^2+(2\omega/v)^2}\right]\right\}
\end{align}
We rewrite these expressions using $p_F^\pm$, thus we get:
\begin{align}
\int \cos(2mv |x|-2\theta)\cos p_\lambda x \cdot e^{-2\omega |x|/v} e^{-ipx}dx = \phantom{aaaaaaaaaaaaaaaaaaaaaaaaaaaaaaaaaaaaaaaaaaa} \\ \nonumber
= \frac{1-\alpha^2}{1+\alpha^2} \cdot \frac{\omega}{v} \left[\frac{1}{(p+2p_F^+)^2+(2\omega/v)^2} + \frac{1}{(p-2p_F^-)^2+(2\omega/v)^2}\right] + \phantom{aaaaaaaaaaaaaaaaaaaaaaaaaaaa} \\ \nonumber
 + \frac{\alpha}{1+\alpha^2} \cdot \left[\frac{p+2p_F^+}{(p+2p_F^+)^2+(2\omega/v)^2} + \frac{p-2p_F^-}{(p-2p_F^-)^2+(2\omega/v)^2}\right] + \phantom{aaaaaaaaaaaaaaaaaaaaa}\\ \nonumber
+ \frac{1-\alpha^2}{1+\alpha^2} \cdot \frac{\omega}{v} \left[\frac{1}{(p+2p_F^-)^2+(2\omega/v)^2} + \frac{1}{(p-2p_F^+)^2+(2\omega/v)^2}\right] + \phantom{aaaaaaaaaa}\\ \nonumber
 + \frac{\alpha}{1+\alpha^2} \cdot \left[\frac{p+2p_F^-}{(p+2p_F^-)^2+(2\omega/v)^2} + \frac{p-2p_F^+}{(p-2p_F^+)^2+(2\omega/v)^2}\right] \nonumber
\end{align}
\begin{align}
\int \cos(2mv |x|-2\theta)\sin p_\lambda x \cdot e^{-2\omega |x|/v} e^{-ipx}dx = \phantom{aaaaaaaaaaaaaaaaaaaaaaaaaaaaaaaaaaaaaaaaaaa} \\ \nonumber
= \frac{1}{i}\left\{\frac{1-\alpha^2}{1+\alpha^2} \cdot \frac{\omega}{v} \left[\frac{1}{(p+2p_F^+)^2+(2\omega/v)^2} + \frac{1}{(p-2p_F^-)^2+(2\omega/v)^2}\right] + \right. \phantom{aaaaaaaaaaaaaaaaaaaaaa}\\ \nonumber
\left. + \frac{\alpha}{1+\alpha^2} \cdot \left[\frac{p+2p_F^+}{(p+2p_F^+)^2+(2\omega/v)^2} + \frac{p-2p_F^-}{(p-2p_F^-)^2+(2\omega/v)^2}\right]\right\} - \phantom{aaaaaaaaaaaaaaaaa}\\ \nonumber
- \frac{1}{i}\left\{\frac{1-\alpha^2}{1+\alpha^2} \cdot \frac{\omega}{v} \left[\frac{1}{(p+2p_F^-)^2+(2\omega/v)^2} + \frac{1}{(p-2p_F^+)^2+(2\omega/v)^2}\right] + \right. \phantom{aaaaaaaa}\\ \nonumber
\left. + \frac{\alpha}{1+\alpha^2} \cdot \left[\frac{p+2p_F^-}{(p+2p_F^-)^2+(2\omega/v)^2} + \frac{p-2p_F^+}{(p-2p_F^+)^2+(2\omega/v)^2}\right]\right\} \nonumber
\end{align}
Using the formula $\cos^2\gamma = (1+\cos 2\gamma)/2$ we can write the momentum space expressions for the non-polarized and SP LDOS components:
\begin{align}
S_x(p) = i (1+\alpha^2) \frac{\omega}{v} \left\{\frac{1}{\left[p+(p_F^--p_F^+)\right]^2+(2\omega/v)^2} - \frac{1}{\left[p-(p_F^--p_F^+)\right]^2+(2\omega/v)^2}\right\} + \phantom{aaaaaaaaaaaaaaaaaa}\\ \nonumber
+ \frac{1}{i}\left\{\frac{1-\alpha^2}{2} \cdot \frac{\omega}{v} \left[\frac{1}{(p+2p_F^+)^2+(2\omega/v)^2} + \frac{1}{(p-2p_F^-)^2+(2\omega/v)^2}\right] + \right. \phantom{aaaaaaaaaaaaaaaaaaaaaaaaaaa}\\ \nonumber
\left. + \frac{\alpha}{2} \cdot \left[\frac{p+2p_F^+}{(p+2p_F^+)^2+(2\omega/v)^2} + \frac{p-2p_F^-}{(p-2p_F^-)^2+(2\omega/v)^2}\right]\right\} - \phantom{aaaaaaaaaaaaaaaaaaaaaa}\\ \nonumber
- \frac{1}{i}\left\{\frac{1-\alpha^2}{2} \cdot \frac{\omega}{v} \left[\frac{1}{(p+2p_F^-)^2+(2\omega/v)^2} + \frac{1}{(p-2p_F^+)^2+(2\omega/v)^2}\right] + \right. \phantom{aaaaaaaaaa}\\ \nonumber
\left. + \frac{\alpha}{2} \cdot \left[\frac{p+2p_F^-}{(p+2p_F^-)^2+(2\omega/v)^2} + \frac{p-2p_F^+}{(p-2p_F^+)^2+(2\omega/v)^2}\right]\right\} \phantom{aaaaa} \nonumber
\end{align}
\begin{align}
S_z(p) = -(1+\alpha^2) \frac{\omega}{v} \left\{\frac{1}{\left[p+(p_F^--p_F^+)\right]^2+(2\omega/v)^2} + \frac{1}{\left[p-(p_F^--p_F^+)\right]^2+(2\omega/v)^2}\right\} - \phantom{aaaaaaaaaaaaaaaaaa}\\ \nonumber
- \frac{1-\alpha^2}{2} \cdot \frac{\omega}{v} \left[\frac{1}{(p+2p_F^+)^2+(2\omega/v)^2} + \frac{1}{(p-2p_F^-)^2+(2\omega/v)^2}\right] - \phantom{aaaaaaaaaaaaaaaaaaaaaaaaaaqaaaa} \\ \nonumber
 - \frac{\alpha}{2} \cdot \left[\frac{p+2p_F^+}{(p+2p_F^+)^2+(2\omega/v)^2} - \frac{p-2p_F^-}{(p-2p_F^-)^2+(2\omega/v)^2}\right] - \phantom{aaaaaaaaaaaaaaaaaaaaaaaaaaaa}\\ \nonumber
- \frac{1-\alpha^2}{2} \cdot \frac{\omega}{v} \left[\frac{1}{(p+2p_F^-)^2+(2\omega/v)^2} + \frac{1}{(p-2p_F^+)^2+(2\omega/v)^2}\right] - \phantom{aaaaaaaaaaaaaa}\\ \nonumber
 - \frac{\alpha}{2} \cdot \left[\frac{p+2p_F^-}{(p+2p_F^-)^2+(2\omega/v)^2} + \frac{p-2p_F^+}{(p-2p_F^+)^2+(2\omega/v)^2}\right] \phantom{aaaaaaaa} \nonumber
\end{align}
\begin{align}
\rho(p) = (1+\alpha^2)\left\{\frac{2\omega/v}{p^2+(2\omega/v)^2} + \left[\frac{\omega/v}{\left[p+(p_F^-+p_F^+)\right]^2+(2\omega/v)^2} + \frac{\omega/v}{(\left[p-(p_F^-+p_F^+)\right]^2+(2\omega/v)^2}\right]\right\} + \\ \nonumber
+ \alpha \left\{\frac{p+(p_F^-+p_F^+)}{\left[ p+(p_F^-+p_F^+)\right]^2+(2\omega/v)^2} - \frac{p-(p_F^-+p_F^+)}{\left[p-(p_F^-+p_F^+)\right]^2+(2\omega/v)^2} \right\} \phantom{aaaaaaaaaaaaaaaaaaaaaaaaaaaa} \nonumber
\end{align}
\subsection{x-impurity}
\begin{align}
\Phi_{\bar 1} (x) = +\frac{J_x}{2}\sum\limits_{\sigma}
	\bpm
		(1+i\sigma)  \left[ \left(E_{\bar 1}-\Delta_s\right)X_0^\sigma(x) + X_1^\sigma(x) \right] \\
		(1-i\sigma)  \left[ \left(E_{\bar 1}-\Delta_s\right)X_0^\sigma(x) + X_1^\sigma(x) \right] \\
		-(1+i\sigma)  \left[ \left(E_{\bar 1}-\Delta_s\right)X_0^\sigma(x) - X_1^\sigma(x) \right] \\
		-(1-i\sigma)  \left[ \left(E_{\bar 1}-\Delta_s\right)X_0^\sigma(x) - X_1^\sigma(x) \right]
	\epm,
\Phi_{1} (x) = +\frac{J_x}{2}\sum\limits_{\sigma}
	\bpm
		-(1-i\sigma)  \left[ \left(E_{1}+\Delta_s\right)X_0^\sigma(x) + X_1^\sigma(x) \right] \\
		(1+i\sigma)  \left[ \left(E_{1}+\Delta_s\right)X_0^\sigma(x) + X_1^\sigma(x) \right] \\
		-(1-i\sigma)  \left[ \left(E_{1}+\Delta_s\right)X_0^\sigma(x) - X_1^\sigma(x) \right] \\
		(1+i\sigma)  \left[ \left(E_{1}+\Delta_s\right)X_0^\sigma(x) - X_1^\sigma(x) \right]
	\epm. 
\end{align}
Using these expressions we can compute the non-polarized and SP LDOS in both coordinate and momentum space. We perform the calculation for the positive-energy state, and we find, omitting index $1$:
\begin{align}
S_x(x) &= -\frac{1+\alpha^2}{2} [2\cos p_\lambda x+\cos (2mv |x|+p_\lambda x-2\theta)+\cos (2mv |x|-p_\lambda x-2\theta)] \cdot e^{-2\omega |x|/v} \\
S_y(x) &= 0 \\
S_z(x) &= -\frac{1+\alpha^2}{2} [2\sin p_\lambda x+\sin (2mv |x|+p_\lambda x-2\theta)-\sin (2mv |x|-p_\lambda x-2\theta)] \cdot e^{-2\omega |x|/v} \\
\rho(x) &= (1+\alpha^2) [1+\cos (2mv |x|-2\theta)] \cdot e^{-2\omega |x|/v}
\end{align}
where $\tan \theta = \alpha$. Momentum space dependence can be derived from the $z$-impurity expressions since everything coincides up to coefficients.

\subsection{y-impurity}
\begin{align}
\Phi_{\bar 1} (x) = +\frac{J_y}{2}\sum\limits_{\sigma}
	\bpm
		(1-\sigma)  \left[ \left(E_{\bar 1}-\Delta_s\right)X_0^\sigma(x) + X_1^\sigma(x) \right] \\
		i(1-\sigma)  \left[ \left(E_{\bar 1}-\Delta_s\right)X_0^\sigma(x) + X_1^\sigma(x) \right] \\
		-(1-\sigma)  \left[ \left(E_{\bar 1}-\Delta_s\right)X_0^\sigma(x) - X_1^\sigma(x) \right] \\
		-i(1-\sigma)  \left[ \left(E_{\bar 1}-\Delta_s\right)X_0^\sigma(x) - X_1^\sigma(x) \right]
	\epm,
\Phi_{1} (x) = +\frac{J_y}{2}\sum\limits_{\sigma}
	\bpm
		-(1+\sigma)  \left[ \left(E_{1}+\Delta_s\right)X_0^\sigma(x) + X_1^\sigma(x) \right] \\
		i(1+\sigma)  \left[ \left(E_{1}+\Delta_s\right)X_0^\sigma(x) + X_1^\sigma(x) \right] \\
		-(1+\sigma)  \left[ \left(E_{1}+\Delta_s\right)X_0^\sigma(x) - X_1^\sigma(x) \right] \\
		i(1+\sigma)  \left[ \left(E_{1}+\Delta_s\right)X_0^\sigma(x) - X_1^\sigma(x) \right]
	\epm, 
\end{align}
after summation over $\sigma$:
\begin{align}
\Phi_{\bar 1} (x) = +J_y 
	\bpm
		+ \left[ \left(E_{\bar 1}-\Delta_s\right)X_0^-(x) + X_1^-(x) \right] \\
		i \left[ \left(E_{\bar 1}-\Delta_s\right)X_0^-(x) + X_1^-(x) \right] \\
		- \left[ \left(E_{\bar 1}-\Delta_s\right)X_0^-(x) - X_1^-(x) \right] \\
		-i \left[ \left(E_{\bar 1}-\Delta_s\right)X_0^-(x) - X_1^-(x) \right]
	\epm,
\Phi_{1} (x) = +J_y 
	\bpm
		- \left[ \left(E_{1}+\Delta_s\right)X_0^+(x) + X_1^+(x) \right] \\
		i \left[ \left(E_{1}+\Delta_s\right)X_0^+(x) + X_1^+(x) \right] \\
		- \left[ \left(E_{1}+\Delta_s\right)X_0^+(x) - X_1^+(x) \right] \\
		i \left[ \left(E_{1}+\Delta_s\right)X_0^+(x) - X_1^+(x) \right]
	\epm.
\end{align}
Using these expressions we can compute the non-polarized and SP LDOS in coordinate space
\begin{align}
S_x(x) &= 0, \\
S_y(x) &= -(1+\alpha^2)[1+ \cos(2 mv |x|-2\theta)] \cdot e^{-2\omega |x|/v}, \\
S_z(x) &= 0, \\
\rho(x) &= +(1+\alpha^2)[1+ \cos(2 mv |x|-2\theta)]  \cdot e^{-2\omega |x|/v}.
\end{align}

\section{The SPDOS for a non-superconducting one-dimensional system in the presence of a magnetic impurity}
The low-energy Hamiltonian in the non-SC regime can be written as
\begin{align}
H_0 = \xi_p \sigma_0 + \lambda (p_y \sigma_x - p_x \sigma_y) = \bpm \xi_p & i \lambda p \\ -i \lambda p & \xi_p \epm
\end{align}
where $\xi_p = \frac{p^2}{2m}-\varepsilon_F$. The corresponding spectrum is given by $\mathcal{E} = \xi_p \pm \lambda p$ and the retarded Green's function reads
\begin{align}
G_0(E,\bs p) = \frac{1}{(E-\xi_p+i0)^2-\lambda^2p^2} \bpm E-\xi_p+i0 & i \lambda p \\ -i \lambda p & E-\xi_p+i0 \epm.
\end{align}
To compute the eigenvalues for a single localized impurity we calculate
\begin{align}
G_0(E,x = 0) = \int \frac{dp}{2\pi}\frac{E-\xi_p+i0}{(E-\xi_p+i0)^2-\lambda^2p^2} \bpm 1 & 0 \\ 0 & 1 \epm = \frac{1}{2} \sum\limits_\sigma \int \frac{dp}{2\pi} \frac{1}{E-\xi_\sigma+i0} \bpm 1 & 0 \\ 0 & 1 \epm,
\end{align}
where $\xi_\sigma = \xi_p + \sigma \lambda p$. For $p>0$ we linearize the spectrum around the Fermi momenta, thus:
$$
\xi_\sigma \approx \left(\frac{p_F^\sigma}{m}+\sigma \lambda \right)(p-p_F^\sigma) = \sqrt{\lambda^2+2\varepsilon_F/m}\;(p-p_F^\sigma) \equiv v (p-p_F^\sigma),
$$
where $p_F^\sigma = m \left[-\sigma \lambda + v \right]$, and thus we get:
$$
\int \frac{dp}{2\pi} \frac{1}{E-\xi_\sigma+i0} \approx \frac{1}{2\pi v} \left[ \int \frac{d\xi_\sigma}{E-\xi_\sigma+i0} +  \int \frac{d\xi_{-\sigma}}{E-\xi_{-\sigma}+i0} \right] = -\frac{i}{v}
$$
This leads to:
\begin{align}
G_0(E,x = 0) = \frac{1}{2} \sum\limits_\sigma \left( -\frac{i}{v} \right)\bpm 1 & 0 \\ 0 & 1 \epm = -\frac{i}{v} \bpm 1 & 0 \\ 0 & 1 \epm
\end{align}
Since there is no energy dependence, there will be no impurity-induced states. The Green's function coordinate dependence is given by the following expression:
\begin{align}
G_0(E,x) = \frac{1}{2} \sum\limits_\sigma \int \frac{dp}{2\pi} \frac{e^{ipx}}{E-\xi_\sigma+i0} \bpm 1 & i\sigma \\ -i\sigma & 1 \epm
\end{align}
To find the coordinate dependence of the Green's function we calculate:
\begin{align}
X_0^\sigma(x) = \int \frac{dp}{2\pi} \frac{e^{ipx}}{E-\xi_\sigma+i0}
\end{align}
\subsection*{Integral calculation}
Below we use the Sokhotsky formula $\frac{1}{x+i0} = \mathcal{P}\frac{1}{x} - i\pi \delta(x)$:
\begin{align*}
X_0^\sigma(x) &= \int \frac{dp}{2\pi} \frac{e^{ipx}}{E-\xi_\sigma+i0} =  \frac{1}{2\pi v} \left[e^{i p_F^\sigma x} \int d\xi_\sigma \frac{e^{i\xi_\sigma x/v}}{E-\xi_\sigma +i0} + e^{-i p_F^{-\sigma} x} \int d\xi_{-\sigma} \frac{e^{-i\xi_{-\sigma} x/v}}{E-\xi_{-\sigma} +i0}\right] 
\end{align*}
We compute explicitly only one of the integrals in the brackets since the other one can be computed in the similar fashion:
\begin{align*}
\int d\xi_\sigma \frac{e^{i\xi_\sigma x/v}}{E-\xi_\sigma +i0} = \mathcal{P}\negthickspace\int\negthickspace d\xi_\sigma \frac{e^{i\xi_\sigma x/v}}{E-\xi_\sigma} - i\pi \negthickspace \int\negthickspace d\xi_\sigma \delta(E-\xi_\sigma) e^{i\xi_\sigma x/v} = -i\pi \left(1+\sgn x\right) e^{i Ex/v}
\end{align*}
Finally we have:
\begin{align}
X_0^\sigma(x) = -\frac{i}{v} \exp \left[ i\left( mv+\frac{E}{v} \right) |x| \right] e^{-i\sigma m \lambda x},\phantom{aaaaaaaaaaaaaaaaaaaaaaaaaaaaaaaaaaaaaaaa}
\end{align}
and the Green's function can be written as:
\begin{align}
G_0(E,x) = \frac{1}{2} \sum\limits_\sigma 
	\begin{pmatrix}
		1 & i\sigma \\
		-i\sigma  & 1
	\end{pmatrix} X_0^\sigma(x).
\end{align}
Below we compute the T-matrix for different types of impurities. Impurity potentials  take the following forms:
\begin{align}
V_{sc} = U \bpm 1 & 0 \\ 0 & 1 \epm, \quad V_z = J_z \bpm 1 & 0 \\ 0 & -1 \epm, \quad V_x = J_x \bpm 0 & 1 \\ 1 & 0 \epm
\end{align}
The corresponding T-matrices are
\begin{align}
T_{sc} = \frac{U}{1+i U/v} \bpm 1 & 0 \\ 0 & 1 \epm, \quad T_z = \bpm \frac{J}{1+iJ/v} & 0 \\ 0 & -\frac{J}{1-i J/v}\epm, \quad \quad T_x = \frac{J}{1+J^2/v^2} \bpm -i J/v & 1 \\ 1 & -i J/v \epm
\end{align}
For each type of impurity we can compute the non-polarized and SP LDOS using Eq. (\ref{DGapp}) and Eqs. (\ref{SPLDOSapp}) where we replace $\bm r$ by $x$. By taking the Fourier transforms of the expressions above we get the the momentum space dependence. Below we denote $\alpha = J/v$.
\subsection{z-impurity}
\begin{align}
S_x(x) &= +\frac{\alpha}{1+\alpha^2} \cdot \frac{1}{\pi v} \left[ \cos (p_\varepsilon |x| - p_\lambda x) -\cos (p_\varepsilon |x| + p_\lambda x) \right] \\
S_y(x) &= 0 \\
S_z(x) &= +\frac{\alpha}{1+\alpha^2} \cdot \frac{1}{\pi v}  \left[ \sin (p_\varepsilon |x| - p_\lambda x ) +\sin (p_\varepsilon |x| + p_\lambda x ) \right] \\
\rho(x) &= -\frac{2\alpha^2}{1+\alpha^2} \cdot \frac{1}{\pi v} \cos p_\varepsilon x 
\end{align}
where we denote $p_\varepsilon = 2\left(mv+E/v\right), p_\lambda = 2m\lambda$. After taking the Fourier transform we get:
\begin{align}
S_x(p) &= +\frac{\alpha}{1+\alpha^2} \cdot \frac{i}{\pi v} \left[\frac{1}{p+p_\varepsilon +p_\lambda} - \frac{1}{p+p_\varepsilon -p_\lambda} - \frac{1}{p-p_\varepsilon +p_\lambda} + \frac{1}{p-p_\varepsilon -p_\lambda}\right] \\
 S_y(p) &= 0 \\
 S_z(p) &= +\frac{\alpha}{1+\alpha^2} \cdot \frac{1}{\pi v} \left[\frac{1}{p+p_\varepsilon +p_\lambda} + \frac{1}{p+p_\varepsilon -p_\lambda} - \frac{1}{p-p_\varepsilon +p_\lambda} - \frac{1}{p-p_\varepsilon -p_\lambda}\right] \\
 \rho(p) &= -\frac{2\alpha^2}{1+\alpha^2} \cdot \frac{1}{v} \left[\delta(p-p_\varepsilon) + \delta(p+p_\varepsilon) \right]
\end{align}
\subsection{x-impurity}
\begin{align}
S_x(x) &= +\frac{\alpha}{1+\alpha^2} \cdot \frac{1}{\pi v}  \left[ \sin (p_\varepsilon |x| - p_\lambda x ) +\sin (p_\varepsilon |x| + p_\lambda x ) \right] \\
S_y(x) &= 0 \\
S_z(x) &= -\frac{\alpha}{1+\alpha^2} \cdot \frac{1}{\pi v} \left[ \cos (p_\varepsilon |x| - p_\lambda x) -\cos (p_\varepsilon |x| + p_\lambda x) \right] \\
\rho(x) &= -\frac{2\alpha^2}{1+\alpha^2} \cdot \frac{1}{\pi v} \cos p_\varepsilon x 
\end{align}
We do not give the Fourier transform for these expressions since they coincide with the ones for a $z$-impurity if we exchange $S_z$ and $S_x$ and change the overall sign.
\subsection{y-impurity}
\begin{align}
S_x(x) &= S_z(x) = 0 \\
S_y(x) &= +\frac{2\alpha}{1+\alpha^2} \cdot \frac{1}{\pi v} \sin p_\varepsilon |x| \\
\rho(x) &= -\frac{2\alpha^2}{1+\alpha^2} \cdot \frac{1}{\pi v} \cos p_\varepsilon x
\end{align}
The corresponding Fourier transform is:
\begin{align}
S_y(p) = \frac{2\alpha}{1+\alpha^2} \cdot \frac{1}{\pi v} \left[ \frac{1}{p+p_\varepsilon} - \frac{1}{p-p_\varepsilon}\right]
\end{align}

\end{document}